\documentclass[aps,rmp,twocolumn]{revtex4}
\usepackage{comment,psfrag,amsmath,appendix,amssymb,dcolumn,verbatim,amsmath,
graphicx,epsfig,bm,epsf,float}

\begin{document}

\title{Vortices in Superfluid Films on Curved Surfaces}

\author{Ari M. Turner$^{*\dagger}$, Vincenzo Vitelli$^{\S}$}
\author{David R. Nelson$^*$}%
\affiliation{%
$^*$Department of Physics, Harvard University, Cambridge MA 02138}
\affiliation{%
$^{\dagger}$Department of Physics, University of California, Berkeley CA 94720}
\affiliation{%
$^{\S}$Department of Physics and Astronomy, University of Pennsylvania,
Philadelphia PA 19104}

\date{\today}
\begin{abstract}We present a systematic study of how vortices in superfluid films interact with the spatially varying Gaussian curvature of the underlying substrate.  The Gaussian curvature acts as a source for a geometric potential that attracts (repels) vortices towards regions of negative
(positive) Gaussian curvature independently of the sign of their topological charge.  Various experimental tests involving rotating superfluid films and vortex pinning are first discussed for films coating gently curved substrates that can be treated in perturbation theory from flatness. An estimate of the experimental regimes of interest
is obtained by comparing the strength of the geometrical forces to the vortex pinning induced by
the varying thickness of the film which is in turn caused by capillary effects and gravity. 
We then present a non-perturbative technique based on conformal mappings that leads an exact solution for the geometric potential as well as the geometric correction to the interaction between vortices. 
The conformal mapping approach is illustrated by means of explicit calculations of the geometric effects encountered in the study of some strongly curved surfaces and by deriving universal bounds on their strength.  
\end{abstract}
\pacs{1729}
\maketitle
\tableofcontents

\section{\label{sec:intr}Introduction}

In superfluid helium, vortices form when the helium is rotated rapidly or
when there is turbulence \cite{Vine,tilleybook}.  Though such vortices are similar
to the vortices that make up a vortex street behind the wings of an airplane
or to the funnel clouds of tornadoes, they are only an Angstrom or two across \cite{Guyon-book}.
A more essential difference is that the vortices in a superfluid do not
need a constant source of energy to survive.  In fact, a vortex is long-lived
because the strength of its flow is fixed by the quantization of angular momentum. Thus, the dissipative mechanisms of a conventional fluid are absent. 

In this article, we focus on forces that the vortices experience as a result of
geometric constraints, with an emphasis on those encountered in
thin layers of liquid helium wetting a \emph{curved} 
substrate with spatially varying Gaussian curvature. As a result of the 
broken translational invariance of the underlying curved space, the energy of a single vortex with circulation quantum number $n_{i}$ 
at position ${\bf u}_{i}$ includes both a divergent term
and a position dependent self-energy, $E_s({\bf u}_i)$, given by \cite{swiss}
\begin{equation}
E_s({\bf u}_i)=- \pi K n_i^2 U_G({\bf u}_i) ,
\label{eq:ecurv-s2bis}
\end{equation}
where $K=\frac{\rho_s \hbar^2}{m^2}$ is the superfluid stiffness expressed in terms of the $^4$He atomic mass, $m$, and the superfluid mass density, $\rho_s$. 
The potential $U_G({\bf u}_i)$
is obtained from solving a covariant Poisson equation with the Gaussian curvature, $G(\mathbf{u_i})$,
acting as a source
\begin{equation}
\nabla^2 U_G(\mathbf{u_i}) = G(\mathbf{u_i}).
\label{eq:geompdeq}
\end{equation}
Vortices (and anti-vortices) are attracted (repelled) to regions of negative (positive) Gaussian curvature. 
These geometric interactions, while more exotic, are similar in origin to boundary-vortex interactions and 
can be suitably treated by the method of conformal mapping. 

Similar ideas naturally arise in a variety of soft-matter systems which have been confined 
in a thin layer wetting a curved substrate. The specific form of the resulting geometric interactions
depends on the symmetry of the order parameter as well as on the shape of the substrate. Examples that have been studied both theoretically and experimentally include 
colloidal crystals on curved interfaces \cite{bowickB,VitelliLucks,Bausch03}, columnar phases of block co-polymers \cite{Santangelo07, Alex-thesis} as well as thin layers of nematic liquid crystals \cite{park,geomgenerate,Vitelli06,Fernandez-Nieves07}. Fueled by the drive towards technological applications based on the notion of self-assembly directed by geometry \cite{Dinsmore02,Nelson02,DeVries07}, the study of these $frustrated$ materials aims at predicting how the non-uniform distribution of curvature of the underlying substrates induces an inhomogeneous phase in the curved monolayer. An understanding of the resulting macroscopic properties can be built from a mesoscopic description cast in terms of the energetics of the topological defects which often exist even in the ground state and play a crucial role in determining how the material melts or ruptures. The advantage of this approach stems from the huge reduction in degrees of freedom achieved upon re-expressing the energy stored in the elastic field in terms of a few topological defects, rather than keeping track of the state of all the microscopic components, e.g. individual particle positions or molecular orientations. This step allows one to carry out efficient computational studies \cite{Hexemer07,bowickB} and provides a suitable starting point for analytical work in the form of effective free energies derived from continuum elastic theory. 

Many of the mathematical techniques employed in this article to study vortices in curved superfluid films find application in the soft matter domain, in particular in those contexts where bond-orientational order is important \cite{swiss}. In flat space both superfluid and liquid crystal films can be described, as a first approximation, by an XY model\cite{Nels-Kost}. 
Both liquid crystal disclinations and vortices are modeled as a Coulomb gas of charged particles interacting logarithmically. 
However, the quantum nature of the problem considered in the present work introduces a fundamental difference between these two classes of systems that is best illustrated by contrasting the angle of the liquid crystal director with the phase of the superfluid's collective wave function. The former represents the orientation of a vector (with both ends identified in the case of a nematic) that lives in the tangent space of the surface while the latter is a quantum mechanical object that transforms like a scalar since it is defined in an internal space. 
This subtle difference resurfaces upon considering the distinct curved-space generalizations of the XY model that apply to each of these two systems.  

The free energy functional ${\cal
F}_{v}$ to be minimized for the case of orientational order on a surface with points labeled by the coordinates ${\bf u}=(u_1,u_2)$ reads
\cite{Davidreview}:
\begin{equation}
{\cal F}_{v} = \frac{K}{2}\int d^{2}u\sqrt{g} g^{\alpha\beta}
(\partial_{\alpha}\theta({\bf u}) - \Omega_{\alpha}({\bf
u}))(\partial_{\beta}\theta({\bf u}) - \Omega_{\beta}({\bf
u}))\!\!\!\! \quad , \label{eq:patic-ener}
\end{equation}
where $g_{\alpha\beta}$ and $g$ indicate the metric tensor and its determinant while $\Omega_{\alpha}({\bf u})$ is a connection that compensates for
the rotation of the 2D basis vectors ${\bf E}_{\alpha}({\bf u})$ (with respect to which $\theta({\bf u})$ is measured)
in the direction of $u_{\alpha}$ \cite{Kami02}.  Since the curl of the field
$\Omega_{\alpha}({\bf u})$ is equal to the Gaussian curvature $G({\bf u})$ \cite{Davidreview},
the integrand in Eq. (\ref{eq:patic-ener}) never vanishes
because $\Omega_{\alpha}({\bf u}) \ne \partial_{\alpha}\theta$ on a surface with $G({\bf u})\ne0$. 
As the
substrate becomes more curved, the resulting energy cost of geometric
frustration can be lowered by generating disclination-dipoles in the ground
state even in the absence of topological constraints. 

The connection $\Omega_{\alpha}({\bf u})$ is a geometric gauge field akin to the electromagnetic
vector potential, with the Gaussian curvature playing the role of a magnetic field. If topological defects are present,
they appear as monopoles in the singular part of $\partial_{\alpha}\theta({\bf u})$. In analogy with electromagnetic theory, their interaction with the Gaussian curvature arises mathematically from the cross-products between $\partial_{\alpha}\theta({\bf u})$ and the geometry 
induced vector potential $\Omega_{\alpha}({\bf u})$, see Eq. (\ref{eq:patic-ener}). As a result of this interaction, disclinations in a liquid crystal
are attracted to regions of the substrate whose curvature
has the same sign as the defect's topological charge \cite{park}, whereas 
vortices in a superfluid favor negatively curved regions independently of their
sense of circulation. The anomalous coupling between vortices and Gaussian curvature introduced in Equations (\ref{eq:ecurv-s2bis}) and (\ref{eq:geompdeq})
originates from the distortion of the flow pattern by the protrusions and wrinkles
of the surface. 

For a disclination with topological index $n_{i}$ (defined by the amount $\theta$ increases along a
path enclosing the defect's core) the geometric potential $E_{v}({\bf u}_{i})$ reads \cite{swiss}
\begin{equation}
E_{v}({\bf u}_{i})= 2 \pi K \!\!\!\!\! \quad n_{i}
\left(1-\frac{n_{i}}{2 } \right) \!\!\!\!\! \quad U_G({\bf
u}_{i}) \!\!\!\! \quad , \label{eq:ecurv-v}
\end{equation}
where $K$ is the elastic stiffness and $U_G$ is the same potential defined in Eq. (\ref{eq:geompdeq}). 
Note that the anomalous coupling also contributes to determine the energetics of liquid crystal disclinations but, in this case,
the gauge coupling, which is linear in $n$, is stronger for small $n$. 

To understand the physical and mathematical origin of these distinct coupling mechanisms, note
that in the ground state of a $^4$He film, the phase $\theta({\bf u})$ can be
constant throughout the surface so that the corresponding energy
vanishes. In a system with geometric frustration, the gauge coupling between defects and the underlying curvature is mediated by the deformed ground state texture that exists in the liquid crystal layer prior to the introduction of the defects simply as a result of geometrical constraints.  Once a defect is introduced it interacts with these preexisting elastic deformations.  Unlike the case of orientational order considered previously, no geometric frustration exists in the superfluid film. 
The superfluid free energy ${\cal F}_{s}$ to be minimized is a simple
scalar generalization of the familiar flat space counterpart
\begin{equation} {\cal F}_{s} = \frac{K}{2}\int
d^{2}u \!\!\!\!\!\! \quad \sqrt{g} \!\!\!\!\! \quad
g^{\alpha\beta}
\partial_{\alpha}\theta({\bf u}) \!\!\!\!\! \quad \partial_{\beta}\theta({\bf u}) \!\!\!\! \quad .
\label{eq:supfl-I}
\end{equation}
The crucial point is that no connection $\Omega_{\alpha}({\bf u})$ is necessary to write down the covariant derivative for this simpler case of a scalar order parameter.  Therefore the ground state is given by $\theta(\mathbf{x})$ equal to a constant.  There is no preexisting texture for a vortex to
interact with, and so another mechanism is required to explain the coupling of vortices to geometry. 

In the following sections, we will employ the method of conformal mapping to demonstrate that when an isolated vortex is placed on a curved surface it feels
a force as if there were a smeared out topological ``image charge," 
jointly proportional to the vortex's own circulation and 
the Gaussian curvature across the substrate. Such an imaginary topological 
charge distribution produces a \emph{real} force analogous to the
force on an electrostatic charge due to its mirror image 
in a conducting surface. 

The method of conformal mapping may seem, prima facie, a surprising route to derive a coupling between vortices and geometry, because the free energy of Eq. (\ref{eq:supfl-I}) is invariant under conformal transformations that introduce a non-uniform compression of the surface while keeping local angles unchanged.   
This
invariance property at first seems to rule out the possibility of a geometrical
interaction!  This apparent contradiction
can be seen by choosing a special set of (isothermal) coordinates that can always bring the two dimensional metric tensor in the diagonal form $g_{\alpha \beta}({\bf u})= e^{2\omega({\bf u})}\delta_{\alpha \beta}({\bf u})$ \cite{Davidreview}. The result of this step is to eliminate
the geometry dependence from the free energy of Eq. (\ref{eq:supfl-I}) since the product $g^{\alpha \beta}({\bf u}) \sqrt{g}=\delta_{\alpha \beta}({\bf u})$ and ${\cal F}_{s}$ reduces to its counterpart for a planar surface, where there
is of course no geometry dependence.

An interaction between vortices and geometry
violates this conformal symmetry 
of the free energy from which it emerges, but in fact the conformal symmetry
is not an exact symmetry when vortices are present.
Analogous subtleties frequently arise in the study of fields that fluctuate
thermally or quantum mechanically, due to the occurrence
of a cut-off length scale below which fluctuations cannot 
occur.  A conformal mapping is a strange type of symmetry that stretches
lengths
and thus does not preserve the microscopic structure of a system. At finite
temperatures, the discreteness of a system, such as a thermally
fluctuating membrane \cite{wends} which is actually made up of a network
of molecules, can have an important effect because the fluctuations
excite modes with microscopic wavelengths. This produces violations
of the conformal symmetry at every point of the surface.
In a superfluid at zero temperature, however, short wavelengths not describable
by the continuum free energy ${\cal F}_{s}$ are excited only in
the cores of vortices.  Obtaining a finite value for
the energy necessitates the removal of
vortex cores of a certain fixed radius in the local tangent plane, 
so a conformal mapping
is not a symmetry in the neighborhood of a vortex.  However the amount by
which this symmetry fails can be calculated in a simple form (intriguingly
independent of the microscopic model of the cores) in terms
of the rescaling function $\omega(\mathbf{u})$ 
evaluated at the locations of the vortices, 
where the symmetry fails.
Rather than ruling out the possiblity of
a geometric interaction, a realistic
treatment of conformal mapping becomes
 a powerful mathematical tool for deriving these interactions, a technique
which is relevant especially to other branches of theoretical physics such as the
study of scattering amplitudes in string theory \cite{threedee}.

While the free energy, ${\cal F}_{s}$, of the curved superfluid layer in Eq. (\ref{eq:supfl-I}) does not exhibit a 
\emph{geometric} gauge field, rotating the sample at a constant angular
velocity leads
to an energy of the same form as Eq. (\ref{eq:patic-ener}).
The resulting forces exerted on the vortices compete with the geometric interactions to determine the 
equilibrium configurations of an arrangement of topological defects. This simple idea is behind some of the experimental suggestions
put forward in this article to map out the geometric potential by progressively increasing the rotational speed while monitoring the equilibrium position
of say a single vortex on a helium coated surface shaped like the bottom of a wine bottle \cite{zieve}. Since the position dependence of the force induced by the rotation is easily calculated, one can read off the geometric interaction by simply assuming force balance. 

\begin{figure}
\includegraphics[width=0.45\textwidth]{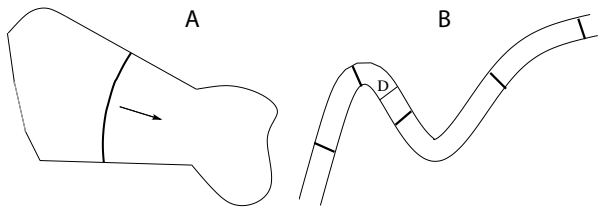}
\caption{\label{fig:slickslopes}
Cross sections of two regions in which one can study vortex energetics.
A) A region with nonparallel boundaries. The vortex is pushed
to the right.  This tendency 
can be interpreted either in terms of a drive toward
a shorter length or as the local induction force due to the curvature in
the vortex enforced by the boundaries.  B) A cross-section
of a constant thickness layer of helium bounded above by air and below by
the substrate.  The vortices keep a constant length and remain straight
while moving around.  Hence there is no local induction 
force/thickness-variation
force to overwhelm the geometrical forces that we focus on.}
\end{figure}

The theory of curved helium films also helps build intuition
for the more
general case of vortex lines confined in a bounded three dimensional region such as
the cavity shown in cross-section in Fig. \ref{fig:slickslopes}A.
The vortex, drawn as a bold black line, can be pinned by the constriction of
the container. The classic problem of understanding 
the interaction of the vortex
with itself and with the bump as the superfluid flows past \cite{vortexhill}, is of crucial importance in 
elucidating how vortices can be produced when a superfluid starts
rotating despite the absence of any friction. A possible mechanism, known as the ``vortex mill" , assumes that vortex rings break
off a pinned vortex line, while the pinned vortex remains in place
\cite{vortexmill,workingmill}.
The common route to studying vortex dynamics in three-dimensional geometries is
the ``local induction approximation" which
assumes that each element of a vortex
experiences a force determined only by its local radius of curvature. This simplifying assumption 
omits any long range forces experienced by the vortices as they interact with the boundaries (or among themselves).
In the opposite limit of films with uniform thickness (which can be
thought of as special types of bulk superfluid regions with two parallel boundaries,
as in Fig. \ref{fig:slickslopes}B), $all$ the forces exerted on the vortices are long-range. This is the regime of interest to our 
investigation.  

This article is organized in two tracks. The first, comprising sections \ref{subsec:effective}$-$\ref{sec:experimental},
is phenomenological in nature and emphasizes intuitive analogies between the
(non-linear) geometric forces and conventional electrostatics, simple illustrations of the main results  and 
experimental ideas. The second track, sections \ref{sec:complex}$-$\ref{sec:geomineq} is more technical and presents
a unified derivation of the geometric potential
by the method of conformal mapping and its application to the study of complex surface morphologies.

The first track starts with a review of superfluid dynamics 
that can be used to relate
the anomalous coupling to hydrodynamic lift.  
In Sec. \ref{subsec:anomalous}, the geometrical force
is evaluated, using a mapping between the geometric potential studied here and the familiar Newton's theorem that allows an efficient calculation of the gravitational field for a spherically symmetric mass distribution. An intriguing consequence of this analogy is that vortices on saddle
surfaces  can be trapped in regions of negative curvature leading to geometrically
confined persistent currents as discussed in section \ref{BS}.  Section 
\ref{earnshaw} relates this observation
to Earnshaw's theorem from electrostatics.
Upon heating and subsequently cooling a curved superfluid film, some of the
thermally generated defects can remain trapped in metastable states located at the saddles of the substrate. The existence of such geometry 
induced vortex hysteresis is conjectured in section \ref{subsec:cande}. 
In section \ref{subsec:Rotation}, we derive the forces experienced by vortices when the vessel containing the superfluid layer is rotated
around the axis of symmetry of a curved surface shaped as a Gaussian bump. The dependence of single and multiple defect-configurations on different angular speeds and aspect ratios of the bump is studied in sections \ref{subsec:single} and \ref{subsec:multiple}. The Abrikosov lattice of vortices on a curved surface is discussed in section \ref{subsec:abr}. In realistic experimental situations the thickness of the superfluid layer will not be uniform and additional forces will drive vortices towards thinner regions of the sample. The strength of these forces is assessed in section \ref{sec:experimental} and related to spatial variations of the film thickness due to gravity and surface tension.
A short discussion of choice of parameters for the proposed
rotation experiments follows in section \ref{subsec:nucleation}. The relevance of our discussion to experiments performed in bounded three dimensional samples is addressed in section 
\ref{sec:unevenexptl}.

The second track starts with a general derivation of the geometric potential
by the method of conformal mapping in section \ref{sec:map}. The computational efficiency of this approach is illustrated in section \ref{sec:bubble} where
the geometric potential of a vortex is evaluated on an Enneper disk, a strongly deformed minimal surface. 
We show that 
changing the geometry of the substrate has interesting effects not only on the one-body geometric potential but also on the two-body interaction between vortices. In section 
\ref{sec:bumps}
we use conformal methods to show how a periodic lattice of bumps can cause
the vortex interaction to become anisotropic.
In section \ref{sec:zucchini}, we demonstrate that  
the quantization of circulation
leads to an extremely long-range force on an elongated surface with the topology of a sphere. The interaction energy is no longer logarithmic, but now grows linearly with the distance 
between the two vortices. As we demonstrate,
the whole notion of splitting the energy in a one body geometric potential and a vortex-vortex interaction is subject to ambiguities on deformed spheres. Section \ref{sec:noodles} provides some guidance on how to perform calculations in this context by choosing a convenient Green's function among the several available. Finally,
in section \ref{sec:geomineq}, we present a discussion of some general upper bounds which constrain the strength of geometric forces.
The conclusion serves as a concise summary and contains a table designed to locate at a glance our main results throughout the article including the more technical points relegated 
to appendices but useful to perform calculations. 

\section{\label{subsec:effective}Fluid Dynamics and Vortex-Curvature Interactions}

We start by writing down the collective wave function of the superfluid as
\begin{equation}
\Psi(\mathbf{ u})=\sqrt{\frac{\rho_{s}(\mathbf{u})}{m_4}} \!\!\!
\quad e^{i\theta (\mathbf{u})} \!\!\!\! \quad,
\label{wave-function}
\end{equation}
where $\mathbf{u}=\{u_{1},u_{2}\}$ is a set of curvilinear coordinates for the
surface, $m$ is the mass of a $^{4}$He atom and $\rho_{s}$ is
the superfluid mass density that we shall assume to be constant in what
follows. To obtain the superfluid current we can use the standard
expression
$
j_{\alpha}(\mathbf{u})=\frac{i \hbar}{2 m_{4}} \left(\Psi
\partial_{\alpha}\Psi^{\ast}-\Psi^{\ast}\partial_{\alpha} \Psi \right) \!\!\!\!
\quad  $
showing that the superfluid velocity is given by 
\begin{equation}
v_{\alpha}(\mathbf{
u})=\frac{\hbar}{m_{4}}  \!\!\!\! \quad
\partial_{\alpha} \theta(\mathbf{u}).
\label{eq:velocity}
\end{equation}
The circulation along a path $C$ enclosing a vortex is given
by
\begin{equation}
\oint_{C}\! d u^{\alpha}  \!\!\!\!\!\! \quad v_{\alpha} = n \kappa
\!\!\!\! \quad , \label{quantization}
\end{equation}
where the quantum of circulation, $\kappa=\frac{h}{m_4}$, is equal
to 9.98 10$^{-8}$ $m^2$ $s^{-1}$ and the integer $n$ is the topological index
of the vortex. The free energy can be cast in the form
\begin{equation}
F = \frac{1}{2} \rho_s\frac{\hbar^2}{m_4^2}
\int_{S} d^{2}u \sqrt{g} \!\!\!\!\! \quad
g^{\alpha\beta}\partial_{\alpha}\theta\partial_{\beta}\theta \!\!\!\! \quad ,
\label{eq:free-energy}
\end{equation}
where $g^{\alpha\beta}$ is the (inverse) metric tensor describing the
surface on which the superfluid layer lies and $g$ is its
determinant. We will often use the superfluid stiffness 
\begin{equation}
K=\frac{\rho_s\hbar^2}{m^2}.\label{eq:stiffness}
\end{equation}

This expression for the free-energy can be parameterized in terms of the vortex positions once the seemingly divergent
kinetic energy near a vortex core is correctly accounted for.  
As is well known, the radius-independence of the circulation about a vortex
implies that the velocity in its proximity
is given by $\frac{\hbar}{m_4r}$, which leads to a logarithmic 
divergence in
(\ref{eq:free-energy}). The energy stored in an
annulus of internal radius $r_{\rm{in}}$ 
and outer radius $r_{\rm{out}}$ reads 
\begin{equation}
E_{\rm{near}}=\pi K\ln\frac{r_{\rm{out}}}{r_{\rm{in}}}.
\label{eq:log-div}
\end{equation}
which diverges as $r_{\rm{in}}\rightarrow 0$. A physical trait of superfluid
helium prevents this from happening: it cannot sustain speeds which are greater
than $v_c$, the critical velocity. Thus the superfluidity is destroyed
below a core radius of $a\sim\frac{\hbar}{m_4v_c}$. This breakdown may be modeled
by excising a disk of radius $a$ around each vortex and by adding a constant core energy
$\epsilon_c$ to account for the energy associated with 
the disruption of the superfluidity in the core.

Starting on the flat plane,
the interaction of two vortices can now be determined.
Superimposing the fields of the two vortices and integrating the cross-term
in the kinetic energy of Eq. (\ref{eq:free-energy})
leads to a Coulomb-like interaction,
$V_{ij}=-2\pi Kn_in_j\ln\frac{|\mathbf{u}_i-\mathbf{u}_j|}{a}$ in addition to vortex
self-energies. In deducing
the force between the vortices from this expression, it is useful to assume that
$a$ does not vary significantly with position. The justification for this simplification is that the 
background flow due to other vortices only gives a fractionally small correction
to the $\frac{1}{r}$ flow near each vortex, and therefore barely affects
where the critical velocity is attained. 

For the more complicated case of a \emph{curved} surface, with a very distant
boundary (see \cite{geomgenerate}
for the discussion of effects due to a boundary
at a finite distance), we found in Ref. \cite{swiss}
that the energy including both
single-particle and two-particle
interactions is,
\begin{equation}
\frac{E(\{q_i,\mathbf{u}_i\})}{K}=\sum_{i<j} 4\pi^2
n_in_j V_{ij}({\bf u}_i,{\bf u}_j)
+\sum_i \left(-\pi n_i^2U_G({\bf u}_i)\right),
\label{eq:particlemodel}
\end{equation}
apart from a position-independent term (given for a distant circular
boundary of radius $R$ by
$\pi(\sum_i n_i)^2\ln\frac{R}{a}+N\frac{\epsilon_c}{K}$, with
$N$ the total number of vortices and $\epsilon_c$ the
core energy of one of them). 
The pair potential $V_{ij}=\Gamma(\mathbf{u_i},\mathbf{u_j})$
is expressed in terms of $\Gamma$,
the Green's function of the covariant Laplacian defined by:
\begin{equation}
\nabla_u^2 \Gamma(\mathbf{u},\mathbf{v})=-\delta_c(\mathbf{u},\mathbf{v})
\label{eq:BasicGreen}
\end{equation}
Note that the covariant delta
function $\delta_c$
includes a factor of $\frac{1}{\sqrt{g}}$ so that its integral with
respect to the ``proper area" $\sqrt{g}du_1du_2$ is normalized. Equation (\ref{eq:BasicGreen})
determines the Green's function up to a constant provided that
we assume additionally that the Green's function is symmetric
between its two arguments. The constant is fixed by assuming
that at large separations, the Green's function approaches the Green's function
of an undeformed plane. This expression shows that vortices behave like
electrostatic particles, with charges given by $2\pi n_i$ and coupling constant
$K$.

The single-particle potential $U_G(\mathbf{u})$ is the ``geometric potential"
defined in Eq. (\ref{eq:geompdeq}).
This potential entails a repulsion $\nabla\pi K U_G$ of vortices of either sign from positive curvature and an attraction to negative curvature. The
following analogy with boundary interactions and image charges is useful.
  The flow-field
is modified by having to conform to the curvature, leading to an image
charge of the vortex which is spread continuously over the surface, with
density $-\frac{n}{2}G(\mathbf{u})$.  Just like the image of a vortex in a circular boundary has an equal and opposite circulation, the continuous image of the
vortex in the curvature has a charge density proportional to the number of
quanta $n$ in the vortex.

\begin{figure*}
\includegraphics[width=\textwidth]{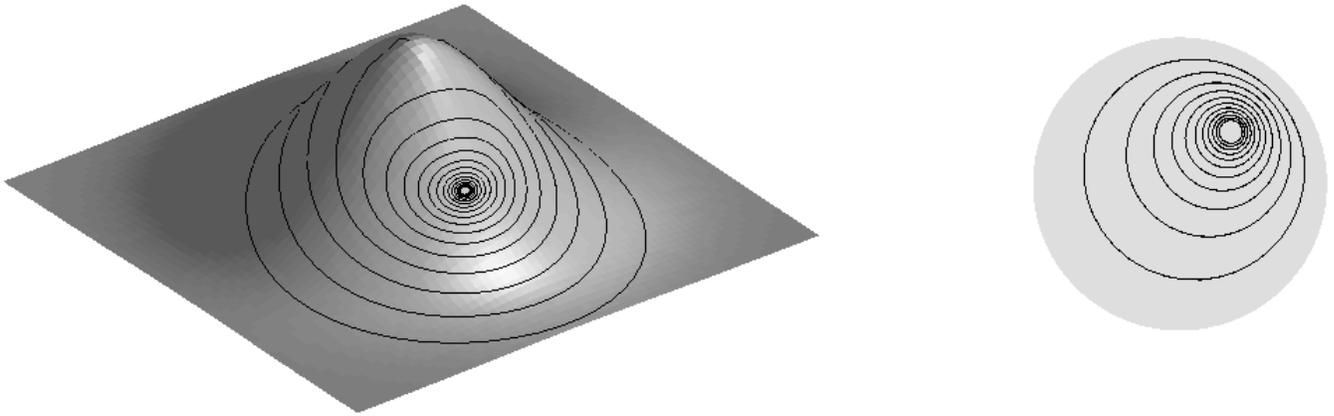}
\caption{\label{fig:volcano}
Left, the flow around a vortex situated on the side of a Gaussian bump,
calculated with the methods described in this paper. The low density of flow
lines above the vortex indicates a lower velocity and thus a higher pressure,
leading to the repulsion represented in Eq. (\ref{eq:particlemodel}),
$-\bf{\nabla} (-\pi K n^2U_G)$. Right, the analogous flow around a vortex in
a disk with a solid boundary. The attraction to the boundary is also seen
to result from high speeds, since the flow lines are compressed in between
the vortex and the boundary.}
\end{figure*}

This point of view may be connected to fluid mechanics by analyzing the streamlines on the bump and in the presence of a circular boundary as illustrated in Fig. \ref{fig:volcano}. Streamlines are tangent to the direction of flow \emph{and} their density is
proportional to the local speed.
Only an incompressible velocity field ($\mathrm{div}\ \mathbf{v}=0$)
may be described by
streamlines, since incompressibility ensures that
any closed curve has an equal number of streamlines
entering and exiting. This condition is satisfied for superfluids (far below the critical speed) since
 minimizing (\ref{eq:free-energy}) leads
to $\mathbf{\nabla}\cdot{\bf \nabla}\theta=0$, or $\mathrm{div}\ \mathbf{v}=0$ 
according to Eq. (\ref{eq:velocity}). Thus the flow is both irrotational (the
circulation around any curve, not enclosing a vortex, is $0$) and
incompressible:
\begin{eqnarray}
\mathrm{div}\ {\bf v}&=&0\nonumber\\
\mathrm{curl}\ {\bf v}&=&0.
\label{eq:veleqns}
\end{eqnarray}
The former relation implies that we may write 
\begin{equation}
{\bf v}=\mathrm{curl}\ \chi \hat{\mathbf{n}};
\label{eq:chidef}
\end{equation}
so that the streamlines are equally spaced level curves of $\chi$. (For example,
around a vortex, the radii of the successive streamlines forms a geometric sequence, 
$r(1-\epsilon)^l$ where $\epsilon$ sets the ratio 
between flowline density and speed.)

Now consider, as an illustration, the flow field on the slope of the bump represented in  
Fig. \ref{fig:volcano}. 
The curves must spread out to go over
the bump, leading to a lower velocity above the vortex. By Bernoulli's
principle (true for irrotational flows), 
this creates a high pressure that pushes the vortex away from
the bump. Note, however, that the actual motion of a vortex is 
more subtle:  Although the gradient of the energy points away from the bump, a vortex (disregarding friction)
always moves at right angles to the gradient of the energy. Thus a
vortex in the absence of drag forces
actually circles around the bump, in the same direction as the fluid
flows around the vortex. Dissipation converts the motion into 
an outward spiral\cite{nelambhalsig}.  We will not study the dynamics.

A convenient mathematical formulation of the problem of determining the flow
pattern of a collection of vortices is obtained by introducing
the scalar function $\chi(\mathbf{u})$ which satisfies
\begin{equation}
\nabla^2\chi(\mathbf{u})=-\sum_i 2\pi n_i \delta_c(\mathbf{u},\mathbf{u_i})
\equiv -\sigma(\mathbf{u}).
\label{eq:flowdeq}
\end{equation}
The sum can be described as a singular distribution of surface charge.
This relation follows from the circulation condition, 
$2\pi n_i=\oint \nabla \theta\cdot\mathbf{dl}$, which can be rewritten
as the integral of the flux of $\nabla\chi$ through the boundary,
 $\oint \nabla\chi\cdot\mathbf{\hat{n}}dl$,
 by 
using Eq. (\ref{eq:chidef}).  
In analogy with Gauss's law, there must 
therefore be delta-function sources for $\chi$ at 
the locations of the vortices, as described by Eq. (\ref{eq:flowdeq}).
Solving Eq. (\ref{eq:flowdeq}) in terms of the Green's function gives:
\begin{equation}
\chi(\mathbf{u})=\sum_i\frac{hn_i}{m}\Gamma(\mathbf{u},\mathbf{u}_i).
\label{eq:green-stream1}
\end{equation}
The flow due to a given vortex is proportional to its ``charge" $2\pi n_i$.
The energy as a function of the
positions of the vortices, Eq. (\ref{eq:particlemodel}),
can now be derived by integrating the kinetic energy in the flow determined
by Eq. (\ref{eq:flowdeq}) for each placement of the vortices.

We will begin by discussing applications of Eq. (\ref{eq:particlemodel}),
saving its derivation until later.
Interestingly, 
the \emph{energy} of the vortices can
be described by a differential equation analogous to 
Eq. (\ref{eq:flowdeq}) for the \emph{flow}. 
We choose one vortex
and fix the positions of all the others. We take the Laplacian of Eq.
(\ref{eq:particlemodel}) with respect to the position of the chosen vortex
and use Eq. (\ref{eq:geompdeq}) and Eq. (\ref{eq:BasicGreen}).
The energy as a function of the chosen vortex $E(\mathbf{u_i})$ satisfies:
\begin{equation}
\nabla^2 \frac{E(\mathbf{u_i})}{2\pi K n_i}=-\sigma_i(\mathbf{u_i})-
\frac{n_i}{2}G(\mathbf{u_i}).
\label{eq:endeq}
\end{equation}
The notation $\sigma_i$ stands for the delta function charge
distributions of all the vortices with the exception of the $i^{\mathrm{th}}$, 
so
that
$\sigma_i(u)=\sum_{j,j\neq i}{2\pi n_j\delta(\mathbf{u}-\mathbf{x}_j)}$.
The self-charge term which we have had to remove (so that $\mathbf{u}_i$ can
be substituted in place of $\mathbf{u}$ as in Eq. (\ref{eq:endeq}))
is replaced here
by a spread-out charge proportional to the removed term.



\subsection{\label{subsec:anomalous}
Anomalous force on rotationally symmetric surfaces}

On an azimuthally symmetric surface the force on a defect can be found
by exploiting Gauss's law. This is analogous to the familiar 
``Newton's Shell" theorem that predicts
 the gravitational field at the surface of a sphere surrounding a
spherically symmetric mass
distribution
by concentrating all the enclosed mass at the center.

\begin{figure}
\psfrag{phi}{$\phi$}
\includegraphics[width=0.45\textwidth]{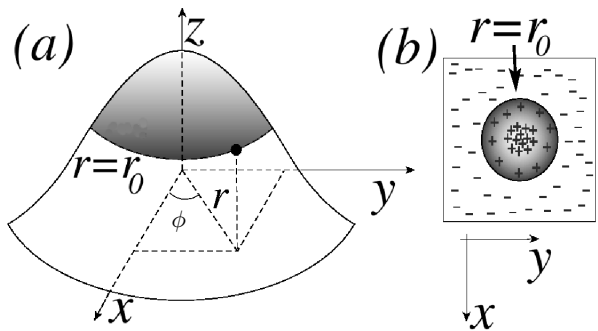}
\caption{\label{fig:bump}(a) A bumpy surface shaped as a Gaussian.
(b) Top view of (a) showing a schematic representation of the
positive and negative intrinsic curvature as a non-uniform
background ``charge" distribution that switches sign at $r=r_{0}$.
The varying density of + and - signs tries to mimic the changing
curvature of the bump.}
\end{figure}

Points on an azimuthally symmetric two dimensional
surface embedded in three dimensional Euclidean space are
specified by a three dimensional vector ${\bf R}(r,\phi)$ given by
\begin{equation}{\bf R}(r,\phi)=\left(\begin{array}{c} r\cos\phi \\ r\sin\phi \\h(r)\end{array}\right)
\!\!\!\! \quad, \label{eq-coord-polar}
\end{equation}
where $r$ and $\phi$ are plane polar coordinates in the $x$$y$
plane of Fig. \ref{fig:bump}, and $h(r)$ is the height as a function of radius;
e.g. $h(r)=h_0e^{-\frac{r^2}{2r_0^2}}$ for the Gaussian bump with height $h_0$ and spatial
extent $\sim r_0$.
It is useful to characterize the deviation of the bump from a
plane in terms of a dimensionless aspect ratio
\begin{equation}
\alpha \equiv \frac{h_0}{r_{0}} \!\!\!\! \quad. \label{aspect ratio}
\end{equation}
The metric tensor, $g_{\alpha\beta}$, is diagonal for this choice
of coordinates. In general, $g_{\phi\phi}=r^2$, $g_{rr}=1+h'(r)^2$, 
and for the Gaussian bump we have
\begin{equation}
g_{\alpha\beta} = \begin{pmatrix} 1+ \alpha^2\frac{r^2}{r_0^2}\exp
\left(-\frac{r^2}{r_{0}^2}\right) & 0 \\
  0 & r^2
\end{pmatrix} \!\!\!\! \quad .
\label{metric-mat}
\end{equation}
Note that the $g_{\phi\phi}$ entry is equal to the flat space
result $r^{2}$ in polar coordinates while $g_{rr}$ is modified in
a way that depends on $\alpha$ but tends to the plane result
$g_{rr}=1$ for both small and large $r$.

The Gaussian curvature for the bump is readily found from the
eigenvalues of the second fundamental form \cite{Dubrovinbook}; e.g., for
the Gaussian bump,
\begin{equation}
G_{\alpha}(r)=\frac{\alpha^2 e^{-\frac{r^2}{r_{0}^2}}}{r_{0}^2
\!\!\!\! \quad \left(1+ \frac{{\alpha}^2 r^2}{r_{0}^2} \exp
\left(-\frac{r^2}{r_{0}^2}\right)\right)^2}
\left(1-\frac{r^2}{r_{0}^2}\right) \!\!\!\! \quad .
\label{Gaussian Gaussian curvature}
\end{equation}
Note that $\alpha$ controls the overall magnitude of $G(r)$ and
that $G(r)$ changes sign at $r=r_{0}$ (see Fig. \ref{fig:bump}b).
The integrated Gaussian curvature inside a cup of radius $r$
centered on the bump vanishes as $r \rightarrow \infty$. The
positive Gaussian curvature enclosed within the radius $r_{0}$
(see Fig. \ref{fig:bump}) approaches $2\pi$ for $\alpha\gg1$, half
the integrated Gaussian curvature of a sphere. We shall show below
that there is always more positive than negative curvature within any
given radius, for an azimuthally symmetric surface. It will follow that
the force on a vortex is
repulsive at any distance.

In general an individual vortex of index $n_{i}$ confined on a curved surface
at position ${\bf u}_{i}$ feels a geometric interaction described by the energy
\begin{equation}
E({\bf u}_i)=- \pi K n_i^2 U_G({\bf u}_i).
\label{eq:ecurv-s2}
\end{equation}
For an azimuthally symmetric surface such as the bump represented
in Fig. \ref{fig:bump}, we can derive Newton's theorem as follows.
Define $\mathbf{E}=-\nabla U_G$ so that the covariant radial
component of $\mathbf{E}$
is
$E_r=-\partial_{r}U_G$. Then 
$-\nabla^2 U_G=\mathrm{div\ }
\mathbf{E}=\frac{1}{\sqrt{g}}\partial_r \sqrt{g}g^{rr}
E_r$,
and if we integrate both sides of Eq. (\ref{eq:geompdeq}) out to $r$,
\begin{equation}
2\pi\sqrt{g}g^{rr}E_r=-\iint {\sqrt{g}drd\phi G(r)}
\label{eq:covariant}
\end{equation}
so that $E_r$ has a simple expression in terms of the net Gaussian curvature
at a radius less than $r$.  Now $E_r$ is the ``covariant component" of the
geometrical ``electric field", not the actual field, which would be obtained by
differentiating with respect to arclength rather than the projected
coordinate $r$.  Therefore
the magnitude of $\mathbf{E}$ is $\frac{E_r}{\sqrt{g_{rr}}}$, which,
rephrasing Eq. (\ref{eq:covariant}), obeys
this generalized version of Newton's theorem:

\noindent\emph{The magnitude of}
 $\mathbf{E}$ \emph {is}
$-\frac{1}{2\pi r}$
\emph{times the integrated Gaussian curvature}.

\noindent(Recall that $g^{rr}=\frac{1}{g_{rr}}$ and
$g=\mathrm{det}\ g_{\alpha\beta}=rg_{rr}$.)
Note that the force on the
vortex is $\mathbf{F}=-\nabla E=-\pi K n_i^2 \mathbf{E}$
according to Eq. (\ref{eq:ecurv-s2}), so
that the geometrical force is proportional
to the integrated Gaussian curvature divided by the distance of the vortex from
the axis of symmetry of the surface;
the expression for the force on the vortex
obtained in the next section by integrating
the Gaussian curvature is
\begin{equation}
F_{geom}=\frac{K\pi}{r}(1-\frac{1}{\sqrt{1+h'^2}}),
\label{eq:aziforce}
\end{equation}
if $n_i=\pm 1$.  Note that this force is always \emph{repulsive} since
the integrated curvature is positive.

\begin{figure}
\psfrag{X}{$E(r)$}
\includegraphics[width=0.45\textwidth]{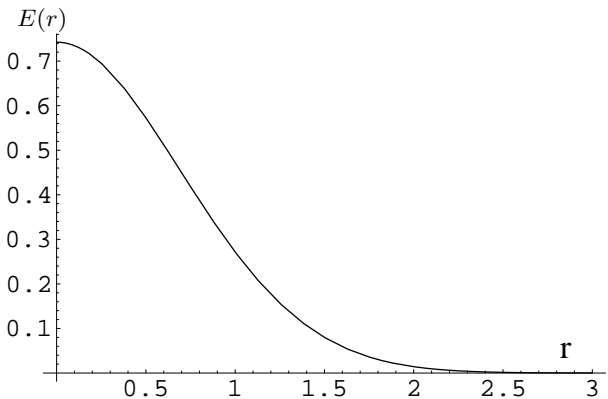}
\caption{Plot of the interaction energy $E(r)=-\pi K U_G(r)$ between a singly
quantized vortex and a Gaussian bump with $\alpha=1$. The energy is measured
in units of $K$ and the radius is measured in units of $r_0$.
Note that the force points away from the bump
and has its maximum strength near $r_0$.} \label{fig:geompot1}
\end{figure}

The geometric potential can now be expressed explicitly by integrating
$\mathbf{E}_r=-\partial_r U_G$ with the aid of Eq. (\ref{eq:covariant}):
\begin{equation}
U_G(r) = - \int_{r}^{\infty} \! dr' \frac{\sqrt{1+ \frac{{\alpha}^2
r^2}{r_{0}^2} \exp \left(-\frac{r^2}{r_{0}^2}\right)}-1}{r'}
\!\!\!\! \quad . \label{potential4-bis}
\end{equation}
The resulting potential $U_{G}(r)$ vanishes at infinity. Its
range and strength are given respectively by the linear size of the
bump and its aspect ratio squared (see Fig. \ref{fig:geompot1})

\begin{figure}
\includegraphics[width=0.48\textwidth]{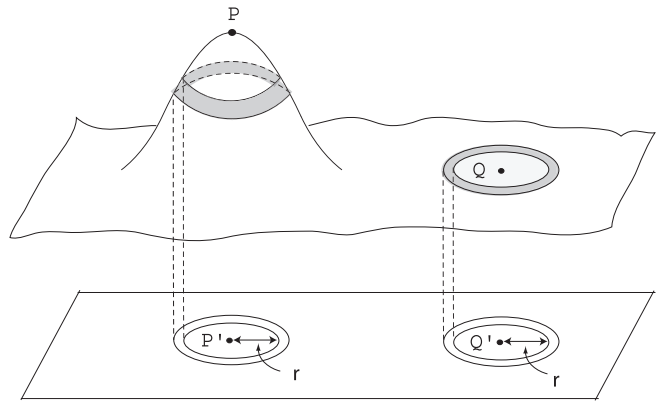}
\caption{\label{fig:intuition} An azimuthally symmetric substrate and its
downward projection on a flat plane. The shaded strip surrounding
$P$ is more stretched than the one surrounding $Q$ despite their
projections onto the plane having the same area. As discussed in the
text it follows that the energy stored
in the field will be lower if the center of the vortex is located at
$Q$ rather than $P$.}
\end{figure}

We now summarize
an intuitive argument that explains why the energy of a vortex on top
of the bump is greater than the energy of a vortex that is far away
\cite{swiss,Halperin-private}.
Fig. \ref{fig:intuition},
reproduced from \cite{swiss},
focuses on a rotationally symmetric bump 
coated by a helium film (of a constant thickness).
We can estimate the energy of a vortex on top of the bump by comparing
the situation to a vortex on a plane, illustrated vertically below
the bump.  Rotational symmetry implies that the superfluid phase is
given by $\theta=\phi$, the azimuthal angle. The velocity
depends on the rate of change on the phase according
to Eq. (\ref{eq:velocity}), so since an infinitesimal
arc of the circle
concentric with the top of the bump has size $rd\phi$, the velocity
is $\frac{\hbar}{mr}$.  Here $r$ is the
radius of the circle measured horizontally to the axis of the bump.
This calculation shows that the velocity, and thus the energy \emph{density},
are the same at any point on the bump and its projection into the plane.
However, the \emph{energy} contained in the tilted annulus on
the bump stretching from $r$ to $r+dr$ is greater than the energy
in the annulus directly below it because, though the energy density is
the same, the annulus's area is greater.  Hence a vortex
on a bump has a greater energy than a vortex in a plane, whether
it is the vortex at $P'$ in the projection plane, or the vortex
at $Q$ which is very far from the bump.  This reasoning indicates
a repulsive force, since
the vortex lowers its energy by moving away from the bump.

The intuitive argument suggests that the Gaussian curvature should appear
in the force law, as in Eq. (\ref{eq:geompdeq}); in fact, it is a widely
known fact that the sign of the Gaussian curvature of a surface
determines how fast
the circumference of a circle on the surface increases, relative
to the circumference of a circle on the plane, as a function of the radius.
Of course,
for less symmetric surfaces, comparing the
energy on a curved surface to that on a flat reference plane directly below it
will be more complicated, since symmetry and the constant circulation
constraint do not force the energy densities to be equal.  Thus, a simple
vertical projection will not set up a monotonic relation between energies.
The conformal mapping technique which we use in Sec. \ref{sec:complex}
 is a variation on the idea of comparing a ``target" substrate to a simple 
``reference" surface
which is in principle applicable to arbitrary surfaces, and furthermore not
only allows one to compare energies, but also to calculate them quantitatively.
The technique can also be used to give a concise derivation of
Eq. (\ref{eq:geompdeq}).

Such an intuitive argument 
applies only for azimuthally symmetric surfaces.
For less symmetric surfaces, comparing the
energy on a curved surface to that on a flat reference plane directly below it
will be more complicated, since symmetry and the constant circulation
constraint do not force energy densities to be equal at corresponding postions.
Thus, a simple
vertical projection will not set up a monotonic relation between energies.
The conformal mapping technique which we use in Section \ref{sec:complex} 
 is a variation on the idea of comparing a ``target" substrate to a simple 
``reference" surface
which is in principle applicable to arbitrary surfaces, and furthermore not
only allows one to compare energies, but also to calculate them quantitatively.
The technique can also be used to give a concise derivation of
Eq. (\ref{eq:geompdeq}).

\subsection{\label{BS}Vortex-trapping surfaces}

In order to illustrate the consequences of the curvature-induced
interaction for different surface morphologies, we study a ``Gaussian saddle"
surface (suggested to us by Stuart Trugman)
 for which the geometric potential has its absolute minimum at the origin. 
\begin{figure}
\includegraphics[width=0.45\textwidth]{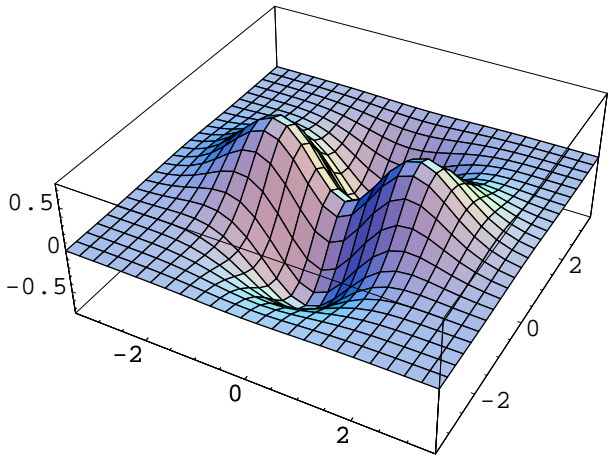}
\caption{Plot of vortex trapping surface. \label{fig:antibump}}
\end{figure}

\begin{figure}
\includegraphics[width=0.45\textwidth]{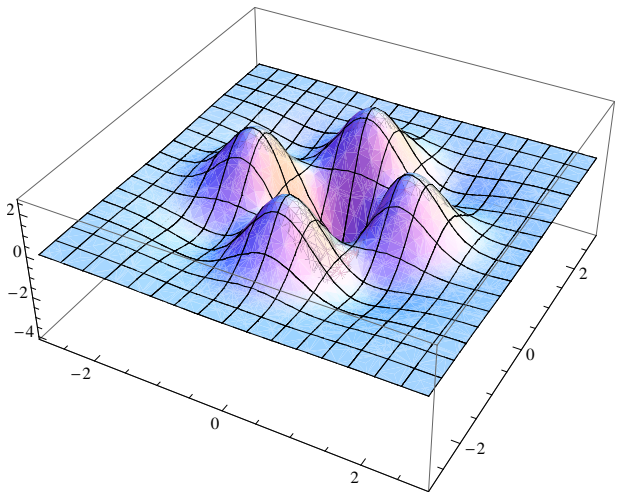}
\caption{Plot of the curvature of the $\lambda=1$ vortex trapping surface.}
\label{fig:antibumpcurv}
\end{figure}

\begin{figure}
\includegraphics[width=0.45\textwidth]{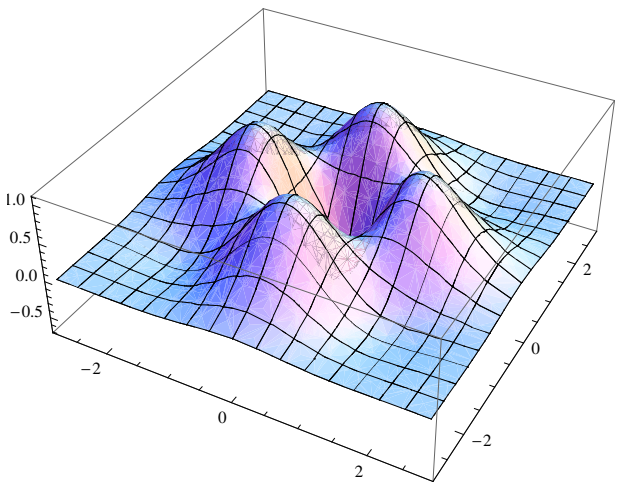}
\caption{Plot of the geometric potential for the $\lambda=1$ vortex trapping
surface.} \label{fig:peacefulvalley}
\end{figure}

First, we show that an alternative design for a vortex trap geometry fails
because of the long-range nature of the curvature-induced interaction. Fig. \ref{fig:bump} shows
that bumps have negative curvature on their flanks; it might seem possible
that a well-chosen bump
would have enough negative curvature to hold
a vortex. However, a vortex cannot be held by nearby negative curvature alone;
it also feels the positive curvature at the center of symmetry because, according
to Gauss's law applied to the azimuthally symmetric region, the force is
due to the \emph{net} curvature contained in any circle
concentric with the top of the bump.  
This curvature is given generally by
\begin{equation}
G=-\frac{1}{r\sqrt{1+h'(r)^2}}\partial_{r}\frac{1}{\sqrt{1+h'(r)^2}},
\end{equation}
\label{eq:tower} and the net curvature within radius $r_v$ is thus
\begin{eqnarray}
\int_0^{r_v}\sqrt{g}\ dr\  d\phi\
G(r)&=&2\pi(1-\frac{1}{\sqrt{1+h'(r)^2}})\nonumber\\
&=&2\pi(1-\cos\theta[r_v])
\label{eq:stuckring}
\end{eqnarray}
where $\theta$ is the angle between the surface at the location of the vortex
and the horizontal plane.
This formula also describes the cone angle of a cone tangent to
the surface at radius $r_v$; it can also be derived from 
the Gauss-Bonnet theorem which implies that the net curvature of a
curved region depends only on the boundary of the region and how
it is embedded in a small strip containing it; thus the net curvature
of the cone (concentrated at the sharp point of the cone)
is the same as the net curvature
of the bump which it is tangent to.
Since this curvature is always positive, the defect is always
repelled from the top of an azimuthally symmetric bump\footnote{More specifically, all bumps with azimuthally symmetric \emph{embeddings} repel vortices from
their tops. The negative curvature
cones discussed in
Sec. \ref{sec:geomineq} have an internal azimuthal symmetry but their three dimensional
embeddings are not symmetric.}.
To find a way to trap a vortex,
one must therefore investigate some non-symmetric surfaces.
The curvature-defect interaction energy on a generic surface
is mediated by the Green's function
of the surface, Eq. (\ref{eq:BasicGreen}), as can be seen by solving Eq. 
(\ref{eq:geompdeq}):
\begin{equation}
E(\mathbf{u}) = K\pi\int\ \!\!\! d^2\mathbf{u} \!\!\!\!\quad \Gamma(\mathbf{u},\mathbf{v})\!\!\!\!\! \quad G(\mathbf{v}),
\label{eq:curvature-defect}
\end{equation}
for a singly quantized vortex.  One such
surface, which we will treat perturbatively in the amount of
deformation from flatness 
(Section \ref{sec:bubble} treats a different confining
surface exactly)
 is the Gaussian saddle 
represented in Fig. \ref{fig:antibump} and  described by
the height function 
\begin{equation}
h_{\lambda}(x,y)=\frac{\alpha}{r_0} (x^2-\lambda y^2)
\!\!\!\!\quad  e^{-\frac{x^2+y^2}{2 r_0^2}} \label{eq:saddlemesa}
\end{equation}
where the exponential factor was included to make the surface flat
away from the saddle \cite{trugman}. Here,
$\lambda$ is a parameter which we later
vary to illustrate the
non-locality of the interaction. 
The leading order contribution to the curvature-defect interaction
(for $\alpha<<1$) is of the same order
$\alpha^2$ as the  curvature corrections to the defect-defect interaction 
(calculated in Appendix \ref{app:calmseas}).  In fact $\Gamma$ is multiplied
in Eq. (\ref{eq:curvature-defect})
by the Gaussian curvature $G(\mathbf{u}')$, of order $\alpha^2$.  
Thus, for a single defect, it is
sufficient to use the flat space Green's function $\Gamma_{flat}$
\begin{equation}
\Gamma _{flat} (x-x',y-y') =
-\frac{1}{2\pi}\log\sqrt{(x-x')^2+(y-y')^2})
\end{equation}
for calculating the defect-curvature interaction.  
Furthermore, the Gaussian
curvature that acts as the source of the geometric potential can be
calculated from the usual second-order approximation:
\begin{equation}
G_{\lambda}(x,y)\approx\frac{\partial^2 h_{\lambda}}{\partial x^2}
\frac{\partial^2
h_{\lambda}}{\partial y^2}-
\left(\frac{\partial^2 h_\lambda}{\partial x\partial y}\right)^2
\end{equation}
This function is plotted in Fig. \ref{fig:antibumpcurv}, and its
 sign is represented in
the middle frame of Figure \ref{fig:occult}.
The graph of the vortex-curvature interaction energy for this surface,
Fig. \ref{fig:peacefulvalley},
shows that a vortex is indeed
confined at the center of the saddle; the energy graphed in this figure
is given by
\begin{eqnarray}
E_{\lambda}(x,y) \approx K\pi\int\ \!\!\! dx'\ dy' \!\!\!\!\quad \Gamma _{flat}
(x-x',y-y') \!\!\!\!\! \quad G_{\lambda}(x',y'),\nonumber\\
\label{eq:smallasp-saddle}
\end{eqnarray}
for $\lambda=1$.
For realistic film thicknesses and $\alpha$ of order unity
(see Sec. \ref{sec:experimental}),
the depth of the well is about $50$ Kelvin! We have
found that the energy associated with a vortex at the origin
is less than for any other position. Of course, the configuration with
a vortex at the origin cannot beat the configuration with no vortices
at all!  The latter has zero kinetic energy; when the vortex is at the
origin, the energy is positive provided that Eq. (\ref{eq:particlemodel})
is supplemented by the position-independent contribution $\pi K \ln\frac{R}{a}$.
This
term is always necessary for comparing configurations with different numbers of
defects, as when one studies the formation of a vortex lattice at increasing
rotational frequencies\cite{campbell}.
\subsection{\label{earnshaw}Negative curvature which does not trap}
\begin{figure}
\includegraphics[width=.47\textwidth]{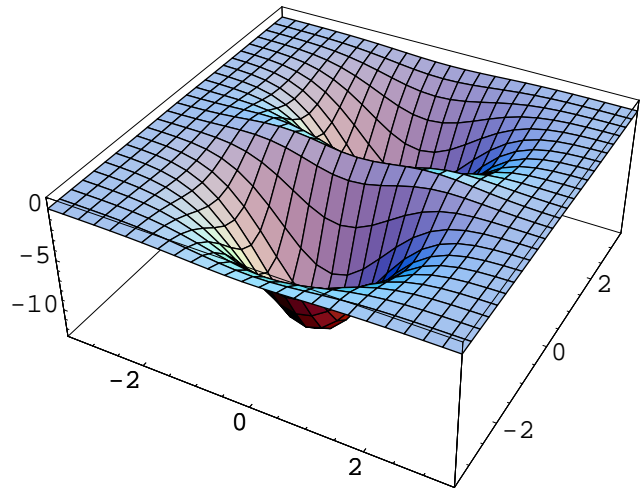}
\caption{\label{fig:brokentrap}
A saddle surface with $\lambda=17$; this parameter value is just
large enough to destabilize a vortex at the center.}
\end{figure}

\begin{figure*}
\includegraphics[width=1.0\textwidth]{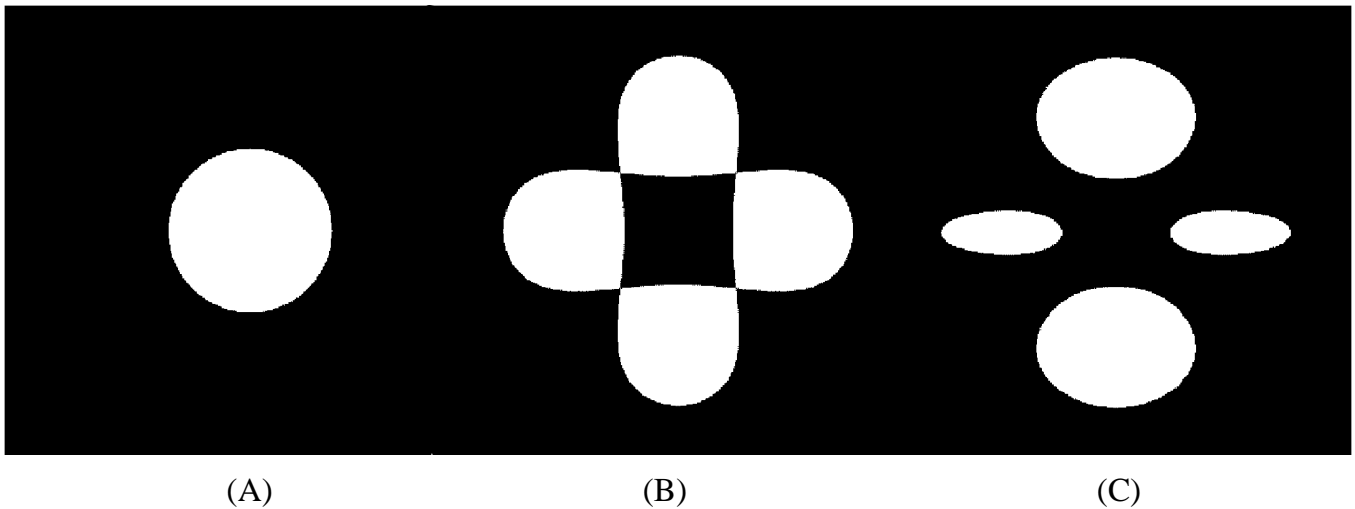}
\caption{\label{fig:occult}
Plots of the sign of the curvature, with white for positive curvature. (A) is for the Gaussian bump, and (B)
and (C) are
for the saddle surfaces with $\lambda=1$ and $17$ respecitvely. 
Because of the lack of symmetry
in the third figure, the center point becomes a saddle point of the energy; the vortex
is pushed away by the strong positive curvature
in the ellipsoidal regions at positive and negative $y$.}\end{figure*}

In this section, we shall discuss what happens when the parameter $\lambda$
of the saddle surface is increased; Fig. \ref{fig:brokentrap} 
illustrates such a
surface corresponding to $\lambda=17$.  To give a hint of what causes the
equilibrium to change its character,
Fig. \ref{fig:occult}
shows the sign of the curvature
for the Gaussian bump, the saddle with $\lambda=1$, and the saddle with
$\lambda=17$.

In the graph of the defect-curvature interaction
energy with $\lambda=17$, one notices that the
origin is an unstable equilibrium position for the vortex.  We will derive
the exact value of $\lambda$ where this instability first occurs below. 
However, symmetry considerations alone show
that the origin \emph{is} a stable
equilibrium point when $\lambda=1$, as Fig. \ref{fig:peacefulvalley}
shows.
One might be tempted to argue from Newton's theorem
that a vortex at a small enough
radius $r$ is always attracted to the origin by the negative curvature
at radii smaller than $r$. However, the
asymmetry of the saddle surfaces invalidates Newton's theorem 
and
positive curvature more distant from the origin than the vortex might be
able to push the vortex toward infinity.
This does not occur for the saddle surface with $\lambda=1$; although
the rotational symmetry needed
for Newton's theorem is absent, the surface does have order four symmetry,
under a $90$ degree rotation combined with the isometry $z \rightarrow -z$.

Upon expanding the defect-curvature interaction energy about the origin,
 we obtain
\begin{equation}
E=E_0+ax+by+cx^2+2dxy+ey^2+\cdots
\end{equation}
This energy must be invariant under the symmetries of the surface (without
a sign change).  Order two symmetry implies that the linear terms vanish,
so the center point is an equilibrium. The order four symmetry
(apparent in Fig. \ref{fig:occult}B) implies that it is either a maximum or a
minimum (a quadratic function with a saddle point has only
$180$ degree symmetry).
In more detail, $90^{\circ}$ rotational symmetry, given by
$x\rightarrow y,\ y\rightarrow -x$,
implies that $c=e,d=0$.
Since the Laplacian of $E$ at the origin is proportional to
minus the \emph{local} curvature, $2c=2e=c+e$
is positive, so the origin is a local minimum. Without the order four
symmetry the negative

curvature only  ensures that $c+e>0$.
Earnshaw's theorem of electrostatics\cite{scott,earnshaw,jeans} 
states that an electric charge 
cannot have a stable equilibrium at a point where the charge density is zero or has the same sign as the
charge.  The charge cannot be confined by electric fields produced
by electrostatic charge distributions in a surrounding apparatus. 
(This theorem
motivated the design of magnetic and electrodynamic traps for trapping
charged particles in plasma physics
and atomic physics.)
The argument provided here can be
generalized to give the following
converse rule based on discrete symmetries (whereas Newton's theorem applies
only for continuous azimuthal symmetry):

If $P$ is a symmetry point of a charge distribution with rotation
angle $\frac{2\pi}{m}$, and $m\geq 3$, and the charge density at $P$ is positive, then $P$ is a point of stable equilibrium for particles of negative charge.

This is the formulation for electrical charges in two dimensions; for vortices,
the sense of the rotation of the vortex does not matter of course, since the
vortex interacts with its own image charge distribution.  Hence if the 
curvature at $P$ is negative, then a vortex
will be trapped there.

Similar reasoning can be used to show that a generalization of 
Eq. (\ref{eq:saddlemesa}), the ``Gaussian Monkey Saddle"
given by $h(x,y)=\frac{\alpha}{r_0^2}\Re(x-iy)^3e^{-\frac{x^2+y^2}{2r_0^2}}$,
traps vortices in an energy well of the form
 $E=E_0+\frac{9\pi K}{4}\frac{r^4}{r_0^4}+(cnst.+cnst. \cos 6\theta)r^6+\dots$.
The reasoning needs to be modified because the curvature at the
origin of the monkey-saddle is zero and the trapping is due to the negative
curvature near the origin.

\begin{figure}
\includegraphics[width=.47\textwidth]{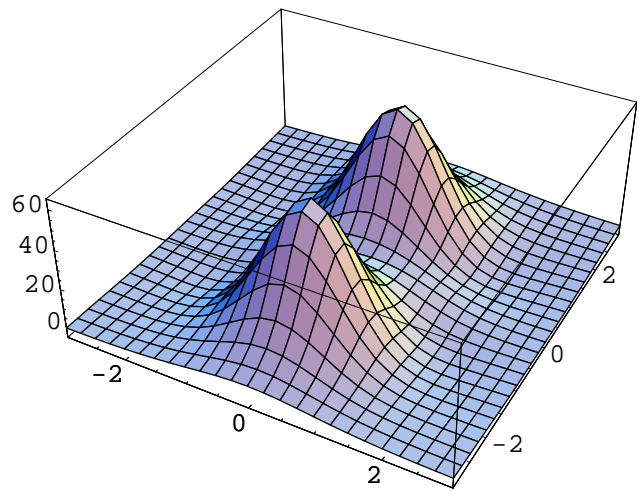}
\caption{\label{fig:dangerous}
The geometry-defect interaction energy of a vortex
on the saddle surface with $\lambda=17$. One notices
a slight instability in the $x$ direction.}
\end{figure}

At a point of low symmetry
 (such as the origin in Eq. (\ref{eq:saddlemesa}) when $\lambda\neq 1$), 
the character of an extremum
depends on the charge distribution elsewhere, since the previous argument
only implies that $c+e>0$. Fig. \ref{fig:occult}C suggests that a vortex
at the origin is destabilized by its repulsion from the positive
curvature above and below the origin, which is not balanced by enough
positive curvature to the left and right.
In fact, more detailed calculations show that
the range of $\lambda$ for which
the origin is an energy minimum is $\sqrt{65}-8<\lambda<\sqrt{65}+8$; the
origin is a saddle point outside this range, as is just barely visible
for the case of $\lambda=17$
in Fig. \ref{fig:dangerous}.  (Likewise, for negative values of
$\lambda$, the origin is a maximum when $\sqrt{65}-8<-\lambda<\sqrt{65}+8$,
but a saddle point outside this range.)

These results follow by changing the integration variables to
$\xi=x-x'$,$\eta=y-y'$ in Eq.(\ref{eq:smallasp-saddle})
and then expanding to second order about the origin $(x,y)=(0,0)$.  The integral expressions for second derivatives of the energy can be evaluated explicitly,
\begin{multline}
E_{\lambda}(x,y)=K\pi[\alpha^2\frac{1+\lambda^2-6\lambda}{16}+
\\+\frac{x^2}{4}(\alpha^2\frac{\lambda^2-1}{4r_0^2}-G_0)
+\frac{y^2}{4}(\alpha^2\frac{1-\lambda^2}{4r_0^2}-G_0)]
\end{multline}
where $G_0=-4\lambda\frac{\alpha^2}{r_0^2}$ is the curvature at the origin.
In Appendix
\ref{app:multipole}, we
determine the geometric potential for arbitrary $x$ and $y$
in (unwieldy) closed form.


\subsection{\label{subsec:cande}Hysteresis of vortices and trapping strength}
The geometrical interaction has its maximum strength when the Gaussian
curvature is the strongest. However, the geometric charge (i.e.,
integrated Gaussian curvature) of any
particular feature on a surface has a strength roughly equivalent at most to
the charge of one or two vortices.  Eq. \ref{eq:aziforce} therefore
suggests that the force on a vortex due to a feature of the
surface is less than the force due to a couple vortices at the same distance.
Precise limits on the
strength of the geometric interaction will be stated and proven in
Section \ref{sec:geomineq}, for arbitrary geometries.

As a consequence
\emph{the geometric interaction has its most significant effects
when the number of vortices is comparable to
the number of bumps and saddles on the surface}, so that the geometrical force is not
obscured by interactions with the other vortices. This is a recurring
(melancholy) theme of our calculations, to be illustrated in  Section
\ref{subsec:abr} for arrangements of vortices in a rotating film. The current
section illustrates the point by discussing hysteresis on a surface with 
multiple saddle points (i.e., traps).  If a vortex-free superfluid film
is heated, many 
vortices form in pairs of opposite signs.  When it is cooled again, positive and
negative vortices can remain trapped in metastable states in the saddles, 
but even with the strongest curvature possible, the argument
above suggests that not more than one vortex can be trapped per saddle.

The effectiveness of the defect trapping by geometry
is determined mainly by the ratio of the saddle density to
vortex density.  As shown in the previous section, the geometric
energy near the center of a vortex trap with $90^{\circ}$ symmetry is given by
\begin{equation}
E(r)\approx \frac{\pi}{4}K|G_0|r^2 .
\end{equation}
The force on the vortex found by differentiating
the energy
reads
\begin{equation}
F(r)\approx -\frac{\pi}{2}K|G_0|r\label{eq:earthcore}.
\end{equation}
Eq. (\ref{eq:earthcore}) shows
that the trap pulls the vortex more and more strongly as the vortex is pulled away from the center,
like a spring, until the vortex reaches the end of the trap at
a distance of the order of $r_0$ where the force starts decreasing.
Since $G_0\sim\frac{\alpha^2}{r_0^2}$ (which is valid for a small aspect ratio $\alpha$), ``the spring breaks down" when the vortex is pulled with a force greater than
\begin{equation}
F_{max}\sim F(r_0)\sim \frac{K\alpha^2}{r_0}.
\label{eq:escapeforce}
\end{equation}

Let us consider a pair of saddles separated by distance $d$. 
It is possible
that one vortex can be trapped in each saddle even for a small $\alpha$ provided that
$d$ is large enough.  Remote vortices
do not interact strongly enough to push one another out
of their traps.  The Coulomb attraction or repulsion of the vortices must be weaker than
the breakdown force of the trap $F_{max}$, i.e., $\frac{K\alpha^2}{r_0}\gtrsim \frac{K}{d}$.
The minimum distance between the
two saddles is therefore
\begin{equation}
d_{min}\sim\frac{r_0}{\alpha^2}.
\label{eq:dmin}
\end{equation}

Let us find the maximum density of trapped vortices that can remain
when the helium film is cooled through the Kosterlitz-Thouless temperature.
Let us suppose there is a lattice of saddles forming a bumpy
texture like a chicken skin.  Suppose bumps
cover the whole surface, so that the
spacing between the saddles is of order $r_0$.  Then not every
saddle can trap a vortex; the largest density of saddles which trap
vortices is of the order of $1/d_{min}^2$, so the fraction
of saddles which ultimately contain vortices is at most
$\frac{r_0^2}{d_{min}^2}\propto\alpha^4$.  Note that not as many
vortices can be trapped if they all have the same sign, since the interactions from
distant vortices add up producing a very large net force.
On the other hand, producing vortices of both signs by heating and then cooling the helium film 
results in screened vortex interactions which are weaker and 
hence less likely to push the defects out of the metastable states in which they are trapped.

\section{\label{sec:Rotation}Rotating Superfluid Films on a Corrugated Substrate}
\subsection{\label{subsec:Rotation}The effect of rotation}
Suppose that the vessel containing the superfluid layer is rotated
around the axis of symmetry of the Gaussian bump with angular velocity
${\bf \Omega} = \Omega \!\!\!\! \quad \mathbf{\hat{z}}$, as might occur at
the bottom of a spinning wine bottle.
The container can rotate independently of the superfluid
in it because there is no friction between
the two. However, a state with vanishing superfluid angular momentum is not
the ground state. To see this, note 
that the energy, $E_{rot}$, in a frame rotating at angular
velocity ${\bf \Omega}$ is given by:
\begin{equation}
E_{rot}=E - {\bf L} \cdot {\bf \Omega} \!\!\!\! \quad .
\label{eq:Erot}
\end{equation}
where $E$ is the energy in the laboratory frame and
$\mathbf{L}$ is the angular momentum. Hence $E_{rot}$ is
lowered when ${\bf L} \cdot {\bf \Omega} > 0$, that is, when the
circulation in the superfluid is non-vanishing. This is achieved
by introducing quantized vortices in the system (see
Eq.(\ref{quantization})), whose microscopic core radius (of the
order of a few \AA) is made of normal rather than superfluid
component. The energy of rotation, ${\bf L} \cdot {\bf \Omega}$,
corresponding to a vortex at position ${x,y}$ on the bump can be
evaluated from
\begin{equation}
L_{z}=  \rho_{s}\int_{S} dx dy  \sqrt{g(x,y)} \left(x v_{y} - y
v_{x} \right) \!\!\!\!\! \quad . \label{eq:Lz}
\end{equation}
Upon casting the integral in Eq.(\ref{eq:Lz}) in polar coordinates
${r,\phi}$ and using the identity
\begin{equation}
\left(x v_{y} - y v_{x} \right) = r \mathbf{\hat{\phi}} \cdot \mathbf{v}
\!\!\!\!\! \quad \!\!\!\! \quad, \label{eq:identity}
\end{equation}
we obtain
\begin{equation}
L_{z}=  \rho_{s}\int_{0}^{R} dr \sqrt{g(r)} \oint_{C}\! d
u^{\alpha}  \!\!\!\!\!\! \quad v_{\alpha}  \!\!\!\!\! \quad .
\label{eq:Lz-b}
\end{equation}
where $R$ is the size of the system. The line integral in
Eq.(\ref{eq:Lz-b}) of radius is evaluated over circular contours of radius
$r$ centered at
the origin of the bump. The circulation vanishes if the vortex of strength
$n$ at
distance $r_v$ is not enclosed by the contour of radius $r$:
\begin{equation}
\oint_{C_r}\! d u^{\alpha}  \!\!\!\!\!\! \quad v_{\alpha} = n
\kappa \theta(r-r_{v}) \!\!\!\! \quad . \label{eq:count-int}
\end{equation}
Upon substituting in Eq.(\ref{eq:Lz-b}), we obtain
\begin{eqnarray}
L_{z}&=& n \rho_{s} \kappa \int_{r_v}^{R} dr \sqrt{g(r)} \nonumber\\
&=& \frac{n \rho_{s} \kappa}{2 \pi} \left(A(R)-A(r_{v})\right)
\!\!\!\!\! \quad , \label{eq:Lz-c}
\end{eqnarray}
where $A(R)$ is the total area spanned by the bump and $A(r_{v})$ is the
area of the cup of the bump bounded by the position of the vortex.
Thus, after suppressing a constant,
the rotation generates an approximately parabolic potential
energy $E_{\Omega}(r)$ (see Fig.(\ref{fig:area})) that confines a vortex of
positive index $n$ close to the axis of rotation as in flat space:
\begin{equation}
E_{\Omega}(r_v)=  n \frac{\hbar \Omega \rho_{s}}{m_{4}} A(r_v) \!\!\!\!\!
\quad, \label{eq:area}
\end{equation}
where a constant has been neglected.
One recovers the flat space result\cite{Vine} by setting
$\alpha$ equal to zero.
Eq.(\ref{eq:Lz-c}) has an appealing intuitive interpretation
as the total number of superfluid atoms beyond the vortex,
$\frac{\rho_{s}}{m} (A(R)-A(r))$,
times a quantum of angular momentum $\hbar$ carried by each of
them. The closer the vortex is to the axis, the more atoms there
are rotating
with the container.

Above a critical frequency $\Omega_1$, the restoring force due to the
rotation (the gradient of Eq. (\ref{eq:area})) is greater than
the attraction to the boundary. The energy of attraction to the boundary is 
approximately
$\pi K\ln(1-\frac{r^2}{R^2})$,
where we assume the aspect ratio of the bump is small
so that the flat space result is recovered. Upon expanding this boundary
potential
harmonically about the origin and comparing to Eq. (\ref{eq:area}),
one sees that
\begin{equation}
\Omega_1\sim\frac{\hbar}{mR^2}.
\end{equation}
Above $\Omega_1$, the origin is a local minimum in the
energy function for a single vortex, though higher frequencies are
necessary to produce the vortex in the first place.  What determines
the critical frequency for producing a vortex is unclear.
There is a higher frequency $\Omega_{1}'\sim \frac{\hbar}{mR^2}\ln\frac{R}{a}$,
at which the single vortex actually has a lower energy (according to Eq. (\ref{eq:Lz-c}))
than no vortex at all, but critical speeds are rarely in agreement with the
measured values \cite{criticalvinen}. In the context of thin
layers, it is likely that a third, much larger critical speed
$\Omega_{\mathrm{crit}}\sim\frac{\hbar}{mRD_0}$,
is necessary before vortices form spontaneously, where $D_0$ is the thickness
of the film (see Sec. \ref{subsec:nucleation}).

\subsection{\label{subsec:single}Single defect ground state}

\begin{figure}
\psfrag{X}{$-U_G(r)$}
\includegraphics[width=0.45\textwidth]{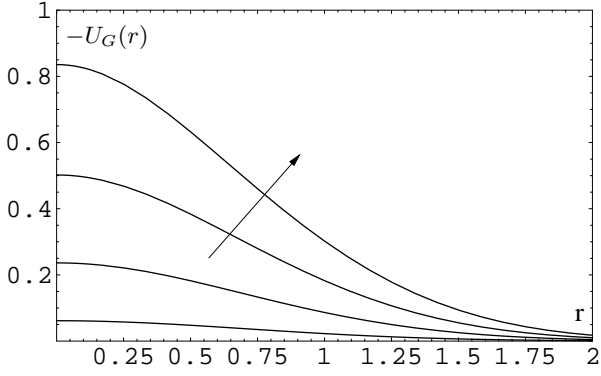}
\caption{Plot of minus the geometric potential $-U_G(r)$ for $\alpha=0.5,
1, 1.5, 2$. The arrow indicates increasing $\alpha$. The radial
coordinate $r$ is measured in units of $\lambda$ and
$r_{0}=\lambda$.} \label{fig:geompot}
\end{figure}

\begin{figure}
\includegraphics[width=0.45\textwidth]{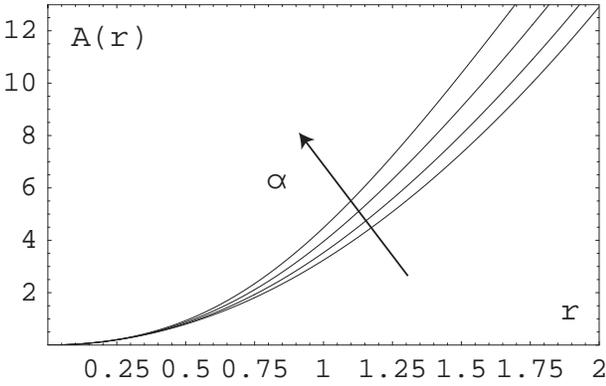}
\caption{Plot of the area of a cup of radius $r$ for $\alpha=0.5,
1, 1.5, 2$. The arrow indicates increasing $\alpha$. The radial
coordinate $r$ is measured in units of $\lambda$ and
$r_{0}=\lambda$.} \label{fig:area}
\end{figure}

The equilibrium position of an isolated vortex far from the
boundary is determined from the competition between the confining
potential caused by the rotation and the geometric interaction
that pushes the vortex away from the top of the bump. The energy
of the vortex, $E(r)$, as a function of its radial distance from
the center of the bump is given up to a constant by the sum of the geometric
potential and the potential due to rotation,
\begin{equation}
\frac{E(r)}{K}= -\pi U_G(r)+
\frac{A(r)}{\lambda ^2} \!\!\!\! \quad , \label{eq:E-tot}
\end{equation}
where we have ignored the
effects of the distant boundary, boundary effects are discussed
in the next section. The ``rotational length"
$\lambda$ is defined as
\begin{equation}
\lambda \equiv \sqrt{\frac{\hbar}{m \Omega}} \!\!\!\! \quad .
\label{eq:bohr}
\end{equation}
A helium atom at radius $\lambda$ from the origin rotating with
the frequency of the substrate has a single quantum of angular momentum.
The geometric contribution to $E(r)$ 
( see Fig. \ref{fig:geompot}) varies strongly as the shape
of the substrate is changed.  The rotation contribution to $E(r)$ confinement
(see Fig. \ref{fig:area}) varies predominantly as the frequency is changed;
near the center of rotation, where the substrate is parallel to
the horizontal plane, the rotational contribution barely changes as $\alpha$
is increased.

\begin{figure}
\includegraphics[width=0.45\textwidth]{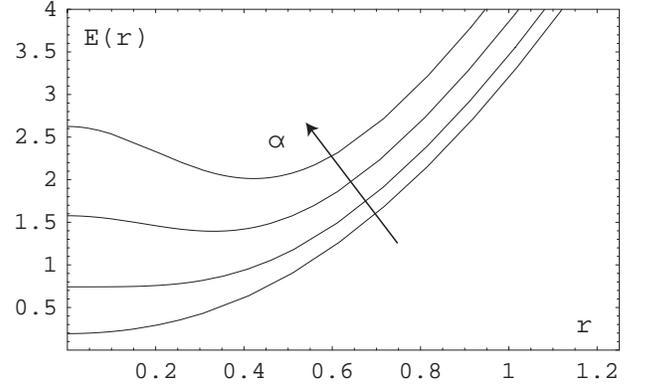}

\caption{Plot of $E(r)$ measured in units of $K=\frac{\hbar ^2
\rho_s}{m^2}$ as $\alpha$ is varied.
 In these units, the thermal energy $k_{B} T$ is less
than 0.1 below the Kosterlitz-Thouless temperature, for $200$\AA$\ $films. 
The radial coordinate $r$ is measured in units of
$\lambda$ and $r_{0}=\frac{\lambda}{2}$. Note that this plot is a
2D slice of a 3D potential. For $\alpha < \alpha _c$, $E(r)$ is
approximately a paraboloid while, for $\alpha > \alpha _{c}$, we
have a Mexican hat potential.} \label{fig:totalpot}
\end{figure}

\begin{figure}
\includegraphics[width=0.45\textwidth]{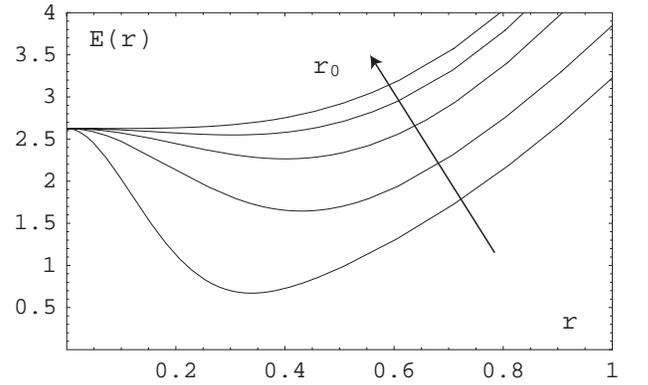}
\caption{Plot of $E(r)$ in units of $\frac{\hbar^2\rho_s}{m^2}$ versus $r$
as $r_0$ is varied.
The aspect ratio is kept fixed at
$\alpha=2$ while the range of the geometric potential
(corresponding to the width of the bump) is varied so that $r_{0}=
0.2, 0.4, 0.6, 0.8, 1$ in units of $\lambda$. 
As $r_{0}$ decreases, the geometric force becomes stronger, so
the system goes
through a transition analogous to the one displayed in Fig.
\ref{fig:totalpot}.} \label{fig:vario}
\end{figure}

\begin{figure}
\includegraphics[width=.45\textwidth]{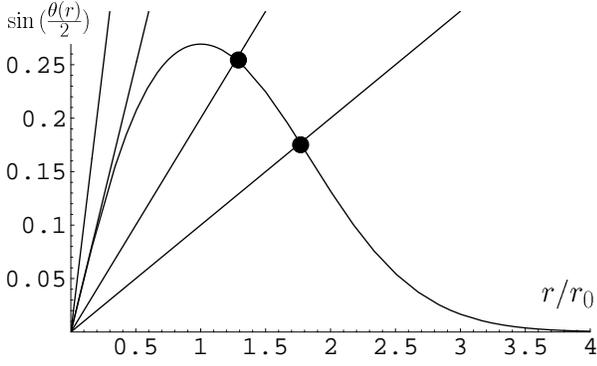}
\caption{\label{fig:constr} Graphical method for determining equilibrium positions
of one vortex. The equilibrium position is at the intersection of
$\sin\frac{\theta(r)}{2}$ and $\frac{r}{\lambda}$. If we fix
$r_0$ and set $\alpha=1$, the rotational frequency
will control the position of the vortex.  The four lines
correspond to $\Omega=\frac{\hbar}{mr_0^2}, \frac{\hbar}{4mr_0^2}$ (which is the critical frequency $\Omega_c$), $\frac{\hbar}{25mr_0^2}
,\ \frac{\hbar}{100mr_0^2}$.}
\end{figure}

As one varies $\alpha$ (fixing $r_0$ and $\Omega$)
there is a transition to an asymmetric minimum.
In fact, Fig. \ref{fig:totalpot} reveals that for $\alpha$ greater than a critical
value $\alpha_c$ the total energy $E(r)$ assumes a
Mexican hat shape whose minimum is offset from the top
of the bump. The position of this minimum is found by
taking a derivative of Eq. (\ref{eq:E-tot})
with
respect to $r$:
\begin{equation}
\pi\frac{dU_G}{dr}=\frac{1}{\lambda^2}\frac{dA}{dr}.
\label{eq:*}
\end{equation}
Now $\frac{dA}{dr}$ can be shown to equal $2\pi r\sqrt{1+h'^2}$ by
differentiating Eq. \ref{eq:Lz-c} and $\frac{dU_G}{dr}$, which
is the same as $F_G\sqrt{1+h'^2}$ can be evaluated by substituting
for $F_G$ from Eq. \ref{eq:aziforce}.  
This
leads to an implicit equation for the position of
the minimum, $r_{m}$, namely
\begin{equation}
\frac{r_{m}}{\lambda} = \sin(\frac{\theta[r_{m}]}{2}) \!\!\!\! \quad .
\label{eq:constr}
\end{equation}
Here $\theta(r)$, defined in Sec. \ref{BS},
 is the angle that the tangent at $r$ to the bump
forms with a horizontal plane.  A simple
construction allows one to solve Eq.(\ref{eq:constr}) graphically by
finding the intercept(s) of the curve on the right-hand side with the
straight line of slope $\frac{1}{\lambda}$ on the left-hand side (see Fig.
\ref{fig:constr}). A brief calculation based on this construction
shows that for $\alpha>\alpha_c=\frac{2r_0}{\lambda}$, there are two
intercepts: one at $r=0$ (the maximum) and one at $r=r_{m}$, the
minimum; whereas for $\alpha < \alpha_{c}$ only a minimum at $r=0$
exists exactly like in flat space.
It is possible to go through this second order transition by
changing other parameters such as the rotational frequency. 
See Figs. \ref{fig:totalpot} and \ref{fig:vario} for
illustrations of how the transition occurs when the shape of
the substrate is varied.
More details
on the choice of substrate parameters are given in Sec.
\ref{sec:experimental}. Once these parameters have been chosen, changing
the frequency would likely be more convenient; 
Fig. \ref{fig:constr} shows how the equilibrium position of the vortex varies.
If the vortex position $r_m$ can be measured precisely as a function of $\Omega$
and if there is not too much pinning, then
the geometrical potential can
even be reconstructed by integrating
$U_G=-\int_{\Omega}^{\Omega_c} {2\frac{m\Omega'}{\hbar}r_m(\Omega')\sqrt{1+h'(r_m(\Omega'))^2}\frac{dr_m}{d\Omega}\big(\Omega'\big)d\Omega'}
+cnst.$ which follows from Eq. \ref{eq:*}. 
\subsection{\label{subsec:multiple} Multiple defect configurations}
As the angular speed is raised, a cascade of
transitions characterized by an increasing number of vortices
occurs just as in flat space. 
In order to facilitate the mathematical analysis we
introduce a conformal set of coordinates $\{\mathcal{R}(r),\phi\}$ (see
\cite{geomgenerate} for details). The function $\mathcal{R}(r)$
corresponds to a nonlinear stretch of the radial coordinate that
``flattens" the bump, leaving the points at the origin and infinity
unchanged:
\begin{equation} \mathcal{R}(r)= r \!\!\!\!\!
\quad e^{U_G(r)} \!\!\!\! \quad , \label{solution2-bis}
\end{equation}
Note the unwonted appearance of the geometric potential $U_G(r)$ playing
the role of the conformal scale factor; this surprise is the starting
point for our derivation of the geometric interaction in Section \ref{sec:map}.
The free energy of $N_{v}$ vortices on a bump bounded by a
circular wall at distance $R$ from its center is given by
\begin{eqnarray}
\frac{E}{4\pi^2 K} &=& \frac{1}{2}\sum_{j \neq
i}^{N_{d}} n_{i} n_{j} \!\!\!\!\! \quad \Gamma^{D}(x_{i};x_{j}) +
\sum_{i=1}^{N_{d}}
\frac{{n_{i}}^{2}}{4 \pi} \ln \left[1-x_{i}^{2}\right] \nonumber\\
&-& \sum_{i=1}^{N_{d}}\frac{{n_{i}}^{2}}{4 \pi}U_G(r_{i}) +
\sum_{i=1}^{N_{d}} \frac{{n_{i}}^{2}}{4 \pi} \ln
\left[\frac{\mathcal{R}(R)}{a}\right] \!\!\!\! \quad . \label{eq:longD1}
\end{eqnarray}
The Green's function expressed in scaled coordinates reads
\begin{eqnarray}
\Gamma^{D}(t_{i};t_{j})=\frac{1}{4 \pi} \ln\left(\frac{1 +t_{i}^{2}
t_{j}^{2} - 2 t_{i} t_{j} \cos \left(\phi_{i}- \phi_{j}\right)
}{t_{i}^{2}+t_{j}^{2}-2t_{i} t_{j} \cos \left( \phi_{i}-
\phi_{j}\right) }\right) \!\!\!\! \quad . \nonumber\\
\label{eq:green-norm1}
\end{eqnarray}
where $\phi_i$ is the usul polar angle and 
the dimensionless vortex position $t_{i}$ is defined by
\begin{eqnarray}
t_{i} \equiv \frac{\mathcal{R}(r_{i})}{\mathcal{R}(R)} \!\!\!\! \quad .
\label{eq:scaled-coord1}
\end{eqnarray}
Eq.(\ref{eq:longD1}) is now cast in a form equivalent to the flat
space expression apart from the third term which results from the
curvature of the underlying substrate and vanishes when
$\alpha=0$. However, we emphasize that the Green's function
$\Gamma^{D}$ also is modified by the curvature of the surface and
thus depends on $\alpha$.

The contributions from the second term and the numerator of the Green's
functions in the
first term account for the interaction of each vortex with its own
image and with the images of the other vortices present on the
bump (see \cite{geomgenerate}). If $R\gg r_0$, and all the vortices
are near the top of the bump (i.e., $r_i\sim r_0$) then these
boundary effects may all be omitted when determining equilibrium positions,
as the forces which they imply are on
the order of $K\frac{r_0}{R^2}$, small compared to the intervortex forces
and geometric forces, which have a typical value of $\frac{K}{r_0}$.

\begin{figure}
\psfrag{a}{A}
\psfrag{b}{B}
\psfrag{c}{C}
\psfrag{d}{D}
\psfrag{e}{E}
\psfrag{f}{F}
\includegraphics[width=.47\textwidth]{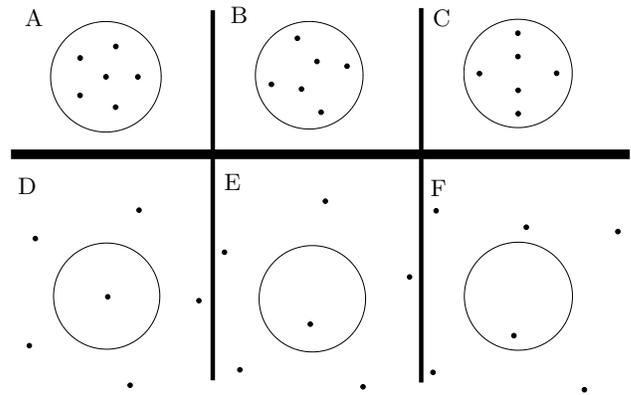}
\caption{\label{fig:colonies} Arrangements of $6$ vortices
that can occur on a curved surface. A circle of radius $r_0$
is drawn to give a sense that the confinement is tighter
in the top row ($\Omega=9\frac{\hbar^2}{mr_0^2}$) than in
the bottom row ($\Omega=\frac{\hbar^2}{mr_0^2}$).
The upper row shows the patterns
which occur at large angular frequencies
($\Omega=9\frac{\hbar^2}{mr_0^2}$). The transition from
the pentagon to the rectangle with two interior points is
discontinuous, and there is a range of aspect ratios $2.1<\alpha<2.7$
where both configurations are metastable.  The third configuration
is nearly degenerate with the second configuration. The lower
row shows the configurations which occur for $\Omega=\frac{\hbar^2}{mr_0^2}$
as $\alpha$ increases. The first transition is continuous and caused
by the central vortex's being repelled from the top
by the geometric interaction. The third configuration is similar to the second
large $\Omega$ configuration
but the effect of the geometric repulsion is seen in its asymmetry.}
\end{figure}

Let us imagine rotating the superfluid, so that each vortex is confined
by a potential of the form Eq.(\ref{eq:area}).
In flat space, the locally stable configurations usually involve concentric 
rings of vortices\cite{campbell}. In particular, there are two
stable configurations of six vortices. The lower energy
configuration has one vortex in the center and five in a pentagon surrounding
it. The other configuration, six vortices in a hexagon, has a slightly higher
energy, and Ref. \cite{yarmchuk} saw the configuration
fluctuating randomly between the two, 
probably due to mechanical vibrations since thermal oscillations
would not be strong enough to move the vortices.  
(The experiment
used a $D_0=2$ cm high column of superfluid; if one
regards the problem as two dimensional by considering flows that
are homogeneous in the $z$ direction, $\rho_s=D_0\rho_3$ is so large
that $K=\frac{\hbar^2}{m^2}\rho_s$ is on the order of millions of degrees
Kelvin.) There are no other stable configurations.  However, on
the curved surface of a bump, there are several more configurations which
can be found by numerically minimizing Eq. (\ref{eq:longD1});
the progression of patterns as $\alpha$ increases depends
on how tightly confined the vortices are compared to
the size of the bump, as illustrated in Fig. \ref{fig:colonies}. If the
vortices are tightly confined, the interactions of the vortices (which are different in curved space) stabilize the new vortex arrangements.  
If the vortices are spaced
far apart, the geometric interaction between the bump and the central
vortex causes a transition akin to the decentering transition in the previous
section.

For example, if $\Omega=9\frac{\hbar^2}{mr_0^2}$,
then at $\alpha=0$, the five off-center vortices start out 
in a ring of radius $.6 r_0$.
This pentagonal arrangement (see Fig. \ref{fig:colonies}A) 
 is locally stable for $\alpha<\alpha_1=2.7$. 
However, for
$\alpha>\alpha_2=2.1$, another arrangement with less symmetry is also stable 
(see
frame B of Fig. \ref{fig:colonies}), and above $\alpha_1$ it takes over
from the pentagon. For $\alpha_2<\alpha<\alpha_1$, both arrangements are
locally stable, with the asymmetric shape becoming energetically favored at some
intermediate aspect ratio. (There is also a third arrangement which coexists
with the less symmetric arrangement
for the larger aspect ratios, seen in the frame C of Fig. \ref{fig:colonies}.)
In the plane, the configuration labelled B, for example, is unstable,
 because 
the outer rectangle of vortices can rotate through angle $\epsilon$, decreasing its
interaction energy with the two interior vortices while keeping the rotational
confinement energy constant. 
(That the interaction energy decreases can be demonstrated
by expanding it in powers of $\epsilon$.) 
Because the Green's functions are different on the
curved surface (they do not depend solely on the distance between the
vortices in the projected view shown), figures B and C are stabilized.

At lower rotational frequencies, the equilibria which occur are
even less symmetric. For $\Omega=\frac{\hbar^2}{mr_0^2}$, the
vortices form a pentagon of radius $r_0$ when the surface is flat. 
This pentagon is far enough away that it has a minor influence
on the central vortex, which undergoes a transition similar to
the one discussed in Sec. \ref{subsec:single}.  At
$\alpha_1'=1.4$, the central vortex moves off-axis (the transition
is continuous), causing only a slight deformation of the pentagon (see
frame D of Fig. \ref{fig:colonies}).
As for the single vortex on a rotating bump, the
geometric potential has pushed the central vortex away from the maximum, and
the other vortices are far enough away that they are not influenced much.
At higher aspect ratios, the figure distorts further, taking a shape
similar to the one which occurs for $\Omega=9\frac{\hbar^2}{mr_0^2}$,
but offset due to the geometric interaction. For these two rotational
frequencies the
hexagonal configuration is less stable than the pentagon; it will not
take the place of the pentagon once the pentagon is destabilized.
The hexagon is of course metastable for nearly flat surfaces.

\subsection{\label{subsec:abr}Abrikosov lattice on a curved surface}

As in Section \ref{subsec:cande} 
the geometric potential will have significant consequences only
when the number of vortices near each
 geometrical feature such as a bump is of order unity. 
As an example, consider the triangular vortex lattice that forms
at higher rotational frequencies ($\Omega\gg \frac{\hbar}{mr_0^2}$
is the criterion for a large number of vortices to reside on top of the Gaussian
bump). In flat space,
 the vortex number density is approximately constant and equal to \cite{tilleybook}
\begin{equation} \nu({\bf u})= \frac{4\pi m\Omega}{\hbar}=\frac{2\Omega}{\kappa} \!\!\!\! \quad .
\label{eq:dens}
\end{equation}
At equilibrium, the force exerted 
on an arbitrary vortex as a result of the rotation exactly
balances the force resulting from the interaction with the other
vortices in the lattice and from the anomalous coupling to the
Gaussian curvature. We can determine the distribution
on a curved substrate by making the continuum approximation to Eq.
(\ref{eq:endeq}).  
The sum of delta-functions $\sigma$ gets replaced by $2\pi\nu(r)$ and the
self-charge subtraction can be neglected in the continuum approximation.  The Gaussian curvature can be neglected because it is small compared to the large density of vortex charge.
Upon
applying Gauss's theorem to the vortex charge distribution
in an analogous way to Section
 \ref{subsec:anomalous}, we
find that the force on a vortex at radius $r$ is given by
\begin{equation}
F_v=\frac{1}{r}\int_0^r 4\pi^2K\nu(r')r'\sqrt{1+h'^2}dr'
\label{eq:field1}
\end{equation}
while the rotational confinement force, obtained by differentiating
Eq. (\ref{eq:Lz-c}), is
\begin{equation}
F_{\Omega}=-\rho_s\frac{2\pi\hbar\Omega}{m} r.
\label{eq:Fomega}
\end{equation}
Balancing the two forces leads to an areal density of vortices,
\begin{equation}
\nu(r)=\frac{ m\Omega}{\pi\hbar\sqrt{1+h'^2}}.
\label{eq:n}
\end{equation}

Eq.(\ref{eq:n}) has a succinct geometric interpretation: the
vortex density $\nu(r)$ arises from distributing the vortices
on the bump so that the projection of this density on
the $xy$ plane is uniform and equal to the flat space result.  
The superfluid tries to mimic a rigidly rotating curved body 
as much
as possible  given that the flow must be
irrotational outside of vortex cores as for the case of a rotating
cylinder of helium \cite{tilleybook}.  To check this, first notice
that the approximate
rigid rotation entails a flow speed of $\Omega r$
at points whose projected distance from the rotation axis is $r$. Hence,
the circulation increases according to the quadratic law
$\oint {\mathbf{v\cdot dl}}=2\pi\Omega r^2$. Since this quantity is proportional
to the \emph{projected} area of the surface out to radius $r$, 
the discretized version
of such a distribution would consist of
vortices, each with circulation $\kappa=\frac{2\pi\hbar}{m}$,
 with a constant \emph{projected}
density $\frac{2 \Omega}{\kappa}$ as in flat space.
This result can be generalized with some effort
to any
surface rotated at a constant rate, whether the surface is
symmetric or not.  

The geometric force has to compete with the interactions among the
many vortices expected
at high angular frequencies. More precisely, the maximum force at radius
$r_0$ according to Eq. (\ref{eq:aziforce}) is of order $\frac{K\pi}{r_0}$ while
the force due to all the vortices Eq. (\ref{eq:field1})
is of order $K\frac{(2\pi)(\pi r_0^2)(2\pi\nu(0))}
{2\pi r_0}=2\pi^2 K r_0 \nu(0)$. The last expression greatly exceeds
$\frac{K\pi}{r_0}$ in the limit of high angular velocity.
The geometrical repulsion 
leads to a small depletion of the vortex density of the order of one vortex in
an area of order $\pi r_0^2$. 

The vortex arrangements produced by rotation are reminiscent of Abrikosov lattices in a superconductor \cite{Vine}.  
In fact an analogy exists between a \emph{thin film}
of superconductor in a magnetic field and a rotating film of superfluid.
A major difference between bulk superfluids and bulk superconductors
is that the vortices in a bulk superconductor have
an exponentially decaying interaction rather than a logarithmic one because of the magnetic field (produced by the vortex current) which screens
 the supercurrent. The analogy is more appropriate
in a thin superconducting film,
where the supercurrents (being confined to the film) 
produce a much weaker magnetic field. In fact, Abrikosov vortices in a 
superconducting film exhibit helium-like unscreened
logarithmic interactions out to length scales of order $\lambda'=\frac{\lambda^2}{D}$ where $\lambda$ is the bulk London penetration depth and $D$ is the film 
thickness (see \cite{pearl} and, for a review, section 6.2.5 of 
\cite{nelsonbook}). Our results on helium superfluids 
without rotation therefore apply
also to vortices in a curved superconducting layer in the absence of an
\emph{external} magnetic field. Curved superconducting layers in external 
magnetic fields can be understood as well by replacing the magnetic field by 
rotation
of the superfluid.
Let us review the analogy between a container of superfluid helium
rotating at angular speed ${\bf \Omega}$ and a superconductor in a
magnetic field ${\bf H}$ \cite{Vine}.  Note that in Eq. \ref{eq:Erot},
$E$ is given by $\frac{1}{2}\rho_s\iint d^2\mathbf{u}\frac{\hbar^2}{m^2}
\left(\nabla\theta\right)^2$ and $\hbar\nabla\theta$ is the \emph{momentum}
in the rest frame, $\bm{p}$, although we are working in the rotating frame
(the frame in which a vortex lattice would be at rest).
For helium, the momentum in
the rest frame ${\bf p}$ is related to the momentum in the
rotating frame ${\bf p'}$ by the ``gauge" transformation
\begin{eqnarray}
{\bf p} \rightarrow {\bf p'} + m \!\!\!\!\! \quad {\bf r} \times
{\bf \Omega} \!\!\!\! \quad . \label{eq:gauge1}
\end{eqnarray}
Similarly, in the case of a superconductor the momentum ${\bf p}$
in the absence of a magnetic field is related to the momentum
${\bf p'}$ in the presence of the field by the familiar relation
\cite{Tinkhambook}
\begin{eqnarray}
{\bf p} \rightarrow {\bf p'} + \left(\frac{e}{c}\right){\bf A}
\!\!\!\! \quad , \label{eq:gauge2}
\end{eqnarray}
where ${\bf A}$ is the vector potential. Comparison of
Eq. (\ref{eq:gauge1}) and Eq. (\ref{eq:gauge2}) suggests a
formal analogy between the two problems,
\begin{eqnarray}
{\bf A} \leftrightarrow \left(\frac{m c}{e}\right){\bf r} \times
{\bf \Omega} \!\!\!\! \quad . \label{eq:gauge3}
\end{eqnarray}
Eq. (\ref{eq:gauge2}) establishes a correspondence between the
angular velocity ${\bf \Omega}$ and the magnetic field ${\bf H}$
that allows to convert most of the relations we derived for helium
to the problem of a superconducting layer, with the identification
\begin{eqnarray}
{\bf \Omega} \leftrightarrow \left(\frac{e}{2 m c}\right) {\bf H}
\!\!\!\! \quad . \label{eq:gauge4}
\end{eqnarray}
Of course, one should keep the external magnetic field small so that
a dense Abrikosov lattice does not form, since (as for superfluids) when there
are too many vortices, the curvature interaction is overcome by the vortex
interactions.

\section{\label{sec:experimental}Experimental Considerations}

Vortices in bulk fluids  are extended objects such as curves connecting opposite boundaries, rings or knots. A vortex interacts with itself and with its image generated by the boundary of the fluid. However,
if the vortex is curved, such forces (the three-dimensional generalization
of the geometric force) are usually dominated by a force which 
depends on the curvature of the vortex called the
``local induction force."  This force 
 has a strength per unit length \cite{saffman} of
\begin{equation}
f_{LIA}=\pi\frac{\hbar^2}{m^2}\rho_3 \kappa \ln\frac{1}{\kappa a},
\label{eq:LIA}
\end{equation}
 where $\rho_3$ is the bulk superfluid density and $\kappa$ is the curvature
of the vortex at the point where this force acts.
This force is in danger of dominating the long range
forces because of the core size appearing in the logarithm.

``Two-dimensional" regions are a special case of three-dimensional
regions in which two of the boundaries are parallel and at a distance $D_0$ much less than the radius of curvature of the boundaries. The two dimensional superfluid density is given by $\rho_s=\rho_3 D_0$,
and the interactions of the vortices should be captured by the
two dimensional theory described in this paper once this substitution is made.  
A discrepancy will occur, however, if the boundaries of the film
are not exactly parallel because the vortices are
forced to curve in order to meet both boundaries at right angles. In this case, there is a force which is
a relic of the local induction force (see Sec. \ref{sec:uneven}),
\begin{equation}
F_{th}=-\frac{\pi\hbar^2}{m^2}\rho_s \frac{\nabla D}{D_0}\ln\frac{r_0}{a},
\label{eq:relic}
\end{equation}
where $r_0$ is the relevant curvature scale.  According to this formula,
vortices are attracted to the thinnest portions of the film.
We will need to ensure that the thickness of the film is uniform
enough so that this force does not dominate over the geometric interactions we are interested
in. 

There is a  maximum film thickness for which the geometric
force is relevant. The most stringent
requirement arises from demanding that
the van der Waals force causes wetting
of the surface with a sufficiently uniform film.
Van der Waals forces compete against gravity, which thickens
the superfluid at lower portions of the substrate, and surface tension, which thickens the superfluid where the mean curvature of the substrate is
negative. Both gravity and surface tension
thin the film on hills and thicken it in valleys, but if the 
film is thin enough, the van der Waals force can keep the nonuniformity very
small.
Section \ref{subsec:nucleation} discusses the critical speeds for the nucleation of
vortices in thin films, which are typically higher 
than those required in long thin rotating cylinders \cite{yarmchuk}.  We assume
that vorticity is not created from scratch, but from
pinned vortices present even before the rotation has begun \cite{tilleybook}.
Finally in Secs. \ref{sec:uneven} and \ref{sec:unevenexptl} a comparison is made between forces on
vortex lines in three-dimensional geometries and on point vortices in
two dimensions. 


\subsection{\label{subsec:thickness}The Van der Waals force and thickness
variation}

We start by providing an estimate of the
variation in the relative thickness 
\begin{equation}
\epsilon \equiv \frac{D_t-D_0}{D_{0}}. \label{eq:epsilon}
\end{equation}
for a liquid layer which wets a bump and apply it to thin helium films. 
$D_t$ denotes the thickness on top of the bump
and $D_0$ is the thickness far away.
The wetting properties of
very thin films ($\sim$ 100 \AA) of dodecane on polymeric fibers
of approximately cylindrical shape have been thoroughly
investigated in \cite{Quer89}. We start by reviewing a
theoretical treatment of the statics of wetting on rough surfaces
by \cite{Ande88}. A film on a solid substrate that
is curved  has a mean curvature determined by the shape of the substrate,
unlike in the case of a large drop of water on a non-wetting surface.  By
choosing an appropriate shape, the drop can adjust its mean curvature
(and thereby balance surface tension against gravity).  The shape
is therefore described by a differential equation.
A thin film on a solid substrate, in contrast,
has approximately the same curvature as
the substrate that it outlines.

\begin{figure}
\psfrag{hL}{$h_L$}
\psfrag{h}{$h$}
\psfrag{D}{$D$}
\psfrag{(x,y)}{$(x,y)$}
\includegraphics[width=.45\textwidth]{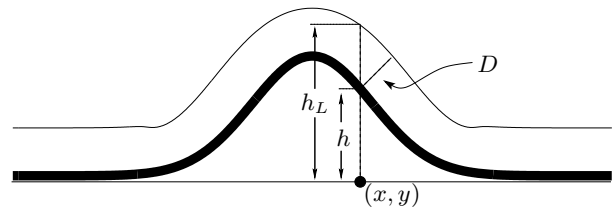}
\caption{\label{fig:paralgau} Definition plot for
a laminating film. $h(\mathbf{x})$ is the height of the
substrate
above the horizontal surface at a point $\mathbf{x}=(x,y)$,
and $h_L(\mathbf{x})$ is the
height of the upper surface of the film. $D(\mathbf{x})$ is the
thickness of the film which (if the film has a slowly varying
thickness) is given by $(h_L(\mathbf{x})-h(\mathbf{x}))\cos\theta(\mathbf{x})$
where $\theta(\mathbf{x})$ is the local inclination angle of the substrate.}
\end{figure}

Consider a film that completely wets a solid surface. The surface
itself is described by
its height function $h({\bf x})$, where {\bf x} denotes a pair of
Cartesian coordinates in the horizontal plane below the surface (see Fig. \ref{fig:paralgau}).
The height function for the liquid-vapor interface $h_{L}(\bf x)$ can
be determined by minimizing the free energy $F$,
\begin{alignat}{1}
F = \iint d^2{\bf x} &[\gamma \sqrt{1+|\nabla h_{L}({\bf
x})|^2} +
\frac{\rho_3 g}{2}(h_L(\mathbf{x})^2-h(\mathbf{x})^2) \nonumber\\&-
\mu \!\!\!\!\! \quad
(h_{L}({\bf x})-h({\bf x})) ]\nonumber\\+
\iint d^2\mathbf{x}&\int_{h_L(\mathbf{x})}^{\infty}
dz \iint d^2\mathbf{x'}\int_{-\infty}^{h(\mathbf{x'})}dz'\nonumber\\
&\ \ w(\sqrt{(\mathbf{x}-\mathbf{x'})^2+(z-z')^2}) \!\!\!\!\!\! \quad ,
\label{eq:wet-1}
\end{alignat}
where $\gamma$, $\rho_3$, and $\mu$ are respectively the liquid-vapor surface
tension, the total mass density, and
the chemical potential (per unit volume). (Note that $\nabla$ here is
not the covariant gradient for the surface; it is the gradient in the $xy$
plane.) The second term describes the gravitational potential energy
integrated through the thickness of the film.
The second and fourth terms model the force between the helium atoms and the substrate assuming for simplicity a non-retarded van der Waals interaction.  The last term involves
an integral over interactions between pairs of points, one
above the helium film and one in the substrate, but with no points 
in the liquid helium itself.   This is equivalent to
including interactions between all pairs of atoms contained in all
combinations of the vapor, liquid and solid regions, as long as
$w(r)=- \!\!\!\!\! \quad \alpha \!\!\!\!\! \quad
r^{-\!\!\!\!\! \quad 6}$ where $\alpha$
is the appropriate combination of parameters for
these phases \cite{Ande88}. 


Minimization of
Eq. (\ref{eq:wet-1}) leads to a differential equation for
$h_{L}({\bf x})$ that is a suitable starting point for evaluating
the profile of the liquid-vapor interface numerically \cite{Ande88}.
In what follows, we will instead work within an
approximation valid when $D_0\ll r_0,h_0$; in this case, the curvature of
the film is fixed. The $local$ film thickness is described by
$D({\bf x})=(h_L({\bf x})-h({\bf x}))/\sqrt{1+|
\nabla h({\bf x})^2)|}$, see Fig.(\ref{fig:paralgau}).
We need to determine
how each contribution to the free energy
per unit area at a point $\mathbf{u}$
is changed by an increase in thickness $\delta D({\bf x})$. 

First let us
consider the variation of the van der Waals energy in order to understand
how this attraction sets the thickness of the film.
When the film thickens by $\delta D$ over a small area $A$ of the film (centered
at $\bm{x},z$),
the change in the van der Waals energy
 is given by $-A\delta D\Pi(\mathbf{x})$ where the disjoining pressure
is
\begin{equation}
\Pi(\mathbf{x})=
\iint d^2\mathbf{x'}\int_{-\infty}^{h(\mathbf{x'})}dz'w(\sqrt{(\mathbf{x}-\mathbf{x'})^2+(z-z')^2}).
\label{eq:wet-4}
\end{equation}
For a film on a horizontal surface at $h=0$,
the surface area and gravitational potential energy
do not increase when the film is thickened.
The equilibrium thickness is determined by balancing the variation of
the chemical potential contribution, $-\mu A\delta D$, against the
disjoining pressure, giving $\mu=-\Pi(D_0)$.
The disjoining pressure obtained by integrating Eq. (\ref{eq:wet-4}) 
for a flat surface is
\begin{equation}
\Pi (D) = - \frac{A_{H}}{6\pi D_0^3}
 \!\!\!\! \quad . \label{eq:wet-3}
\end{equation}
$A_{H}=\pi^2\alpha$ is the Hamaker constant for the solid and the vapor
interacting across a liquid layer of thickness $D_0$
\cite{Israelachvili-book}. One sees that a negative value of $A_H=\pi^2\alpha$
is necessary for wetting.
The equilibrium thickness is
\begin{equation}
D_{0}=\sqrt[3]{\frac{A_H}{6 \pi \mu}} \!\!\!\! \quad.
\label{eq:wet-5}
\end{equation}
(For example, liquid $^4$He on a CaF$_2$ surface has $A_H\approx -10^{-21}$ J,
and has a liquid-vapor surface tension of $3\times 10^{-4}$ J/m$^2$.)
When there is a bump on the surface, Eq. (\ref{eq:wet-5}) gives
the equilibrium thickness far from
the bump. Note that both $A_H$ and $\mu$ are negative in this
expression. Increasing $\mu$ therefore increases the thickness of the film
as expected.

Now let us continue by considering the effects of gravity and surface 
tension for a curved substrate. 
The increase in gravitational potential energy is
$\rho_3 g h\delta D$, just because there is an additional mass per unit area of
the fluid
$\rho_3 \delta D$ at height $h$. (The additional elevation from adding the fluid at
the top of the fluid that was already present can be ignored if the layer 
is very thin.)
The variation of the surface tension energy can
be related to the mean curvature \cite{Kami02}
using the fact that
the area of a small patch of the liquid vapor interface $A({\bf
x})$ (at a distance $D$ from the substrate) is related to the
corresponding area of the solid surface $A_{0}({\bf x})$ by the
relation \cite{Hide-book}
\begin{equation}
A({\bf x}) = A_{0}({\bf x}) \left[1 + 2 H({\bf x}) D({\bf x})+
G({\bf x}) D^2({\bf x}) \right] \!\!\!\! \quad . \label{eq:wet-6}
\end{equation}
The second term is proportional to the mean curvature $H=\frac{1}{2}(\kappa_1+\kappa_2)$
of the surface, where we use the convention that the principal curvatures 
$\kappa_1,\ \kappa_2$
are positive when the surface curves away from the outward-pointing 
normal. The last term, proportional to the Gaussian curvature, can be
ignored relative to the previous term since it is smaller by a factor of
$\frac{D_0}{r_0}$. The mean curvature of the upper surface of the fluid is
nearly the same as for the substrate,
so the energy required to increase the thickness
of the film is $2\gamma H({\bf x})\delta D$. For example, at the top of
the bump, an increased thickness leads to an increased area, so surface
tension prefers a smaller thickness there. Gravity also
thins the film at the top of the bump so that vortices are attracted to the top.

Now we must balance these forces against the disjoining pressure.  The flat space form
of the disjoining pressure is not significantly altered by the
curvature of the substrate for very thin films.
According to Eq. (\ref{eq:wet-4}), the disjoining pressure 
is the sum of all the van der Waals interaction energies between the points of 
the substrate and a fixed point at the surface of the film. The integral 
(evaluated in Appendix \ref{app:vdW})
for a point at a distance
$D$ away from the substrate shows
\begin{equation}
\Pi[D({\bf x})] \approx \frac{-A_{H}}{6 \pi D({\bf x})^3} \left(1
- \frac{3}{2} \!\!\!\! \quad H({\bf x}) D({\bf x}) \right)
\!\!\!\! \quad . \label{eq:wet-7}
\end{equation}
The curvature correction in the second term of Eq.(\ref{eq:wet-7})
arises (when $H>0$ as at the top of the bump)
because the surface bends away from the vapor
molecules which interact only with the very nearest atoms of
the solid substrate. This effect is small if $D_0<<r_0$ and
will be neglected in our estimates.

We can now collect the various contributions to set up a pressure
balance equation that allows us to estimate the relative change in
layer thickness $\epsilon$ defined in Eq. (\ref{eq:epsilon}). This
equation reads:
\begin{equation}
\frac{A_{H}}{6 \pi D({\bf x})^3} + 2 \!\!\!\!\! \quad \gamma
\!\!\!\!\! \quad H({\bf x}) + \rho_{3} \!\!\!\!\! \quad g
\!\!\!\!\! \quad h({\bf x}) - \mu = 0 \!\!\!\! \quad .
\label{eq:wet-8}
\end{equation}
Apart from the lengths $r_{0}$, $h_{0}$ and $D_0$ inherited
from the geometry of the system, it is convenient to define three
characteristic length scales $\delta$, $\varrho$ and $l_c$,
obtained by pairwise balancing of the first three terms of
Eq. (\ref{eq:wet-8}):
\begin{eqnarray}
\delta &\equiv& \sqrt{\frac{-A_{H}}{6 \pi \gamma}}  \nonumber\\
\varrho &\equiv& \sqrt[4]{\frac{-A_{H}}{6 \pi \rho _{3} g }} \nonumber\\
l_{c} &\equiv& \sqrt{\frac{\gamma}{\rho_{3} g }} \!\!\!\! \quad .
\label{eq:wet-9}
\end{eqnarray}
The last relation in Eq.(\ref{eq:wet-9}) defines the familiar
capillary length \cite{Guyon-book} below which surface tension dominates
over gravity while the first and the second
give the length scales involving the disjoining pressure. 
For $^4$He on CaF$_{2}$, $\delta \simeq 10$ \AA  (for most liquids it
is one order of magnitude less), $\varrho \simeq 0.7 \mu m$ and
$l_c \simeq 0.4 mm$.

Upon substituting Eq. (\ref{eq:wet-5}) in Eq. (\ref{eq:wet-8}), we obtain
an approximate relation between $D({\bf x})$ and $D_{0}$,
\begin{eqnarray}
\frac{D({\bf x})}{D_0} \approx 1- \frac{D_{0}^3}{3} \left(\frac{2
H({\bf x})}{\delta ^2} + \frac{h({\bf x})}{\varrho ^4} \right)
\!\!\!\! \quad . \label{eq:wet-10}
\end{eqnarray}
This relation leads to an estimate of the relative change in layer thickness, $\epsilon$, valid for thin
films (if we take $\alpha\sim 1$), namely
\begin{eqnarray}
\epsilon \sim \frac{D_{0}^3}{r_{0} \delta ^2} \left(1 +
(\frac{r_{0}}{l_{c}})^2 \right)\!\!\!\! \quad , \label{eq:wet-11}
\end{eqnarray}

Now the thickness-variation force (Eq. (\ref{eq:relic}))
is small compared to the coupling to the geometry only if 
\begin{equation}
|\epsilon|<\frac{\alpha^2}{2\ln\frac{r_0}{a}}.
\label{eq:epsilonmanners}
\end{equation}
where we estimated the maximum of the geometrical force to be $\frac{K\pi\alpha^2}{2r_0}$, see Eq. (\ref{eq:aziforce}).  
This limit on $\epsilon$ leads in turn to an upper bound for the film thickness $D_0$ for each choice
of $r_0$. Assuming $\alpha\sim 1$ and
splitting into cases according to the size of the bump gives the limits:
\begin{eqnarray}
D_0\lesssim \frac{r_0^{\frac{1}{3}}\delta^{\frac{2}{3}}}
{(\ln\frac{r_0}{a})^{\frac{1}{3}}}\ \ \mathrm{for}\ r_0<l_c
\label{eq:badsurften}\\
D_0\lesssim \frac{\rho^{\frac{4}{3}}}
{(r_0\ln\frac{r_0}{a})^{\frac{1}{3}}}\ \ \mathrm{for}\ r_0>l_c.
\label{eq:badgravity}
\end{eqnarray}
For smaller bumps, surface tension plays the main role in creating
thickness variation and for larger bumps, gravity has the largest
effects.  In order for the vortices to be easily observable,
$D_0$ should be as large as possible.  Eqs. (\ref{eq:badsurften}) and
(\ref{eq:badgravity})
show that the film
can be made thickest (while retaining its approximate uniformity) when
gravity and surface tension have comparable effects, $r_0\simeq l_c$.  
The optimal thickness
(calculated with $\alpha=1$ and for the Gaussian bump; similar numbers
are optimal for the saddle surface) is about $150$ \AA\ at 
$r_0=.5$ mm.\footnote{Interestingly, even 
without rotating the bump, there are two radii
where the vortex could rest for a film
thickness of $200$ \AA.  Then the confinement due to the
varying film thickness and
the geometric repulsion are comparable in magnitude, producing
an equilibrium off-center position for the vortex.} 
A smaller
value of $r_0$ might be required if there is a lower
limit $\Omega_{min}$ on the rotational frequency as discussed in the next
section, requiring a slightly thinner film, about $100$ \AA.

The restriction on the film thickness is the most serious obtacle to studying
two-dimensional superflows experimentally. The method of observing vortices
described in \cite{yarmchuk} requires the vortices to be long enough to
be able to trap an observable number of electrons. If
the method of \cite{yarmchuk} turns out to be unsuitable for thin films and
an alternative method cannot be found, one might also study a saturated
superfluid layer confined
between two solid surfaces. In
this case, none of the considerations on wetting are relevant, and the 
``film" could have a large thickness. 
The two solid surfaces would have to be parallel to one another, and hence
not congruent.  (Congruent surfaces displaced
by a fixed distance in the \emph{vertical} direction lead to a
thickness varying 
as $\cos\theta(\mathbf{x})$ where $\theta$ is again the inclination angle.) The two surfaces would have to be very accurately
shaped in order to make the film uniformly thick.

Another concern is that a vortex may be pinned to an irregularity on
the substrate strongly enough that it will not
move to the location favored by the geometrical force.
On the other hand, the geometrical force is much stronger than
random forces due to thermal energy. 
Even with a film as thin as hundreds of Angstroms, the geometric force is very
strong. With $\rho_s=\rho_3 D_0=.2\rm{g}/\rm{cc} D_0$,
which assumes that the superfluid density
is not depleted too much by thermal effects or by the thinness of the film,
the value of $K=\frac{\rho_s\hbar^2}{m^2}$ is
about $40$ Kelvin. Since the potential wells which trap the vortices
on the saddle surface or on the rotating Gaussian bump
have depths on the order of $K\alpha^2$ the geometric force
will be strong enough 
to prevent the vortex from wandering out of the trap due to thermal Brownian
motion except very close to the Kosterlitz-Thouless transition where
$\rho_s$ is depleted.

\subsection{\label{subsec:nucleation}Parameters for the rotation experiment}

A practical consideration can limit the thickness of the film further.
In the experiments of \cite{yarmchuk}, uniform rotations slower than
$\Omega_{min}=1$ rad/s or so were hard to attain. The maximum
value of $\Omega$ for which the geometrical force can displace
the vortex as in Sec. \ref{subsec:single} can be determined by taking the limit
 $r_m\rightarrow 0$ in Eq. (\ref{eq:constr}); this shows
that for a given $r_0$,
the vortex is displaced from the top of the bump only if
\begin{equation}
\Omega<\frac{\hbar\alpha^2}{4mr_0^2},
\label{eq:rotnvsgeom}
\end{equation}
Hence, if details of the rotational apparatus require
that $\Omega> 1 $ rad/s, then the size of the bump should
be less than $.1$ mm (for $\alpha=1$). Consequently Eq. (\ref{eq:badsurften})
requires even thinner films than for the case discussed in
the previous section where the bump size is determined purely
by the natural forces of gravity and surface tension
and is equal to the capillary length.

An additional issue is whether it is possible to create just one
vortex reliably.
In practice, the number of vortices in a rotating film is a property
of a history-dependent metastable state whose dynamics are not 
completely understood (see Chapter 6 of Ref. \cite{tilleybook}). 
We can discuss this in the context
of a flat rotating disk. The critical frequency
for local stability of a single vortex at the
center of the disk is $\Omega_{stab1}=\frac{\hbar}{mR^2}$.
There is a barrier to creating such a 
vortex, so the frequency initially would have to be raised up to a higher
frequency in order for a vortex to form at the boundary and then move in to the center
of the rotating helium. The critical frequency for creating
vorticity at the boundary is
$\Omega_{\rm{crit}}$.
It is not clear what this critical speed 
is.  Perhaps
$\Omega_{\rm{crit}}$ for a thin disk of helium is determined
by the height $D$ of the disk;
the critical \emph{linear} speed\footnote{At
the critical speed, vortices are believed to form by breaking away from
pinned vortices.  This implies that the critical velocity is determined
by the shortest dimension $H$ of the chamber which is perpendicular to
the flow of the normal fluid.  The critical velocity is estimated in the remanent vorticity
theory\cite{vortexmill}
by assuming that ring vortices break off of pinned vortices stretching across this
shortest dimension and
then expand until they hit the surfaces of the superfluid and break into
two line vortices with opposite circulations. The critical speed is the speed
at which a ring vortex of size $H$
would expand due to the motion of the fluid rather than contract.
The critical velocity is determined by balancing the Magnus force
of the fluid moving past the vortex (which expands the vortex) against the attraction of the vortex
for itself, and it is $v_c=\frac{\hbar}{mH}\ln{\frac{H}{2a}}$. 
Thus the critical speed for a narrow cylinder is determined
by setting $H$ equal to the radius $R$ of the cylinder so
$\omega_c=v_c/R=\frac{\hbar}{mR^2}\ln\frac{R}{a}$, as seen experimentally
in \cite{yarmchuk}.  
For a thin film on a disk, the shortest
dimension is the height, so $H=D$.}
would
then be $R\Omega_{\rm{crit}}\sim \frac{\hbar\ln\frac{D}{a}}{mD}$
\cite{criticalvinen}.
Now because $\Omega_{\rm{crit}}$ is so much greater
than $\Omega_{stab1}$, a lattice of many vortices will form. According
to Eq. (\ref{eq:n}), there will be about $\frac{m\Omega R^2}{\hbar}\sim\frac{R}{D}\ln\frac{D}{a}$ of them. By slowing down to just above
$\Omega_{stab1}$, where the rotational confinement is not
strong enough to confine more than one vortex (even metastably), one would
hope to retain just a single vortex.

\subsection{\label{sec:uneven}Films of varying thickness from
the three-dimensional point of view}

In a chamber of arbitrary shape the length of a vortex line
crossing through the chamber changes as the
vortex moves, and therefore knowledge of the core energy per unit length
is crucial for determining the forces experienced by the vortex.  The energy
of a length $D$ isolated vortex
in a cylinder of radius $R$ including the core energy is given by
\begin{eqnarray}
E_{line}&=&\rho_3 D\frac{\pi\hbar^2}{m^2}\ln\frac{R}{a}+\epsilon_{c3}D\nonumber\\
&=&\rho_3D\frac{\pi\hbar^2}{m^2}\ln\frac{R}{\tilde{a}}
\label{eq:linetension}
\end{eqnarray}
where $\tilde{a}=ae^{-\frac{m^2\epsilon_{c3}}{\rho_3\pi\hbar^2}}$ is the
core radius rescaled to account for the effects of the core energy \emph{per unit length}, 
$\epsilon_{c3}$.

\begin{figure}
\psfrag{d}{$D$}
\psfrag{X'}{$X'$}
\psfrag{X}{$X$}
\psfrag{S}{$\mathcal{S}$}
\psfrag{S1}{$\mathcal{S'}$}
\psfrag{beta}{$\beta$}
\includegraphics[width=.47\textwidth]{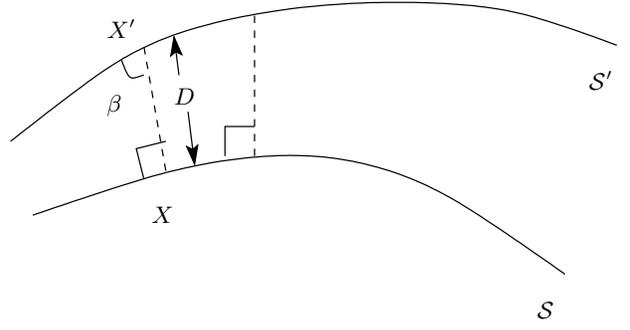}
\caption{\label{fig:railroad}
Illustration of the definition of $D(\mathbf{x})$, for nonparallel surfaces.  The
distance between $\mathcal{S'}$ and $\mathcal{S}$ is measured along the segment which is
perpendicular to $\mathcal{S}$ at $\mathbf{x}$.  The segment meets $\mathcal{S'}$ (obliquely) at 
a point which we call $\mathbf{x'}$. A segment displaced to the right is also drawn, illustrating
the derivation of Eq. (\ref{eq:ezekiel}).}
\end{figure}

A vortex line connecting two approximately parallel parts of the boundary feels
a force as a consequence of the variation of $E_{line}$.
Let the two nearly parallel boundaries be
$\mathcal{S}$ and $\mathcal{S'}$ as illustrated
in Fig. \ref{fig:railroad}. 
Let us define $D(\mathbf{x})$
to be the length of the line which is perpendicular to $\mathcal{S}$ at
$\mathbf{x}$ and which extends to $\mathbf{x'}$ on $\mathcal{S'}$ (see Fig. \ref{fig:railroad}). 
If a vortex is attached to $\mathcal{S}$ at $\mathbf{x}$
then
since
the fluid flow is fastest at the vortex, the energy depends primarily
on the thickness of the helium film where the vortex is located. This
thickness 
is approximately given by $D(\mathbf{x})$ no matter how the vortex
connects the two surfaces (unless it is extremely wiggily).
A useful fact is that, 
if the two surfaces are at a constant separation, then the line between
$\mathbf{x}$ and $\mathbf{x'}$ is perpendicular to both surfaces.
This can be derived by differentiating,
 $D(\mathbf{x})^2=|\mathbf{x}-\mathbf{x'}|^2$. Upon displacing
$\mathbf{x}$ slightly, $2 D\delta D=2(\mathbf{x}-\mathbf{x'})\cdot(\delta\mathbf{x}-\delta\mathbf{x'})=2(\mathbf{x'}-\mathbf{x})\cdot\delta\mathbf{x'}$, since the line connecting $\mathbf{x}$
to $\mathbf{x'}$ was chosen to be
perpendicular to $\mathcal{S}$.  Expressing the inner product in terms of the
angle $\beta$ indicated in Fig. \ref{fig:railroad}, this reads,
\begin{equation}
\delta D=|\delta\mathbf{x'}|\cos\beta\approx|\delta\mathbf{x}|\cos\beta.
\label{eq:ezekiel}
\end{equation}

In order to calculate the kinetic energy of a vortex in a curved film of
varying thickness, we will derive a more detailed form of
Eq. (\ref{eq:linetension}) for a curved film.
First, the core
energy due to the local disruption of superfluidity has the same form,
$\epsilon_{c3} D(\mathbf{x})$ where $D(\mathbf{x})$ is the thickness
of the film at the location $\mathbf{x}$ of the vortex.  The major
component of the energy is given by $\frac{\hbar^2}{2m^2}
\rho_3\iiint{d\zeta rdrd\phi\frac{1}{r^2}}$ where it is assumed that the flow is
parallel to the boundaries and roughly independent of $\zeta$,
the coordinate normal to $\mathcal{S}$ and approximately
normal to $\mathcal{S'}$.
We integrate over $\zeta$  and divide the remaining two-dimensional
integration into two parts:
\begin{alignat}{1}
E=\epsilon_c D(\mathbf{x})+&\frac{\hbar^2}{2m^2}
\big(\rho_3\iint_{r<L_{th}} D(r,\phi)\frac{drd\phi}{r}
\nonumber\\&+\rho_3
\iint_{r>L_{th}} D(r,\phi)\frac{drd\phi}{r}\big),
\label{eq:epsilonc}
\end{alignat}
where we are using polar coordinates centered on the location
of the vortex.
Here, $L_{th}$ is the distance over which $D$ varies appreciably
so that $L_{th}\sim\frac{D}{\nabla D}$. We regard the thickness
as a constant in the second term because the integral starts far enough away
from the vortex that it does not depend on the specific thickness of the
film at the location of the vortex. 
The first term may
be approximated by replacing $D(r,\phi)$ by $D(\mathbf{x})$ 
(the thickness of the film
at the location of the vortex, where the energy is very big).
Therefore, 
\begin{alignat}{1}
E\approx\epsilon_{c3} D(\mathbf{x})
+\frac{\pi\hbar^2}{m^2}&\rho_3 D(\mathbf{x})\ln\frac{D}{a|\nabla D|}\nonumber\\
&+\mathrm{
energy\ of\ the\ distant\ flow}.
\label{eq:youtalktoomuch}
\end{alignat}
The force obtained by taking the gradient of the kinetic energy
reads
\begin{equation}
\mathbf{F}=-\epsilon_c\nabla D-\frac{\pi\hbar^2}{m^2}\rho_3\nabla
D\ln\frac{D}{a\nabla D}+\mathbf{F_G}
\label{eq:shrinkforce}
\end{equation}
where $\mathbf{F_G}$ represents the variation in energy of the long range
portion of the flow, which includes the geometric forces.

We now show that the thickness variation force can be interpreted as the Biot-Savart self-interraction
of the curved vortex.  The net Biot-Savart force points towards the center of curvature of the vortex,
and so favors shrinkage of its length.  We will use the ``local induction approximation" to the Biot-Savart self-interaction, given in Eq. (\ref{eq:LIA}).   This expression results from integrating the interaction of a particular element of the vortex line with nearby elements; the peculiar dependence on the core radius $a$ arises from the need to place a cut-off in the diverging interaction for nearby elements.

\begin{figure}
\psfrag{S}{$\mathcal{S}$}
\psfrag{S1}{$\mathcal{S'}$}
\psfrag{beta}{$\beta$}\includegraphics[width=.47\textwidth]{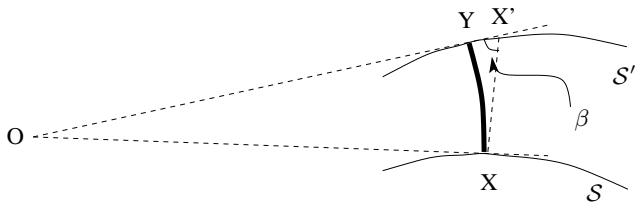}
\caption{\label{fig:crevice}A figure for reference in determining the
curvature of a vortex. If the upper and lower surface $\mathcal{S}$ and
$\mathcal{S'}$ are nearly but not quite parallel, the curvature of a vortex
connecting them can be estimated simply. Extend the tangents at
the end-points $X,Y$ of the vortex until they intersect at the center $O$
of curvature of the vortex (which happens because the vortex
is perpendicular to $\mathcal{S,S'}$). The radius of curvature, $OX$, can then be
found by using Eq. (\ref{eq:ezekiel}) and noticing that $OX=XX'\tan\beta$ and $\sin\beta\approx 1$.
}
\end{figure}

To see this, note that 
if the opposing boundaries $\mathcal{S},\mathcal{S'}$ are not at a constant
distance, the vortex must
curve in order to connect them, because it meets both boundaries at right
angles.  The resulting curvature of the vortex, $\kappa$, is equal to
$\frac{\nabla D}{D}$ (see Fig. \ref{fig:crevice}), so that the force per 
unit length given by
Eq. (\ref{eq:LIA}), multiplied by $D$, agrees with the force derived by the energy method, Eq. 
(\ref{eq:shrinkforce}). 
On the other hand, for films of constant thickness, the surfaces $\mathcal{S},\mathcal{S'}$
can be connected by straight vortices.  According to the Biot-Savart law a straight vortex has zero
interaction with itself.  In addition, there is still an interaction energy with its image, which
is where the geometric force, the third term in Eq. \ref{eq:shrinkforce}, comes from.
When the helium has boundaries, a distribution of  ``image vorticity" beyond the boundaries
can be supplied to simulate the effects of the boundary conditions \cite{saffman}.  The integral of the 
Coulomb interaction with the distribution of surface curvature (Eq. (\ref{eq:curvature-defect}) is reminiscent of the
long-range integral of the Biot-Savart interaction with the distribution of vorticity beyond the surface. 

\subsection{\label{sec:unevenexptl}Vortex Depinning}
\begin{figure}
\psfrag{A}{$A$}
\psfrag{B}{$B$}
\psfrag{C}{$C$}
\psfrag{D}{$D$}
\includegraphics[width=.2\textwidth]{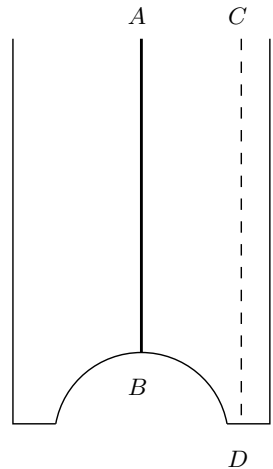}
\caption{\label{fig:last}
Sketch of the experimental geometry from Ref. \cite{zieve}.  This
experiment may give some indication about the difference in energy
between a vortex stretching from $A$ to $B$ and one
stretching from $C$ to $D$.  A wire is placed along $AB$ and
a vortex is formed around it.  Ref. \cite{zieve} found
that this vortex depins more easily than a vortex in a tube with a flat bottom.
This suggests that the energy of a vortex (when there is no wire) would
decrease by moving from $AB$ to $CD$, but estimates seem to rule out this
explanation.}
\end{figure}

Because of the divergence of the force described by Eq. (\ref{eq:relic}) in the limit 
$a\rightarrow 0$, it will be hard to see effects of the geometric force without
films of very nearly uniform thickness.
In \cite{zieve} an experiment is described in which a vortex extending
along 
a wire in a helium-filled  tube leaves the wire 
more easily
when the wire
is connected to a bump on the bottom of the tube
than when the bottom is flat. In this case, any geometric
repulsion from the bump should be negligible in comparison to the extra
energy associated with the stretching of the vortex.  
(See Fig. \ref{fig:last}.)
Geometric energies should be of order $R\rho_3\frac{\hbar^2}{m^2}$ 
where $R$ is the radius of the tube and
approximately the height of the bump,
 but the vortex has to stretch an additional
length on the order of $R$  in order to leave the bump.  Therefore,
the extra kinetic energy from Eq. (\ref{eq:linetension}),
$R\rho_3\frac{\hbar^2}{m^2}\ln\frac{R}{a}$, probably overwhelms the geometric
effects. Thus, the energy barrier for detachment of the vortex 
line should increase when the wire is attached to a bump.

This suggests
that  depinning is not simply caused by thermally activated
 barrier crossings (which would occur at the rate proportional
to $e^{-\Delta E/T}$).  In fact,
for the length scale of the bump in the experiment (a few millimeters), 
the additional energy
barrier due to the stretching of the vortex line
is millions of degrees kelvin at the temperature where the vortex
depins! The depinning may instead depend on
``remanent vorticity" \cite{tilleybook} in the form of extra pinned
vortices stretching from the top of the bump to the cylinder's walls.
This would decrease the energy barrier because the vortex could leave
the wire by attaching itself to some of the vortices that already exist.

In a more nearly two-dimensional geometry,
Eq. \ref{eq:shrinkforce} shows that
observing the equilibrium positions of vortices in a film of
varying thickness could shed light on the value of the core
size and core-energy and whether they are fixed functions of temperature
as most models imply.
The structure of vortex cores is still
not well understood, and there are several alternative models
\cite{tilleybook}.  
Experiments on the geometric force in contexts where the two contributions
to Eq. (\ref{eq:shrinkforce}) are comparable could give information
about the effective core radius $\tilde{a}$. By allowing
the film thickness to vary
by about $\epsilon = \frac{\Delta D}{D}\sim\frac{1}{\ln\frac{r_0}{a}}$, see Eq.(\ref{eq:epsilonmanners}), one ensures
that the vortex core size has a decisive effect on the 
equilibrium positions of the vortices (comparable to the effects of 
the long range forces).

\section{\label{sec:complex}Complex Surface Morphologies}
Up to this point, our discussion has been confined to rotationally symmetric 
surfaces and slightly deformed
surfaces for which the electrostatic analogy and perturbation theory can be successfully employed to determine
the geometric potential. To investigate geometric effects that arise for
 strong deformations and for surfaces with the topology of a sphere, we adopt a more
versatile geometric approach based on the method of  conformal mapping, often employed in the study of complicated boundary problems in electromagnetism and fluid mechanics. 
This approach also sheds light on  the physical origin
of the geometric potential. A concrete goal is to solve for the energetics (and the associated flows)
of topological defects on a complicated substrate $T$, the target surface, whose metric tensor we denote by $g_{Tab}$. 
This is accomplished by means of a conformal
map $C$ that transforms the target surface into a reference surface $R$, with metric tensor $g_{Rab}$. 
The computational advantages result from choosing the conformal map so that $R$ is a simple surface (e.g. an infinite flat plane, a flat disk, or a regular 
sphere) that preserves the topology of the target surface. Figure \ref{fig:splat} represents a complicated planar domain denoted by $T$ which can be mapped conformally onto a simple annulus labeled by $R$. We will introduce all the basic concepts in the context of this simple planar problem before turning to the conformal mapping between target and reference surfaces which is represented schematically in Fig. \ref{fig:emental}. Such mappings
can always be found in principle \cite{Davidreview}.

The conformal transformation will map the original positions of the defects on $T$, denoted by $\mathbf{u}$, onto a new set of coordinates on $R$ denoted by $\mathbf{\mathcal{U}}=C(\mathbf{u})$. In what follows,
capital calligraphic fonts always indicate coordinates on
the reference surface. For sufficiently small objects near a point $\mathbf{u}$ the map will act as a similarity transformation; that is, an infinitesimal length, $ds_T =\sqrt{ g_{Tab} du^{a} du^{b}}$, will be rescaled by a scale factor $e^{\omega(\mathbf{u})}$ which is independent of the orientation of
the length on $T$:
\begin{equation}
ds_R=e^{\omega(\mathbf{u})} ds_T ,
\label{eq:confdef}
\end{equation}
where $ds_R=\sqrt{g_{RAB}d\mathcal{U}^Ad\mathcal{U}^B}$.  This result
in turn implies a simple relation between the metric tensors of the two 
surfaces:
\begin{equation}
g_{RAB}=e^{2 \omega(\mathbf{u})} g_{TAB},
\label{eq:confdef2}
\end{equation}
where we have assumed for simplicity that the coordinates used on
the target surface are chosen so that corresponding points on the two
surfaces have the same coordinates $\mathcal{U}^A$.

We will demonstrate that, once the geometric quantity $\omega(\mathbf{u})$ is calculated, the geometric potential of an
isolated vortex interacting with the curvature is automatically determined. For multiple vortices,
the energy consists of single-vortex terms and vortex-vortex interactions. On a deformed sphere or plane,
the geometric potential reads
\begin{equation}
E_1(\mathbf{u_i})=-\pi n_i^2 K\omega(\mathbf{u}_i),
\label{eq:conformalself-energy}
\end{equation}
where $K$ is the stiffness parameter defined in Eq. (\ref{eq:stiffness}).
 For a deformed disk, there are boundary
interactions not included in Eq. (\ref{eq:conformalself-energy}).
(We will not consider multiply connected surfaces here, but
 the single particle energy on a multiply connected surface has 
additional
contributions which cannot be described by a local Poisson equation.)
The interaction energy is
\begin{equation}
E_2(\mathbf{u}_i,\mathbf{u}_j)=-2\pi n_i n_j K\ln\frac{\mathcal{D}_{ij}}{a},
\label{eq:conformalpair-energy}
\end{equation}
where $\mathcal{D}_{ij}$ is the distance between the two \emph{image} points on the
reference surface. When the reference
surface is an undeformed sphere (the other possibilities are a plane or disk), 
$\mathcal{D}_{ij}$ is the 
distance between
the points along a chord rather than a great circle\cite{lube92}. 
We will show below that on a deformed plane $\omega$ is equal to $U_G$,
but from now on we will use $\omega$ instead;  the two functions
are conceptually different, and are not even equal on a deformed sphere.

Equation (\ref{eq:conformalself-energy}) is derived
by the method of conformal mapping in section \ref{sec:map} and its computational efficiency is illustrated in section \ref{sec:bubble} where
the geometric potential of a vortex is evaluated on an Enneper disk, a minimal surface 
that naturally arises in the context of soap films, but whose geometry is
distorted enough compared to flat space
that it cannot be analyzed with perturbation theory.  Changing the geometry of the substrate has interesting effects not only on the one-body geometric potential but also on the two-body interaction between vortices. In section 
\ref{sec:bumps}
 we use conformal methods to show how a periodic lattice of bumps can cause
the vortex interaction to become anisotropic.
In section \ref{sec:zucchini}, we demonstrate that  
the quantization of circulation
leads to an extremely long-range force on an elongated surface with the topology of a sphere. The interaction energy is no longer logarithmic as in Eq.(\ref{eq:conformalpair-energy}), but now grows linearly with the distance 
between the two vortices. Indeed the charge neutrality constraint imposed by the compact topology of a sphere, blurs the distinction between geometric potential and vortex-interaction drawn in Equations (\ref{eq:conformalself-energy}) and (\ref{eq:conformalpair-energy}). This is most easily seen by bypassing 
vortex energetics on the reference surface (which is after all an auxiliary concept) and opting for a more
direct restatement of the problem in terms of Green's functions on the actual
target surface coated
by the helium layer. The interaction energy now reads
\begin{equation}
E_2'(\mathbf{u}_i,\mathbf{u}_j)=
4\pi^2 K n_i n_j \Gamma(\mathbf{u}_i,\mathbf{u}_j),
\label{eq:greenpair-energy}
\end{equation}
and the single-particle energy takes the form of a self-energy
\begin{equation}
E_1'(\mathbf{u}_i)=-\pi n_i^2 KU_G(\mathbf{u}_i)=\pi n_i^2 K
\iint G(\mathbf{u'})\Gamma(\mathbf{u_i},\mathbf{u'})
d^2\mathbf{u'},
\label{eq:greenself-energy}
\end{equation} 
where $\Gamma$ is a Green's function for the surface that generalizes the logarithmic potential familiar from two dimensional electrostatics. For a deformed
plane the two descriptions of the interaction energy are equivalent
since the Green's
function on a deformed plane can be obtained by conformal mapping,
\begin{equation}
\Gamma(\mathbf{u}_i,\mathbf{u}_j)=-\frac{1}{2\pi}\ln\frac{\mathcal{D}_{ij}}{a}.
\label{eq:greentransf}
\end{equation}
We will see that the expressions for the single-particle
energies are also equivalent. In contrast, for a deformed sphere, we show
in section \ref{sec:noodles} and appendix \ref{app:CG}
that
 the two formulations do not agree term by term
( $E_1'\neq E_1$ and
$E_2'\neq E_2$), although the \emph{combined} effect of one-particle and 
interaction
terms is the same (up to an additive constant). Both self-energies and interaction energies
include effects of the geometry and explicit formulas are provided on an azimuthally symmetric deformed sphere. Finally,
in section \ref{sec:geomineq}, we present a discussion of some general upper bounds to which the strength of geometric forces is subjected (even in the regime of strong deformations) which are useful in experimental
estimates and which illustrate a major difference between electrostatic and geometric forces:  The former can always be increased by piling-up 
physical charges but the latter are generated by the Gaussian curvature that can grow only at the price of ``warping" the
underlying geometry of space. 
Too much warping either leads to self-intersection of the surface
or a dilution of the long-range force.

\subsection{\label{sec:map}Using conformal mapping}

We start by proving a simple relationship between the total
energies (including self- and interaction parts)
$E_T$ and $E_R$ of two corresponding vortex-configurations on the target and reference surfaces respectively:
\begin{equation}
E_T=E_R-\pi K\sum_{i=1}^N\omega(\mathbf{u}_i).
\label{eq:anom}
\end{equation}
The right-hand side of Eq. (\ref{eq:anom}) can be calculated for 
the reference surface and then subsequently decomposed into single-vortex and vortex-vortex
interactions; several examples are worked out in detail in Sections \ref{sec:bubble} and \ref{sec:bumps}. 
\begin{figure}
\psfrag{Reference}{Reference}
\psfrag{Target}{Target}
\includegraphics[width=0.45\textwidth]{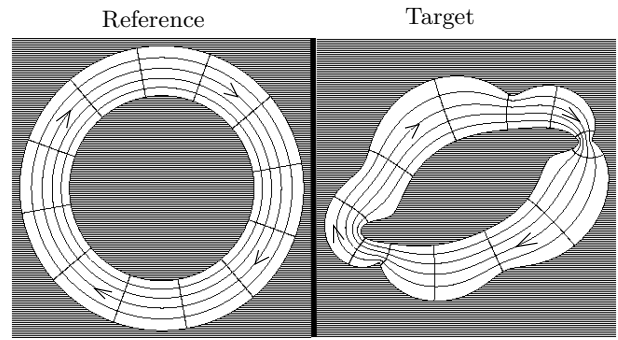}
\caption{\label{fig:splat}
The flow in a wiggily annular region (the target substrate $T$), 
obtained by conformally
mapping the flow from a circular annulus (the reference substrate $R$). 
Since every pair of radial spokes in the
first picture comprises the same energy, this is true in the conformal image
as well.  The small region in the constriction on the right manages this
by compensating for its small area by having a high flow speed.}
\end{figure}
The general approach is best illustrated by considering the planar flow in the complicated annular container
shown in Fig. \ref{fig:splat}.  which can be tackled by conformally mapping it
to a simpler circular annulus.

The flow in the reference annulus is clearly circular, and it has
the same $\frac{1}{r}$ dependence as for a vortex. 
A crucial property
of conformal transformation allows us to transplant this understanding of the reference flow
to the target annulus. This property concerns
the following mapping of the stream function $\chi$ (see Eq. (\ref{eq:chidef}))
from the reference surface to the target surface:
\begin{equation}
\chi_T(\mathbf{u})=\chi_R(\mathbf{\mathcal{U}})
\label{eq:mapping}
\end{equation}
Once the coordinate change $\mathbf{\mathcal{U}}=C(\mathbf{u})$ is found, the prescription to determine $\chi_T$ provided in Eq.(\ref{eq:mapping}) guarantees that the
corresponding flow will be irrotational and incompressible, as required. Visually, all we are doing is  taking the streamlines on
the reference surface, which are level curves
of $\chi_R$, and mapping them by $C^{-1}$ to the target surface.
We can informally state this first property of conformal maps as follows:

\indent{{\it 1) The conformal image of a physical flow pattern is still a physical pattern.}}

Note that any multiple of the mapped stream function,
$\alpha\chi_R(C(\mathbf{u}))$, corresponds to an irrotational
and incompressible flow but in this case the rates of flow
in the target and reference substrate are different. Only the choice $\alpha=1$ ensures that
both flows have the same number $n$ of
circulation quanta around the hole (or around each
vortex if some are present.) This
follows from another basic property of conformal maps:

\indent{{\it 2) In flow patterns related by a conformal map, according to
Eq. (\ref{eq:mapping}), circulation integrals around corresponding
curves are the same.}}

This property follows from the fact that the contribution to the circulation from each element of the contour, $\mathbf{v\cdot dl}$, is conformally invariant. The
infinitesimal length $\mathbf{dl}$ and the velocity $\mathbf{v}$ scale inversely to each other under conformal transformations and the angle between them is preserved by the map.
To understand why, note that
flow lines are compressed together (stretched apart) when they are
mapped to the target space if the conformal parameter
$\omega$ is greater (less) than zero.
As a result, the velocity increases (decreases) by the same factor as distances
are decreased (increased). This heuristic argument is confirmed
by noting that the velocity is given by the $covariant$ curl of the stream function, 
see Equations (\ref{eq:chidef}) and (\ref{eq:mapping}). By definition, the covariant curl carries a multiplicative factor of $g^{-1/2}=e^{\omega}$ \cite{Dubrovinbook}, hence
\begin{equation}
v_T=e^{\omega}v_R.
\label{eq:mapvelocity}
\end{equation}
On the other hand, $|\mathbf{dl}_T|= e^{-\omega} |\mathbf{dl}_R|$ rescales in the opposite way, as indicated in Eq. (\ref{eq:confdef}).

The problem of
finding the \emph{energy}
in the deformed annulus can now be reduced to a simple-rotationally
symmetric problem by appealing to a third property of the conformal mapping:

\indent{{\it 3) The kinetic energies in corresponding regions are equal, provided
the regions do not contain vortices.}}

The proof of this statement relies on the previous discussion: 
the kinetic energy contained in an element of the area of the surface $dA$
is $\frac{1}{2}\rho_s \mathbf{v}^2dA$. By Eq. (\ref{eq:mapvelocity}), $\mathbf{v}^2$ scales as $e^{2\omega}$ whereas $dA$ scales by $e^{-2\omega}$ making the energy conformally invariant.
Figure \ref{fig:splat} illustrates pictorially that the energy
density in the original flow on the target surface 
is smoothed out and simplified,
its variations being replaced by variations in the conformal scale factor.

We now return to the energetics of flows containing point vortices.
The starting point of our analysis follows from the defining property of a conformal map, namely that
a conformal image of a small figure has the same shape as the original
figure, while a larger shape becomes distorted (consider Greenland, which
has an elongated shape, but appears to round out at the top in a Mercator
projection, which is itself a conformal map). 
To quantify the size limits, note that if a shape has size $l$, $\omega$ changes
by about $l\nabla\omega$ across the shape. Thus,
as long as
\begin{equation}
l<<\frac{1}{|\nabla\omega|},
\label{eq:columbus}
\end{equation}
the mapping rescales the shape uniformly. The right-hand side
is ordinarily of order $L$, the curvature scale of the surface. As a result we can conclude
that

\indent{{\it 4) The circular shape of the streamlines near a microscopic vortex core on a substrate of slowly varying curvature is preserved.}}

On a deformed substrate with a flow induced by vortices, 
the flow speeds will increase
or decrease not just depending on the distance to the vortex, but also
depending on the shape of the surface. For example, the vortex on top of
the bump in the example of Sec. \ref{subsec:anomalous} has a flow that decays
more slowly with distance than in flat space.  Also, for a vortex
well off to the side of a bump, if the bump's height $h$
is larger than its width $2r_0$, it turns out that
the flow pattern penetrates only
up to an elevation of about $r_0$ up the side of the bump.

\begin{figure*}
\includegraphics[width=\textwidth]{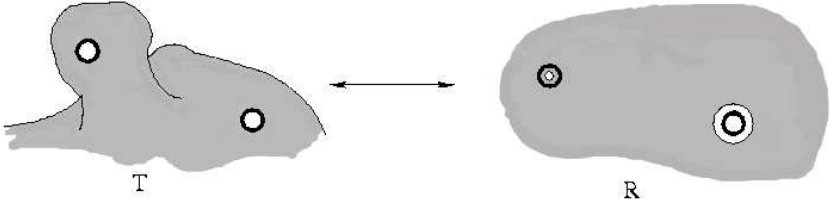} 
\caption{\label{fig:emental} Comparing the energy 
of $T$ and $R$ by splitting $T$ up into portions $I$ and $O$ and
using different maps to map them to $R$.
Left, a target surface, which has
the topology of an infinite plane and is distorted by a three dimensional
lump. Right, the reference plane (shown in the plane of the page). $I$
consists of the interior of radius $l$
disks in $T$,  and is mapped rigidly to the heavily demarcated disks in $R$.
$O$
consists of the gray portion of $T$, and is mapped conformally to the
gray portion of $R$.
$I$ and $O$ contain the same kinetic energy as their images in $R$, but
the images do not fit together perfectly.  Thus the difference in energy
between the flows in $T$ and in $R$ is determined by calculating how much
energy is contained in the annuli which are either mismatched or covered
twice.}
\end{figure*}

The method of conformal mapping elucidates these geometrical rearrangements of the flow pattern. To find the flow pattern around the vortices at positions
$\mathbf{u}_i$, we find
the flow pattern around vortices at the corresponding positions
$\mathbf{\mathcal{U}}_i$ on the reference surface and then map these
streamlines onto the target surface
by Eq. (\ref{eq:mapping}). The energies are \emph{not}
equal in this case, in spite of property 3.
Property
3 does not apply to a region containing vortex cores, because
we would have to
suppose the area of the cores on the reference surface
to be greater by $e^{2\omega}$
and the energy in the cores to be smaller by a factor of $e^{2\omega}$, in order for the conformal relation Eq. $\frac{1}{2}\rho_s v_T^2dA_T=
\frac{1}{2}\rho_s v_R^2dA_R$ to continue to hold.  
In contrast, the core radius is fixed by the short-distance correlations of the 
helium atoms and the core energy is related to the interaction energy of 
the atoms.

The vortex cores are not significantly affected by
the curvature of the substrate; moreover, the whole flow pattern
in the vicinity of the core is nearly independent of the location of a vortex.
We observe that
each vortex has a ``dominion," a region where the flows are forced by
the presence of the vortex to be
\begin{equation}
v=n\frac{\hbar}{mr}+\delta v
\label{eq:alsouseful}
\end{equation}
The leading term has the same form as one expects for
a vortex in a rotationally symmetric situation, and the effects
of geometry are accounted for by $\delta v$ ; by dimensional
analysis, this error is of the order of $\frac{\hbar}{mL}$ where
$L$ is the radius of curvature of the substrate (or possibly the
distance to another vortex or to the boundary, whichever is shortest).
Therefore we can introduce any length $l<<L$ and note
that $l$ is then a distance below which the
effects of curvature do not have a significant effect (compared
to the diverging velocity field).
The geometry correction gives a contribution
to the energy within this radius that is also small, as seen
by integrating the kinetic energy over the annulus
between $a$ and $l$ (using Eq. (\ref{eq:alsouseful})):
\begin{equation}
\pi K\ln\frac{l}{a}+ \epsilon_c + O(K)\frac{l^2}{L^2}
\label{eq:dominionenergy}
\end{equation}
where $a$ is the core radius and $\epsilon_c$ the core energy. 
The error term is \emph{quadratic}
in $\frac{l}{L}$ because the integral over the cross-term from
squaring Eq. (\ref{eq:alsouseful}) cancels.

When we wish to find the kinetic energy of the superflow, these near-vortex
regions are thus the simplest to account for, as their energy is nearly
independent of their positions relative features such as bumps on the substrate
as long as
\begin{equation}
l<<L.
\label{eq:useful}
\end{equation} Since the target and reference configurations have
the same number $N$ of vortices, the energies contained within
radius $l$ of the vortices are the same:
\begin{equation}
E^T_{<l}=E^R_{<l}=\pi NK\ln\frac{l}{\tilde{a}}
\label{eq:inert}
\end{equation}

In order to find the forces on a set of vortices, we need to account
for all the energy of the vortices in regions away from the
vortices where the flow \emph{has}
been affected by the curvature. Let us imagine cutting the target surface
up into an inner region $I$ (the union of the radius $l$ disks around each vortex) and an outer region $O$ (consisting of everything else) as illustrated
in Fig. \ref{fig:emental}.  We can
map $I$ to the reference surface by simply translating each of the disks
so that they surround the vortices on $R$.
The modifications to the flow are all in $O$; 
for example, streamlines there are deformed from their circular shape.
These irregularities can be removed (or at least simplified) by using
the
conformal mapping to map $O$ to $R$, just as in the case of the annulus
illustrated above. The inhomogeneity of the conformal
map compensates for the irregularity of the flow.
Some portions of $O$ are expanded and some are contracted,
but its circular boundaries are small enough (see Eqs. (\ref{eq:useful}) ,
(\ref{eq:columbus}))
that they
are simply rescaled into circles of different radii (property 4):
\begin{equation}
l_i=e^{\omega(\mathbf{u}_i)}l
\label{eq:scaleregion}
\end{equation}

Now we have mapped $I$ and $O$ from the target surface to regions
on the reference surface which contain the same energy.
But these
images of the regions on $R$ do not fit together.
The conformal map on $O$ stretches or contracts each hole in it, to
circles of radii $e^{\omega(\mathbf{u}_i)}l$. These stretched
edges (corresponding to
the solid black circle in Fig. \ref{fig:emental}) do not fit together with
the images of $O$, which have been moved rigidly from the target surface.
The energies are related by
\begin{equation}
E_T=E^T_{<l}+E^T_{>l}=E^R_{<l}+E^R_{>l_i}
\label{eq:mismatch}
\end{equation}
We must correct for the gaps and overlaps between the two image regions
on $R$ in order to relate the last expression to $E_R$.
If $\omega(\mathbf{u}_i)>0$, there is an annular gap near vortex $i$;
using Eq. (\ref{eq:alsouseful}) (since this gap is part of the flow
controlled by this vortex):
\begin{eqnarray}
\Delta E_i&=&
-\frac{1}{2}\rho_s\int_{l}^{l_i}2\pi rdr\frac{n_i^2\hbar^2}{m^2r^2}\\
&=&-\pi Kn_i^2\omega(\mathbf{u}_i)
\end{eqnarray}
Summing all these contributions gives our desired result, Eq. (\ref{eq:anom}).

We emphasize
the energy difference is not produced within the cores, or anywhere
near the vortices.  In fact, the fact that the energies on $T$ and $R$
differ is a result of assuming that there is \emph{no change} in the flow
within a macroscopic distance $l$ of the vortex.  The scale 
$l\ll L$ only has to be small compared to the geometry of
the system and has no relation to the atomic structure of the core.
On the other hand, taking $l$ as small as possible has an elegant
consequence: Eq. (\ref{eq:anom}) actually has an error which is
$O\left(K\left(\frac{a}{L}\right)^2\right)$, smaller than
$O\left(K\left(\frac{l}{L}\right)^2\right)$ as predicted at first.
Taking smaller values of $l$ gives a more accurate result, since
the conformal mapping (which does not suffer from the error in Eq.
(\ref{eq:dominionenergy}))
is used to calculate the energy of a larger portion of the flow pattern.

Now Eq. (\ref{eq:anom}) shows that the position-dependent scale 
factor, $\omega(\mathbf{u})$, plays
the role of a single-particle
energy.  Additionally, this single particle energy can be
regarded as the ``geometric potential," since it turns out
to be related to the curvature in
a way analogous to how the electrostatic potential is related
to the charge.  The function $\omega$ depends on the 
shape of the boundaries and on the curvature of the surface. 
A varying scale factor is necessary to map between
surfaces with different distributions of curvature (such as planes
with
and without bumps). (A constant scale
factor only rescales the curvature.)  The curvature therefore depends on the
\emph{variation} of $\omega$. In fact it can be shown that
\begin{alignat}{1}
G_T(u,v)=e^{2\omega(u,v)}&G_R(U(u,v),V(u,v))\nonumber\\
&+\frac{1}{\sqrt{det(g_{T,cd})}}
\partial_{a}\sqrt{det(g_{T,cd})}g_T^{ab}\partial_{b}\omega.
\label{eq:Greenland}
\end{alignat}
The second term is the Laplacian of $\omega$ as a function on the
target surface\footnote{As a check of this
identity, imagine reversing the roles
of the reference and target surfaces. Then $\omega$ should
 be regarded as a function on the reference
surface. This changes the Laplacian
by a factor of $e^{2\omega}$ (because $g_R$ is replaced by $g_T$).  Also,
the sign of $\omega$ should be reversed. Rearranging the equation now
brings it back into the original form with $T$ and $R$ switched.}.  The
correspondence with electrostatic potentials, with $G_T(u,v)$ as a source, 
is clear if the reference surface is a plane
or disk and $G_R=0$.  Then Eq. (\ref{eq:Greenland}) reduces to
Eq. (\ref{eq:geompdeq}).

For a deformed sphere or
plane, single particle potentials come entirely from the second term in 
Eq. (\ref{eq:anom}).
The reference energy, corresponding to vortices on a sphere or plane cannot favor
one position over another because of the homogeneity of these reference surfaces.  The first term, $E_R$, leads to 
the vortex-vortex interactions which depend
only on the separation of the vortices \emph{on the reference surface},
again by symmetry; this energy is
\begin{equation}
E_R=4\pi^2 K\sum_{i<j}n_in_j\Gamma(\mathbf{\mathcal{U}_i},\mathbf{\mathcal{U}_j})
\label{eq:refenergy}
\end{equation}
where $\Gamma$ depends on the reference surface:
\begin{equation}
\Gamma_{plane}(\mathbf{\mathcal{X}},\mathbf{\mathcal{Y}})=
-\frac{1}{2\pi} \ln\frac{|\mathbf{\mathcal{X}}-\mathbf{\mathcal{Y}}|}{a}
\end{equation}
and
\begin{eqnarray}
\Gamma_{sphere}(\mathbf{\mathcal{X}},\mathbf{\mathcal{Y}})&=&
-\frac{1}{2\pi} \ln\frac{2R\sin\frac{\gamma}{2}}{a}\\
&=&-\frac{1}{2\pi}\ln\frac{|\mathbf{\mathcal{X}}-\mathbf{\mathcal{Y}}|}{a}
\label{eq:sphereen}
\end{eqnarray}
where $\gamma$ is the angle between the two points. ($2R\sin\frac{\gamma}{2}$
is the \emph{chordal} distance between the points, not the geodesic distance
along the surface, as one might have guessed for the natural generalization
of a Green's function to curved space.)
The detailed derivation of the second formulation
of the vortex interaction energies in terms of the Green's functions
on the deformed surface (see Eqs. 
(\ref{eq:greenpair-energy}) and (\ref{eq:greenself-energy}))
is contained in Appendix 
\ref{app:CG}.

\subsection{\label{sec:bubble}Vortices on a ``Soap Film" Surface}
There are experimental and theoretical motivations for
studying substrates shaped as minimal surfaces.
An example of a minimal surface is easy to make 
by dipping a loop of wire in soap; 
the  spanning soap film tries to minimize its area. Vortices can 
be studied on a helium film coating a \emph{solid} substrate whose
surface has
the shape of such a film.
Such surfaces are characterized by a vanishing mean curvature, $H({\bf x})$, 
so
the contribution of the surface tension $2\gamma H(\mathbf{x})$
to the thickness variation equation,
Eq. (\ref{eq:wet-8})
is drastically reduced.  
From the mathematical point of view, there is a
widely known parametrization due to Weierstrass \cite{Hide-book} which readily leads to an  
\emph{exact} expression for the geometric potential of a vortex on such
a surface.

Weierstrass's formulae, which give a
minimal surface for each choice of an analytic function $R(\zeta)$, read:
\begin{eqnarray}
x(\zeta)&=&Re\int_0^\zeta {R(\zeta')(\zeta'^2-1)d\zeta'}\nonumber\\
y(\zeta)&=&Im\int_0^\zeta {R(\zeta')(\zeta'^2+1)d\zeta'}\nonumber\\
z(\zeta)&=&Re\int_0^\zeta {R(\zeta')2\zeta'd\zeta'}
\label{eq:weierstrass}
\end{eqnarray}
The correspondence between this parametric surface
and the complex variable $\zeta=\mathcal{X}+i\mathcal{Y}$ is a
conformal map, and
the conformal factor can be expressed in terms of $R(\zeta)$.
Therefore, the analysis of
vortices on such a surface is not difficult at all when the
$\mathcal{X},\mathcal{Y}$-plane is used as the reference surface.

As an example, let $R(\zeta)$ be equal to $L\zeta$ where $L$ controls the
size of the target surface.
Then the surface produced is given in parametric form by
\begin{eqnarray}
x&=&L\left[\frac{\mathcal{X}^3}{3}-\mathcal{X}\mathcal{Y}^2
-\mathcal{X}\right]\nonumber\\
y&=&L\left[\frac{-\mathcal{Y}^3}{3}+\mathcal{Y}\mathcal{X}^2+\mathcal{Y}\right]
\nonumber\\
z&=&L(\mathcal{X}^2-\mathcal{Y}^2)
\label{eq:enneper}
\end{eqnarray}

\begin{figure}
\psfrag{x}{$x$}
\psfrag{y}{$y$}
\psfrag{z}{$z$}
\includegraphics[width=.47\textwidth]{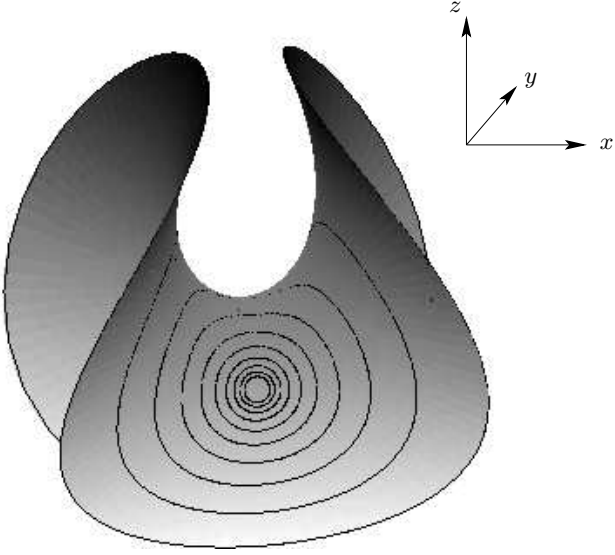}
\caption{\label{fig:escaping?}Vortex and its streamlines on an ``Enneper Disk".}
\end{figure}
We consider a superfluid film coating only a circle of radius $A$ about the origin of the
$\mathcal{X,Y}$ plane because the complete
surface has self-intersections. 
This surface can be called the Enneper disk and is illustrated in Fig. 
\ref{fig:escaping?}.  The figure illustrates that the left and right hand sides
of the saddle fold over it and would pass through each other if allowed to
extend further while
the front and the back would eventually
intersect each other underneath the saddle.
The former pair of intersection
curves correspond to the two branches
of the hyperbola $\mathcal{X}^2=3(\mathcal{Y}^2+1)$. When the
reference surface is curved into the Enneper surface, the $X$ axis
bends upward so that the branches map to the same
intersection curve in the $yz$ plane. 
(The other intersection curves are obtained
by exchanging $\mathcal{X}$ and $\mathcal{Y}$.)
 Since the points where these
hyperbolae are closest to the origin are $(\pm\sqrt{3},0),(0,\pm\sqrt{3})$,
a non-self intersecting portion of the Enneper surface results as long
as $A<\sqrt{3}$.

Now we explicitly calculate how a single vortex interacts with the curvature of such a surface
by using Eq. (\ref{eq:anom}). (We will use conformal
mapping instead of Eq. (\ref{eq:curvature-defect}) since
the latter equation does not hold on a surface with the topology
of a disk, as $\omega$ does not satisfy the
Dirichlet boundary conditions which are implied by such an expression.)
The metric obtained from (\ref{eq:enneper}) is given by
\begin{equation}
dx^2+dy^2+dz^2=L^2
(d\mathcal{X}^2+d\mathcal{Y}^2)(1+\mathcal{X}^2+\mathcal{Y}^2)^2.
\end{equation}
(Surprisingly, this \emph{metric} is rotationally symmetric. This implies
that the surface may be slid along itself without stretching, but with changing
amounts of bending.)
Hence
\begin{equation}
\omega_{Enneper}=-\ln L(1+\mathcal{R}^2),
\end{equation}
where $\mathcal{R}^2=\mathcal{X}^2+\mathcal{Y}^2$.  According to Eq. 
(\ref{eq:anom}),
this indicates that the vortex should be attracted to the middle
of the surface,
but of course this force competes with
the boundary interaction $K\pi \ln\frac{A^2-\mathcal{R}^2}{aA}$
which tries to pull the vortex to the edge. This expression
for the boundary interaction is obtained from the familiar formula for the energy of a vortex interacting with its image
in a flat
reference disk\cite{geomgenerate}.
The total energy is then
\begin{equation}
E=K[\pi\ln\frac{L}{aA}+ \pi\ln (A^2-\mathcal{R}^2)(1+\mathcal{R}^2)].
\end{equation}
As long as  $A>1$,
the central point of the saddle is a local minimum and this condition is compatible
with the requirement $A<\sqrt{3}$ for non-self-intersecting
disks. Fig. \ref{fig:escaping?}
shows the flow lines of
a vortex forced by the geometric interactions towards the center of an Enneper surface with $A=1.5$.

In general, conformal mapping
allows us to express the energy of a single vortex on a deformed surface
with a boundary
in the form:
\begin{equation}
E=\pi K[\ln\frac{A^2-\mathcal{R}(\mathbf{u})^2}{aA}-\omega(\mathbf{u})],
\label{eq:netenergy}
\end{equation}
where $\mathcal{R}(\mathbf{u})$ refers to the image of a defect at $\mathbf{u}$
under a conformal map to a flat, circular disk of radius $A$.
The Green's function method cannot be used to determine the energy
of defects
on a surface with a boundary.  Although 
the conformal factor $\omega(\mathbf{u})$
satisfies the Poisson equation, Eq. (\ref{eq:geompdeq}),
it cannot be expressed
as the integral of the curvature times the Green's function (as
in Eq. (\ref{eq:greenself-energy})),
since $\omega$ does not satisfy simple boundary conditions.
In any case, the first term in Eq. (\ref{eq:netenergy})
has no general expression in terms of $\omega$ either.
Interestingly, the \emph{total}
 single-particle
energy satisfies a \emph{nonlinear} differential equation (the Liouville
equation):
\begin{equation}
\nabla_{\mathbf{u}}^2E(\mathbf{u})=-\pi K G(\mathbf{u})-
\frac{4\pi K}{a^2}e^{-\frac{2E(\mathbf{u})}{\pi K}}.
\label{eq:liouvilledisk}
\end{equation}
This result can be derived by using Eq. (\ref{eq:Greenland})
to calculate the Laplacian of the first term and using
$\nabla^2=e^{2\omega(\mathbf{u})}\frac{1}{\mathcal{R}}
\frac{\partial}{\partial\mathcal{R}}\mathcal{R}
\frac{\partial}{\partial\mathcal{R}}$
to calculate the Laplacian of the second term.
$E(\mathbf{u})$ also satisfies an asymptotic boundary condition:
\begin{equation}
e^{\frac{E(\mathbf{u})}{\pi K}}\rightarrow \frac{2d}{a}
\end{equation}
as $d$, the distance from $\mathbf{u}$ to the boundary, approaches 0.
Together the differential equation and the boundary condition
should determine the total geometrical
and boundary energy of a single vortex, although the nonlinear Eq. 
(\ref{eq:liouvilledisk})
is difficult to solve.
\subsection{\label{sec:bumps}Periodic surfaces}

In this section, we illustrate how a periodically curved substrate distorts the flat space
interaction energies between vortices,
besides generating the single-particle geometric potential.
This effect is shown to be a consequence of the action of a  conformal map
 which generally will map the target surface into
a periodic reference
substrate with different lattice vectors from the vectors of the
target substrate.
According to the general relation Eq. (\ref{eq:greentransf}), 
the long-distance behavior of
the Green's function is given by the logarithm of a distorted distance.

\begin{figure}
\psfrag{l1}{$\lambda_1$}
\psfrag{m1}{$\mu_1$}
\psfrag{o1}{}
\psfrag{l2}{$\lambda_2$}
\psfrag{m2}{$\mu_2$}
\psfrag{o2}{}
\psfrag{x}{$x$}
\psfrag{y}{$y$}
\includegraphics[width=.45\textwidth]{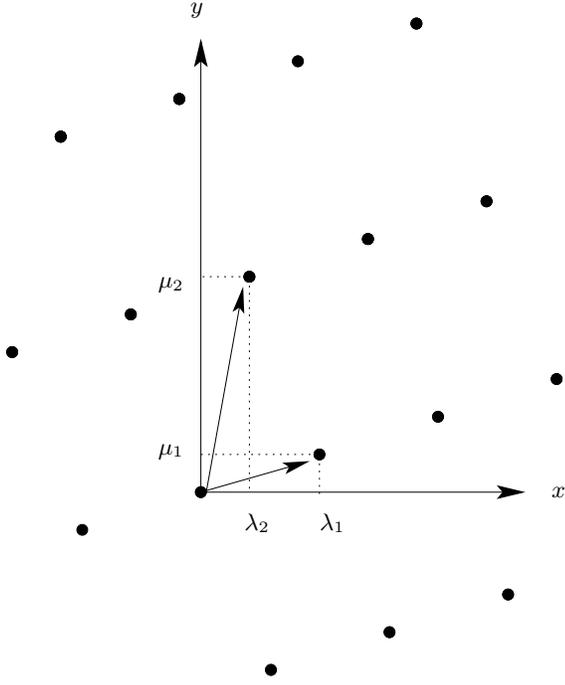}
\caption{\label{tulipbulbs} An illustration of the period
lattice of the target surface, projected into the $xy$-plane.
The coordinates $\lambda_1,\mu_1$ and $\lambda_2,\mu_2$
describe the basis vectors,
which we do not assume to be orthogonal.
}
\end{figure}

Consider a surface with a periodic height function
$z=h(x,y)$, i.e., say $h$ satisfies
\begin{equation}
h(x+\lambda_i,y+\mu_i)=h(x,y) \quad \ \mathrm{for\ }i=\{1,2\} ,
\label{eq:goingtoseed}
\end{equation}
where $i$ labels the two basis vectors,
 which are not assumed to be orthogonal.
Figure (\ref{tulipbulbs}) shows the corresponding periods $(\lambda_i,\mu_i)$. A conformal mapping can be chosen to preserve the fact that the substrate
is periodic but not the actual values of the periods, which are therefore
given on the reference substrate by two new pairs denoted by $(\Lambda_i,M_i)$.
In other words, we suppose that a tesselation of
the target substrate by congruent unit cells is
mapped to a set of congruent unit cells on the reference
substrate.  Then 
the map transforming the original coordinates $(x,y)$ into the target coordinates $(\mathcal{X},\mathcal{Y})$ satisfies
\begin{eqnarray}
\mathcal{X}(x+\lambda_i , y+\mu_i)=\mathcal{X}(x, y)+\Lambda_i \nonumber\\
\mathcal{Y}(x+\lambda_i , y+\mu_i)=\mathcal{Y}(x, y)+M_i .
\label{eq:periods}
\end{eqnarray}

\begin{figure}
\includegraphics[width=.47\textwidth]{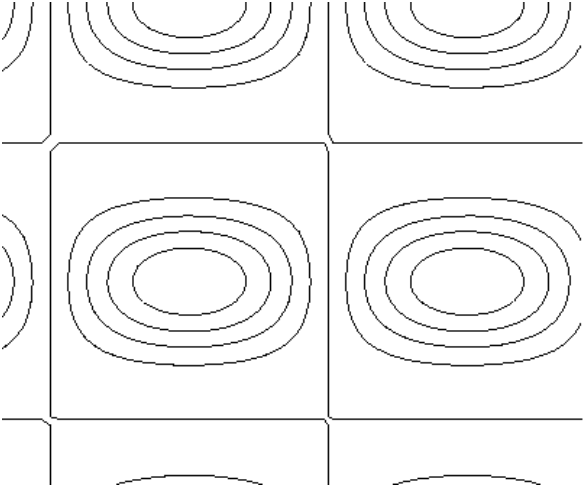}
\caption{\label{fig:squaredaway}A periodic surface with a square lattice
but not square symmetry.
This surface is illustrated by its contour plot.
It has the two reflectional symmetries but
not the $90^{\circ}$ symmetry of a square lattice; hence,
generically, the conformal image will have a \emph{rectangular} lattice.}
\end{figure}

There is no simple formula for
the new set of primitive lattice vectors $(\Lambda_i,M_i)$ for the reference space. In some cases, though, precise information can be derived
from the fact that
 the $(\Lambda_i,M_i)$
share the symmetry of the topography of the original substrate.
For example, if the lattice is composed of bumps which have
a $90^{\circ}$ rotational point symmetry, then the reference lattice
will be square. On the other hand, the topology of the periodic surface with a
$square$ lattice shown in the contour plot of Fig. (\ref{fig:squaredaway}) does not posses a $90^{\circ}$ rotational symmetry and hence its
conformal image will have a $rectangular$ lattice.

To get an idea how the conformal mapping behaves
macroscopically, we try to decompose it into a linear 
transformation $\mathcal{L}$ 
with matrix coefficients $\{A, B; C, D\}$ and a periodic modulation
captured by the functions $\xi(x,y)$ and $\eta(x,y)$ that
distort the $\mathcal{X}$ and $\mathcal{Y}$ axes respectively:
\begin{eqnarray}
\mathcal{X}(x,y)=Ax+By+\xi(x,y) \nonumber\\
\mathcal{Y}(x,y)=Cx+Dy+\eta(x,y) .
\label{eq:shiftperiods}
\end{eqnarray}
(This decomposition is justified by the self-consistency of the 
following calculations.) 
The matrix coefficients of the linear transformation can be determined by 
requiring
consistency with Eq. (\ref{eq:periods}). Start by evaluating the 
left hand sides of the two Equations (\ref{eq:shiftperiods}) at the positions $\{x+\lambda_i,y+ \mu_i \}$ which are
shifted by the two pairs of periods $\{\lambda_i,\mu_i\}$, so that the right hand sides become $\mathcal{X}(x,y)+\Lambda_i$ and $\mathcal{Y}(x,y)+M_{i}$, according to Eq. (\ref{eq:periods}). Then subtract the resulting equations from the corresponding unshifted Equations (\ref{eq:shiftperiods}), for each value of $i$. We then obtain two  pairs of equations
\begin{eqnarray}
A \lambda_i + B \mu_i&=&\Lambda_i \nonumber\\
C \lambda_i + D \mu_i&=&M_i \quad \ \mathrm{for\ }i=\{1,2\} ,
\label{eq:int1}
\end{eqnarray}
where we have used the fact that the periodic functions $\xi(x,y)$ and $\eta(x,y)$ are unchanged when shifted by the periods. We can now solve the four
equations of Eq. (\ref{eq:int1})
simultaneously for $A,\ B,\ C,\ \mathrm{and\ }D$
to see that the linear transformation matrix $\mathcal{L}$ reads
\begin{equation}
\mathcal{L}=\left(\begin{array}{cc}A&B\\C&D\end{array}\right)=
\left(\begin{array}{cc}\Lambda_1&\Lambda_2\\M_1&M_2\end{array}\right)
\left(\begin{array}{cc}\lambda_1&\lambda_2\\\mu_1&\mu_2\end{array}\right)^{-1}.
\label{eq:affine}
\end{equation}
(Now we can justify the original decomposition, Eq. (\ref{eq:shiftperiods}),
by \emph{defining}
$\mathcal{L}$ by Eq. (\ref{eq:affine}) and \emph{defining} $\xi(x,y)$ and
$\eta(x,y)$ as the discrepancy between the conformal
map $\mathcal{X}(x,y),\mathcal{Y}(x,y)$ and the linear map $\mathcal{L}(x,y)$
as in Eq. (\ref{eq:shiftperiods}).  We can then check that $\xi(x,y)$
and $\eta(x,y)$ \emph{are} periodic functions of the coordinates.)

The linear transformation can be used to calculate approximately the long distance behavior of the Green's function
\begin{eqnarray}
\Gamma(x,y;x',y')&=&-\frac{1}{4\pi}
\ln[(\Delta\mathcal{X})^2+(\Delta\mathcal{Y})^2]\nonumber\\
&\approx&-\frac{1}{4\pi}\ln[(A\Delta x+B\Delta y)^2\nonumber\\
&&\ \ \ \ \ \ +(C\Delta x+D\Delta y)^2] ,
\label{eq:ellipsie}
\end{eqnarray}
where we used the fact that the periodic functions $\xi(x,y)$ and $\eta(x,y)$ are bounded and hence negligible
in comparison to $\Delta x$ and $\Delta y$ at long distances. This
expression illustrates the fact that
the matrix $\mathcal{L}$ captures the long distance lattice distortions
induced by the conformal mapping,
apart from the additional waviness described by $\xi(x,y)$ and $\eta(x,y)$. The linear transformation determined by $\mathcal{L}$
is by itself typically \emph{not conformal}, meaning
that it generates an anisotropic deformation of the target lattice which does not preserve the angle between the original
lattice vectors.

The deformation of the lattice is controlled by the curvature of the substrate. To spell out this connection and allow an explicit evaluation of the long-distance
Green's function in Eq. (\ref{eq:ellipsie}), we explicitly
evaluate the matrix elements $L_{ij}$ in terms
of the height function $h(x,y)$ of
a gently curved (or low-aspect-ratio) surface, one for which
 $h(x,y)\ll \{\lambda_i,\mu_i\}$, $\{\xi(x,y),\eta(x,y)\}\ll1$
and $\{A-1,D-1,B, C\}\ll1$.
The new set of (isothermal) coordinates $\mathcal{X}$ and $\mathcal{Y}$,
used to implement the conformal transformation, are found by solving the Cauchy-Riemann Equations
(\ref{eq:tack})
\begin{eqnarray*}
&&\partial_x\mathcal{X}
=\sqrt{g}g^{yy}\partial_y\mathcal{Y}+\sqrt{g}g^{xy}\partial_x\mathcal{Y}
\nonumber\\
&&\partial_y\mathcal{X}
=-\sqrt{g}g^{xx}\partial_x\mathcal{Y}-\sqrt{g}g^{xy}\partial_y\mathcal{Y},
\end{eqnarray*}
which, upon substituting from Eq. (\ref{eq:shiftperiods}) and
making the small aspect ratio approximation discussed in Appendix
\ref{app:calmseas}, reduce to
\begin{eqnarray}
&&A+\partial_x\xi=D+\partial_y\eta+\frac{1}{2}(h_x^2-h_y^2)\\
&&B+\partial_y\xi=-C-\partial_x\eta+h_xh_y.
\label{eq:alphabet}
\end{eqnarray}

We now proceed to show that these equations do not have solutions unless the lattice is distorted, that is to say
the matrix $\mathcal{L}$ cannot be the identity for a generic
periodic function $h(x,y)$. Note that
the periodicity of $\xi(x,y)$ and $\eta(x,y)$ implies that the integral
of either one
over any unit cell, e.g.,
\begin{equation}
\iint_{\mathrm{cell}}dxdy\ \xi(a+x,b+y)
\end{equation}
is independent of the quantities $(a,b)$ by which the unit cell is shifted. Upon differentiating the integral with respect to $a$, one obtains
\begin{equation}
\iint_{\mathrm{cell}}dxdy\ \xi_x(x,y)=0.
\label{eq:zero}
\end{equation}
Similarly, the averages of $\eta_x,\eta_y,$ and $\xi_y$ are also equal to zero.
Hence, upon averaging Eq. (\ref{eq:alphabet}) over a unit cell we obtain the key relations
\begin{eqnarray}
A-D&=&\frac{1}{2}<h_x^2-h_y^2>\nonumber\\
B+C&=&<h_xh_y>,
\label{eq:approxperiods}
\end{eqnarray}
which prove our assertion that $L$ cannot be a simple dilation or rotation. Some shear is naturally introduced by the non trivial metric (or height function) of the
underlying surface.

The $\mathcal{L}$ matrix is undetermined
up to a dilation and a rotation but this
is of no consequence to the determination of the Green's function. In fact,
to
find the Green's function, note that Eq. (\ref{eq:approxperiods}) allows
us to
write the
matrix coefficients in terms of two undetermined constants $\epsilon_1$ and $\epsilon_2$ that will drop out of the final answer:
\begin{eqnarray}
&&A=1+\epsilon_1+\frac{1}{4}<h_x^2-h_y^2>\\
&&D=1+\epsilon_1+\frac{1}{4}<h_y^2-h_x^2>\\
&&B=\epsilon_2+\frac{1}{2}<h_xh_y>\\
&&C=-\epsilon_2+\frac{1}{2}<h_xh_y> ,
\end{eqnarray}
so that consistency with Equations (\ref{eq:approxperiods}) is guaranteed.
(The variables $\epsilon_1$ and $\epsilon_2$ parameterize an overall
infinitesimal
scaling (by $1+\epsilon_1$) and a rotation (by angle $\epsilon_2$)
respectively.)
Substitution of these equations into Eq. (\ref{eq:ellipsie}) gives the desired long-distance behavior
of the Green's function purely in terms of derivatives of the height function, which we assume to be known:
\begin{eqnarray}
\Gamma(x,y;x',y')&\approx& -\frac{1}{4\pi}\ln [\Delta x^2+\Delta y^2\nonumber\\
&&\ \ \ \
+\frac{1}{2}<h_x^2-h_y^2>((\Delta x)^2-(\Delta y)^2)\nonumber\\
&&\ \ \ \ \ +2<h_xh_y>\Delta x\Delta y] .
\end{eqnarray}
This is the central result of this section; it can also be applied
to interactions between disclinations in liquid crystals \cite{geomgenerate} 
and dislocations in crystals \cite{VitelliLucks}.
The anisotropic correction to the Green's function, captured by the second and third term, suggests that a distorted version of the triangular
lattice of vortices expected on a flat substrate may form
when the helium-coated surface is rotated slowly enough that there
is only one vortex to several unit cells.  However, the
actual ground state is likely to be difficult to observe, as the geometric potential will try to trap the vortices near saddles as discussed in Section
\ref{BS}.

\subsection{Band-flows on elongated surfaces\label{sec:zucchini}}
\begin{figure}
\includegraphics[width=.2\textwidth]{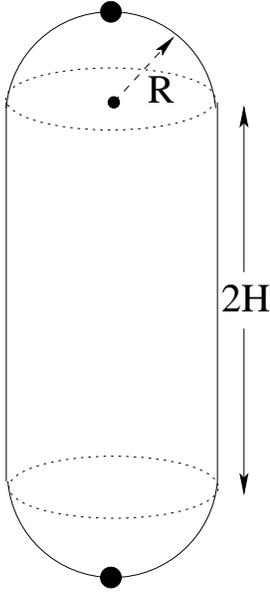}
\caption{\label{fig:goodnplenty}A capped cylinder; a cylinder of length
$2H$ is closed off by hemispheres at the north and south poles
of radius $R$. The circulation around
every lattitude is the same.}
\end{figure}

In this section, we show that
the quantization of circulation can induce
an extremely long-range force on a stretched-out sphere (such as a
surface with the shape of
a zucchini or a very prolate spheroid).  We
first demonstrate the main result in the context of a simple
example before presenting a general formula for the forces experienced
by vortices on azimuthally
symmetric surfaces. Details are presented in
Appendix \ref{app:deimos}.
Consider
a cylinder
of length $2H$ and radius $R<<H$ with hemispherical caps of radius $R$
at the ends,
depicted in Fig. \ref{fig:goodnplenty}, and imagine a symmetric
arrangement of a vortex ($n$=1) and an anti-vortex ($n$=-1) at the
north and south poles respectively.
Extrapolating our intuition from flat space
suggests that the energy of the vortex and
anti-vortex
is $2\pi K\ln\frac{D}{a}$, where $D$ is the distance between the vortices.
However, more careful reasoning shows that the energy
grows \emph{linearly} rather than logarithmically with $D$. 
The reason is that, unlike in flat space, the velocity field
does not fall off like the inverse of the distance from each vortex.
Note that the azimuthal symmetry of the arrangement of the
vortices implies that the flow
is parallel to the lines of latitude of the surface.
Since the circulation around each latitude must be $\frac{h}{m}$, 
the flow speed on the cylindrical part reads 
\begin{equation}
v=\frac{h}{2\pi m R}.
\label{eq:farina}
\end{equation}
The 
kinetic energy of this part of the flow is
 $[4\pi RH]\frac{1}{2}\rho_s v^2=2\pi\frac{KH}{R}$. Since this
cylindrical part
of the flow forms the main contribution to the kinetic energy when $H>>R$
we find that the energy of a vortex-antivortex pair situated
at opposite poles is linear,
\begin{equation}
E_{poles}\approx 2\pi\frac{KH}{R}.
\label{eq:lateral}
\end{equation}
(The exact expression also includes a near-vortex energy of approximately 
$2\pi K\ln\frac{R}{a}$.)

In contrast, when
the vortices forming the neutral pair
are across from each other on the same latitude, the aforementioned long-range
persistence of the velocity field is absent because the vorticity is
screened within a distance of order $R$. The resulting
kinetic energy follows the familiar logarithmic growth
\begin{equation}
E_{equator}\approx 2\pi K\ln\frac{2R}{a}.
\label{eq:opposite}
\end{equation}

\begin{figure}
\includegraphics[width=.45\textwidth]{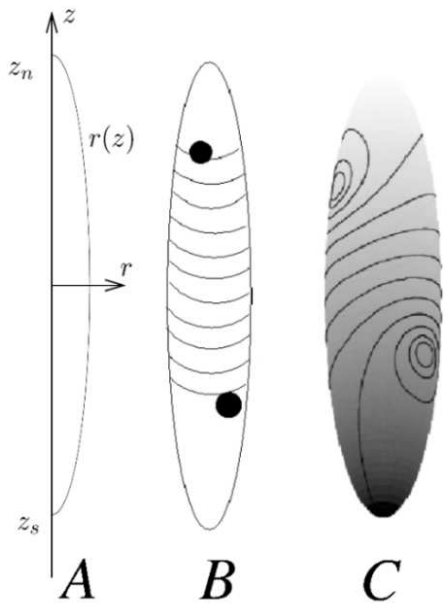}
\caption{\label{fig:kitestring}The flow on an azimuthally symmetric
surface, described by the coordinates $(z,r)$ and an azimuthal
angle $\phi$ (not shown). 
A) The surface is defined as the surface of
revolution of the curve $r(z)$ in the $r-z$ plane.
The other two images show the flow on an ellipsoid and
compare 
the flow pattern predicted by the band model (B)
and the exact solution determined by conformal mapping (C).
The flow lines in the band between the two vortices
become close to horizontal and are approximately azimuthally
symmetric.
Beyond the vortices, they are spaced far apart indicating a
vanishingly small speed for a greatly elongated surface.
}
\end{figure}

More generally, consider an azimuthally symmetric surface described by the
radial distance, $r(z)$, as a function of height, $z$, as indicated in Fig. 
\ref{fig:kitestring} A. If the north and south poles of
the surface are at $z_s$ and $z_n$,
then $r(z_s)=r(z_n)=0$ since the surface closes at the top and bottom. 
A point
on such a surface can be identified by the coordinates
$(\phi,\sigma)$ where $\phi$ is the azimuthal angle and $\sigma$ is the
distance to the point from the north pole along one of the longitudes such
as the one shown
in Fig. \ref{fig:kitestring}A.
In Appendix \ref{app:deimos}, we develop an approximation scheme which
rests on the observation that the flow pattern becomes mostly
azimuthally symmetric
if $\frac{dr}{dz}<<1$, even if the vortices break the azimuthal symmetry
of the surface because they are not at the poles.
If a pair of $n=\pm 1$ vortices are present at different heights $z_{1,2}$,
then
the fluid in the band between them
flows almost horizontally and at a nearly $\phi$-independent speed
(except for irregularities near the vortices) while the fluid beyond
them is approximately stagnant (see Fig. \ref{fig:kitestring}). 
 Along any latitude inside the band the circulation is \emph{exactly}
$\frac{h}{m}$, while it is zero above and below it. 
These properties approximately determine the flow away from
the vortices since the asymmetric irregularities near the
vortices decay
exponentially, giving a speed of
\begin{equation}
\frac{h}{2\pi m r(z)}
\label{eq:simplebandspeed}
\end{equation}
in between $z_1$ and $z_2$, the locations of the vortices,
and zero elsewhere.  We describe this approximation as the ``band model".
This expression shows that constrictions in the surface cause the speed
to increase.  For the cylinder with spherical caps
and arbitrarily placed vortices, 
Eq. (\ref{eq:simplebandspeed}) shows that the speed
is approximately constant within the band. (It increases within a distance
on the order of $R$ from the vortices, which are on the edges of the band.)

The kinetic energy can be determined approximately by noting that
the energy in a thin ring on the surface between the vortices 
(extending from the longitudinal arclength $\sigma$ to $\sigma+d\sigma$)
is
\begin{equation}
[2\pi r(z)d\sigma][\frac{1}{2}\rho_s (\frac{\hbar}{m r(z)})^2]=\pi K
\frac{d\sigma}{r(z)}
\nonumber
\end{equation}
where the first factor represents the area of the ring since
$\sigma$ is the geodesic distance along the surface and
the second factor represents the included kinetic energy. 
The flow is zero
past the two vortices, so the total energy is
\begin{equation}
E=K\pi \int_{\sigma_1}^{\sigma_{2}}\frac{d\sigma}{r}.
\label{eq:bandenergy1}
\end{equation}
For the capped cylinder with a constant $r(z)$ the integral is 
$\pi\frac{KD}{R}$, where $D$ is the vertical distance between the vortices.
This expression generalizes the result for vortices at the poles of the surface,
Eq. (\ref{eq:lateral}). However, when the vortices are on opposite ends of
an equator, they are too close for the asymmetries to be neglected. The
energy in this case is calculated in Appendix \ref{app:deimos}.
The force $F_{1,band}$, experienced by
vortex $1$ can be determined by taking the
gradient of Eq. (\ref{eq:bandenergy1}). If vortex 1 is moved downward, the 
band
of moving fluid shrinks so the energy drops. 
The force on the vortex therefore reads
\begin{equation}
F_{1,band}=\frac{\pi K}{r(\sigma_1)}
\bm{\hat{\sigma}}.
\label{eq:bandforce1}
\end{equation}
On the capped cylinder, this force
is independent of the positions of the vortices.  Even on
an arbitrary elongated surface, a noteworthy feature is
 that the force on vortex $1$ does
not depend on the position of vortex $2$!
This force can be explained with the
familiar phenomenon of lift: the vortex is on the boundary between
stationary and moving fluid, so there is a pressure difference
due to the Bernoulli principle. 

Approximating the flow pattern generated by multiple vortices in
a similar fashion
requires only minor modifications of the previous argument.
In a low-resolution snapshot of the flow,  the point-vortices
would appear as circles of discontinuity in the velocity
field that go all the way around the axis (the analogue for a layer
of superfluid
of a two
dimensional
vortex sheet).
If the vortices are
labeled in order of decreasing $z$ a loop just below
the $l^{\mathrm{th}}$ vortex contains 
\begin{equation}
N_l=\sum_{i=1}^l n_i
\label{eq:balancesheet}
\end{equation}
units of circulation above it. 
Approximate azimuthal symmetry of the flow then implies that,
\begin{equation}
\mathbf{v}(z,\phi)_{band}=N_l\frac{\hbar}{m r(z)}\mathbf{\hat{\phi}}\ \ \ 
(\mathrm{for\ }z_l<z<z_{l+1}),
\label{eq:bandspeed}
\end{equation}
a natural generalization of Eq. (\ref{eq:simplebandspeed}) that is proved
in Appendix \ref{app:deimos}.

Conformal mapping can be employed to justify (without detailed calculations)
the
decay of the nonazimuthally symmetric parts of the flow that are not
determined by the quantization condition.
We sketch the basic reasoning here by focusing
(for simplicity) on the flow pattern \emph{near the
equator} of the surface, at a distance
$\sigma_{eq}$ from the north pole.  The conformal 
transformation that maps the elongated sphere onto a regular 
reference sphere with coordinates $\Theta,\Phi$ reads
(see Appendix \ref{app:deimos})
\begin{eqnarray}
&&\sin\Theta=\mathrm{sech}\int_{\sigma}^{\sigma_{eq}}
{\frac{d\sigma'}{r(\sigma')}}\nonumber\\
&&\Phi=\phi.
\label{eq:deimosmap}
\end{eqnarray}
The upper and lower
halves of the elongated sphere can be
 mapped to the upper and lower hemispheres
by choosing appropriately between the two values of $\Theta$
that correspond to a given value of $\sin \Theta$. 
Near the equator the integral can be approximated
by $\frac{\sigma-\sigma_{eq}}{r_{eq}}$ since $r$ varies slowly.
Suppose the vortices are far from the equator, 
 at a distance greater than $kr_{eq}$ for a large $k$. Then the vortices
above the equator are mapped
exponentially close (at a distance less than $e^{-k}$)
to the sphere's north pole. Likewise vortices
on the southern half of the surface map exponentially close
to the sphere's south pole. We have thus reduced the task of
finding the flow due to a complicated arrangement of
vortices to a symmetric case. In fact, after mapping the
flow on the long, thin surface to the reference sphere, nothing can
be resolved beyond a pair of
multiply quantized vortices at the north and south poles
containing $N$ and $-N$
units of circulation respectively, where $N$ is the 
total circulation number of
all the vortices above the equator. 
Since the image vortices are very close to the poles,
their flow pattern on the reference surface is approximately azimuthally
symmetric near the equator.  When mapped back to
the elongated surface, the flow retains its approximate azimuthal symmetry
in the region around the equator, completing our argument.
A similar argument proves the approximate azimuthal symmetry
of the flow
near lines of latitude other than the equator; 
one simply adjusts the conformal map
in Eq. (\ref{eq:deimosmap})
so that another latitude of the target surface
is mapped to the equator of the reference sphere.


Now the geometrical force derived in Eq. (\ref{eq:bandforce1}) 
can compete with physical forces such as those induced by rotating an ellipsoid about its long axis
with angular velocity $\bm{\Omega} = \Omega \!\!\!\! \quad \bm{\hat{z}}$. 
Let us extend the
treatment of rotational forces on curved substrates, introduced in section \ref{sec:Rotation}, to the case of an
ellipsoid described by the radial function
\begin{equation}
r(z)=R\sqrt{1-\frac{z^2}{H^2}}.
\label{eq:prolate}
\end{equation}
Let us 
use the aspect ratio $\alpha=\frac{H}{R}$ to describe how elongated this
ellipsoid is, and determine
how a vortex-anti-vortex pair
is torn apart by the rotation as
the angular frequency is increased. As in Section \ref{sec:Rotation}, metastable
vortex configurations can often be found, so we will consider
transitions between different \emph{local} minima of the vortex-energy function.
Recall that the effect of the rotation on the superfluid energy is expressed in terms of the angular momentum $L_z$  by an extra term $-\Omega L_z$ in Eq. (\ref{eq:Erot})
which must be evaluated for (at least)
a pair of opposite-signed
vortices to satisfy the topological constraints imposed by the spherical
topology of the 
surface. We find, in analogy to Eq. (\ref{eq:Lz-c}), that the
extra contribution to the vortex energetics is additive and reads
\begin{equation}
-\Omega L_z=n_1\hbar \Omega \frac{\rho_s}{m} [A(\sigma_1)-A(\sigma_2)],
\label{eq:Lzvortex}
\end{equation}
where $A(\sigma)$ represents the area of the ellipsoid out to
a distance $\sigma$ from the north pole while $\sigma_1$ and $\sigma_2$ represent the positions of the two defects ($\sigma_1<\sigma_2$). 
If $n_1=1$, the energy in 
Eq. (\ref{eq:Erot}) is
decreased by moving the positive vortex closer to the north pole and the negative one closer to the south pole. To understand this, 
note that the sense of rotation
of the superfluid around a vortex is defined by an observer facing
the surface. Hence, a negative vortex on the southern half of the surface
rotates in the same direction as a positive one on the northern half
(relative to the positive $z$-axis), and both agree with the sense of rotation
of the substrate.
The rotational force on vortex 1
derived from Eq. (\ref{eq:Erot}) is
\begin{equation}
\mathbf{F_{\Omega}}=-\frac{n_1\hbar\rho_s}{m}2\pi r(\sigma)\mathbf{\hat{\sigma}} .
\label{eq:Frotsequel}
\end{equation}

As $\Omega$ increases, \emph{pairs} of positive and negative vortices will
appear in this geometry. 
As each vortex pair is created, the positive vortex will move to the top side
of the surface, and the negative one to the bottom. There is a critical frequency,
$\Omega_b$, at which a single pair of vortices, once created, can
exist metastably in a configuration symmetric about the $xy$-plane.  
As the angular
frequency is increased, the vortices are gradually pulled apart
until at the frequency $\Omega_a$, they reach the poles. (The latter
transition is
analogous to the center/off-center transition for a single vortex described
in Section \ref{subsec:single}.)
%
When $\Omega_{a}>\Omega>\Omega_b$, the  equilibrium condition, obtained by balancing the forces in
Equations (\ref{eq:bandforce1}) and (\ref{eq:Frotsequel}), reads
\begin{equation}
\frac{\pi K}{r_1}=2\pi\frac{\hbar\rho_s}{m}\Omega r_1
\label{eq:heartgrowsfonder}
\end{equation}
Since $r$ and $z$ are connected by the equation of the ellipsoid,
the vortices are located at heights
\begin{equation}
\pm z\approx \pm\alpha\sqrt{R^2-\frac{\hbar}{2m\Omega}}.
\label{eq:prize}
\end{equation}

\begin{figure}
\psfrag{Energy/K}{Energy/$K$}
\psfrag{sigma/R}{$\sigma/R$}
\includegraphics[width=.45\textwidth]{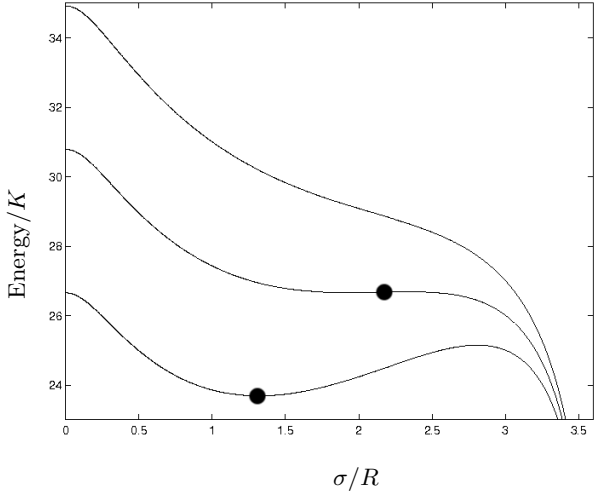}
\caption{\label{fig:spndl}The rotational and fluid energy
(units of $K$) as a function of $\sigma_1=\sigma_{tot}-\sigma_2=\sigma$
(units of $R$) for $H=3.5 R$ and $\frac{m\omega}{\hbar}=.49,.61,.74 R^{-2}$.
The middle curve, roughly at $\omega=\omega_b$,
shows the last position where the vortex is stable as $\omega$ is decreased.}
\end{figure}

The vortices first become metastable when force balance is achieved with both vortices close to the equator. Upon substituting the equatorial value
$r_1=R$ 
into Eq. (\ref{eq:heartgrowsfonder}) an estimate of $\Omega_b$ is obtained
\begin{equation}
\Omega_b\approx\frac{\hbar}{2mR^2}.
\label{eq:newadventure}
\end{equation}
When the pair first appears, there will actually 
be a non-zero defect separation,
although substituting Eq. (\ref{eq:newadventure}) into Eq. (\ref{eq:prize})
suggests otherwise. Imagine slowing the
rotation speed through $\Omega_b$. 
The vortices will approach each other gradually; within the 
large vortex-separation approximation 
of Eq. 
(\ref{eq:bandforce1}), the attraction between them will \emph{decrease}
as they become closer because $r(z)$ increases. However, when the
vortices become close enough, the attraction between them
starts increasing and the vortices are suddenly pulled together.
The minimum $z$-coordinate for metastable vortices is
derived along these lines
in Appendix \ref{app:deimos} (which also discusses what
happens at $\Omega=\Omega_a$) and reads
\begin{equation}
z_1=-z_2=z_b\approx R\ln\alpha .
\label{eq:logs}
\end{equation}
The transition through $\Omega_b$ is illustrated pictorially 
in Fig. \ref{fig:spndl} which shows
how the local minimum in the energy function disappears as
the frequency decreases. 

\subsection{\label{sec:noodles}Interactions on a closed surface}
To understand interactions between vortices on an arbitrary 
deformed sphere one must come to
 terms with the neutrality constraint on the total
circulation of a flow.  On any compact surface,
\begin{equation}
\sum_i n_i=0.
\label{eq:neutrality}
\end{equation}
This constraint on the sum of the circulation indices $\{n_i\}$ always holds:
if the surface is divided into two pieces by a curve, the sum
of the quantum numbers
on the top and bottom half must be equal and opposite (because they
are both equal to the circulation
around the dividing curve).
As we shall see, this relation implies that there are multiple ways of splitting up the energy into single-particle energies and two-particle interaction
energies, despite the fact that the total
energy is well-defined.
The behavior of the one-particle
and interaction terms depends on how the splitting is carried out.
%
%
To illustrate this ambiguity, multiply Eq. (\ref{eq:neutrality}) by 
$4\pi^2 K n_1 f(\bm{u}_1)$ and separating out the $i=1$ term, to obtain
\begin{equation}
4\pi^2 n_1^2 f(\mathbf{u}_1)=-\sum_{i\neq 1} 4\pi^2 n_1 n_i f(\mathbf{u}_1).
\end{equation}
Hence, a portion $4\pi^2 K f(\mathbf{u}_1)$ of the ``geometrical
energy" of vortex 1 can be reattributed to this vortex's interaction with all
the other vortices.
This can be seen explicitly by checking
that the \emph{net}
energy according to Eqs. (\ref{eq:greenself-energy}) and
(\ref{eq:greenpair-energy}),
\begin{alignat}{1}
E(\{n_i,\bm{u}_i\})=\sum_{i<j}&4\pi^2 Kn_in_j\Gamma(\bm{u}_i,\bm{u}_j)\\
&-\sum_i\pi n_i^2 KU_G(\bm{u}_i)
\label{eq:redouble}
\end{alignat}
is not changed by the following transformation:
\begin{eqnarray}
&&\Gamma'(\mathbf{u}_1,\mathbf{u}_2)=\Gamma(\mathbf{u}_1,\mathbf{u}_2)
-f(\mathbf{u}_1)-f(\mathbf{u}_2)\\
&&U_G'(\mathbf{u})=U_G(\mathbf{u})+4\pi f(\mathbf{u})
\label{eq:ces}
\end{eqnarray}

\begin{figure}
\psfrag{u1}{$\mathbf{u}_1$}
\psfrag{u2}{$\mathbf{u}_2$}
\psfrag{P2}{$Q_2$}
\psfrag{P1}{$Q_1$}
\psfrag{P1*}{$Q_1^*$}
\psfrag{P2*}{$Q_2^*$}
\includegraphics[width=.5\textwidth]{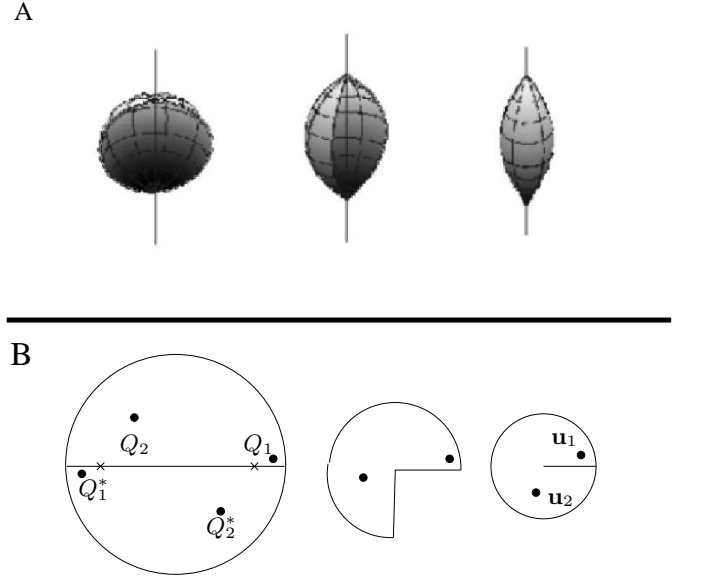}
\caption{\label{fig:noodlemachine} A) The process of folding a hemisphere
into a pointed sphere, bounded at the north and south poles by two $180^{\circ}$
disclinations.  
The north and south poles move outward along the axis, while
the latitudes stay horizontal. The Gaussian curvature is
invariant because the decreasing curvature of the lines of longitude
is compensated by the tighter curvature around the lines of lattitude.
B) A top view of vortices during the furling-up of one hemisphere.
The first stage shows both the $(0<\alpha<\pi)$
hemisphere that is furled up and the other
hemisphere together with two defects
and their images. The hemisphere is furled up into the pointed sphere
so that the left and right halves of the cut (which appears as a horizontal
diameter in this top view) are brought together
to form the seam on the pointed sphere; simultaneously, the defects $Q_1,Q_2$ 
move to positions $\mathbf{u}_1$ and $\mathbf{u}_2$ on the pointed sphere.
The furling process leads to a continuous flow pattern
on the pointed sphere.  For example, the two points marked with an x on the cut
in the original sphere are sealed together.  Both points feel
a strong flow, the one on the right because it is close to the
vortex at $Q_1$ and the one on the left 
because it is close to this
vortex's image.
}
\end{figure}

This flexibility 
is reflected in the possibility of choosing different Green's functions
$\Gamma'(\mathbf{u}_1,\mathbf{u}_2)$  for the covariant Laplacian on a deformed
compact surface. A detailed discussion of Green's functions
is given in Appendix \ref{app:CG}.
Here we higlight this ambiguity by performing explicit calculations using two distinct choices of Green's functions on a model surface formed
from a unit sphere.  First cut the sphere in halves along a great circle.
Choose one of the hemispheres and bring opposite sides of the great
circle bounding it together and glue them.  The result is
a pointed sphere resembling a conchigliette noodle (i.e., a shell noodle)
sealed shut (see Fig. \ref{fig:noodlemachine}A). The surface closes up
smoothly since opposite sides of the seal have matching curvatures;
the surface also turns out to be rotationally symmetric.  (A similar
shape forms when a pollen grain with a weak sector
(such as the pollen of a lily) dries out \cite{microscopic}.)

The geometrical interaction on this surface can be given an appealing
interpretation in terms of the method of images from electrostatics.  
When one uses the method of images to study charges in a
half-space bounded by a conducting plane, one completes the space,
with the other half-space.  Then one introduces charges into this
fictitious region to ensure that the right boundary conditions (orthogonality
of the field lines to the original boundary plane) are satisfied.
For the pointed sphere, we first complete the surface by opening it
up again and
adding back the second hemisphere.  We can describe points on the pointed
sphere by $(\phi,\sigma)$ (where $\sigma$ is the distance from
the north point and $\phi$ is the azimuthal angle, as 
for the general case described in Sec. \ref{sec:zucchini}).
Points on the completed sphere are naturally labeled by 
the standard spherical
coordinates $(\alpha,\theta)$ that represent 
the azimuthal and polar angles respectively.  The mapping from the
pointed sphere that returns each point to its position
on the original sphere is
given by 
\begin{eqnarray}
\alpha=\frac{\phi}{2}\nonumber\\
\theta=\sigma.
\label{eq:half-rations}
\end{eqnarray}  
As $\phi$
ranges from $0$ to $2\pi$, $\alpha$ goes from $0$ to $\pi$, across
the hemisphere used to make the pointed sphere. 
Note that $\theta=\sigma$ since
the pointed sphere is formed by isometrically bending the hemisphere 
(the angle $\theta$ is equal to the geodesic distance to the north pole
of the hemisphere).

Now for each vortex at $\mathbf{u}=(\phi,\sigma)$
on the pointed sphere, we introduce two vortices on the sphere,
one at $Q$ with $(\alpha,\theta)=(\frac{\phi}{2},\sigma)$ and one
at 
$Q^*=(\frac{\phi}{2}+\pi,\sigma)$.  The latter is the image vortex
of the former, obtained by rotating $Q$ by $180^{\circ}$ around
the $z$-axis.

The energy of a defect configuration on the pointed sphere is derived by halving
the energy of the flow pattern produced by the doubled set of vortices
on the full sphere.  In analogy with the electrostatic problem,
the purpose of situating
the image defects in the way just described
is to preserve the continuity of flows across the seam. Imagine
drawing the flow pattern of all the vortices on the
sphere.  Focus on the hemisphere $0<\alpha<\pi$. 
Because the vortices are placed
symmetrically about the sphere's axis, the flow near
$\alpha=0$ will match the flow near $\alpha=\pi$ when the
surface is sealed.  (See Fig. \ref{fig:noodlemachine}B.)

The flow pattern on the pointed sphere results from rolling up
half of the flow pattern on the sphere.
Once the positions of the image defects are chosen, the flow pattern
is found by deriving it from the stream function $\chi(\bm{u})$
introduced in Sec. \ref{subsec:effective}.  The stream function 
at a point $\bm{u}$ on the pointed sphere can
be expressed in terms of the Green's function of the sphere, according
to Eq. (\ref{eq:green-stream1}):
\begin{equation}
\chi(\mathbf{u})
=\sum_i{\frac{n_ih}{m}[\Gamma^{sphere}(Q,Q_i)+\Gamma^{sphere}(Q,Q_i^*)]}
\label{eq:darkandlight}
\end{equation}
where $Q$ is the point on the sphere corresponding to $\mathbf{u}$
on the pointed sphere and
$Q_i$ and $Q_i^*$ are the locations of the $i^{\mathrm{th}}$ pair of vortices
on the sphere. 


The energy of the vortices on the sphere
takes up the familiar electrostatic form of a 
sum of the interactions between all pairs of
defects and/or their images.  The energy stored in the 
flow pattern on the pointed sphere (which is half as large as
on the complete sphere) reads
\begin{align}
\frac{E_N}{K}&=\frac{1}{4}\sum_{i\neq j} 4\pi^2n_in_j[\Gamma^{sphere}(Q_i,Q_j)
+\Gamma^{sphere}(Q_i,Q_j^*)\nonumber\\
&\ \ \ \ \ \ \ \ \ +\Gamma^{sphere}(Q_i^*,Q_j)
+\Gamma^{sphere}(Q_i^*,Q_j^*)]\nonumber\\
&\ \ \ +\frac{1}{2}\sum 4\pi^2n_i^2\Gamma^{sphere}(Q_i,Q_i^*)\nonumber\\
&=\frac{1}{2}\sum_{i\neq j} 4\pi^2 n_in_j[\Gamma^{sphere}(Q_i,Q_j)
+\Gamma^{sphere}(Q_i,Q_j^*)]\nonumber\\
&\ \ \ +\sum 2\pi^2n_i^2\Gamma^{sphere}(Q_i,Q_i^*)
\label{eq:shadows}
\end{align}
In the second expression, we note that the terms in the first line
are equal in pairs, so that a factor of $\frac{1}{2}$ cancels.
This energy is given the same form as Eq. (\ref{eq:redouble})  
by separating out
the part which depends on the positions
of \emph{two} vortices, proportional to
\begin{equation}
\Gamma_s(\mathbf{u}_1,\mathbf{u}_2)
=\Gamma^{sphere}(Q_1,Q_2)+\Gamma^{sphere}(Q_1,Q_2^*)
\label{eq:standnoodleg}
\end{equation}
and the part which depends on one vortex,
\begin{equation}
U_s(\mathbf{u})=-2\pi\Gamma^{sphere}(Q,Q^*).
\label{eq:standnoodlev}
\end{equation}
The function in Eq. (\ref{eq:standnoodleg}) is
a Green's function for the pointed sphere, as
the placement of the images guarantees that this function
is well-defined on the \emph{pointed sphere}.  It appears
in the stream function, Eq. (\ref{eq:darkandlight}), as well as in
the energetics, as expected for a Green's function.

The potential which describes the single-particle
energy of a vortex
becomes singular as the 
vortex approaches the apex of the cone at the north or south pole,
since then the vortex $Q$ approaches its image $Q^*$.  
This is in accord with the result, Eq. (\ref{eq:greenself-energy}),
that the Gaussian curvature is the source of the single-particle energy since
the pointed sphere has delta-function concentrations of curvature
at its north and south poles:
\begin{equation}
G(\mathbf{u})=1+\pi\delta_N(\mathbf{u})+\pi\delta_S(\mathbf{u}).
\label{eq:ouch}
\end{equation}
where $\delta_N(\mathbf{u})$ and $\delta_S(\mathbf{u})$ are the appropriate
delta funnctions.
The geometric repulsion from the positive curvature
points arises from the repulsion between vortices
and their images!
We can check step-by-step that $U_s$ is sourced 
by the Gaussian curvature,
\begin{equation}
U_s(\mathbf{u})=-\iint \Gamma_s(\mathbf{u},\mathbf{u'}) G(\mathbf{u'}) d^2\mathbf{u'}.
\label{eq:selfconsistent}
\end{equation}
We substitute for $G(\mathbf{u'})$ from Eq. (\ref{eq:ouch}) and
for $\Gamma_s$ from Eq. (\ref{eq:standnoodleg}) which can be written
in the form,
\begin{align}
&\Gamma_{s}(\sigma_1,\phi_1;\sigma_2,\phi_2)=
\Gamma^{sphere}(\sigma_1,\frac{\phi_1}{2};\sigma_2,\frac{\phi_2}{2})\nonumber\\
&\ \ \ \ \ \ \ \ \ \ \ \ \ \ \ \ \ +\Gamma^{sphere}
(\sigma_1,\frac{\phi_1}{2};\sigma_2,\frac{\phi_2}{2}+\pi)\nonumber\\
&=-\frac{1}{4\pi}\ln
4[(1-\cos\sigma_1\cos\sigma_2)^2
\nonumber\\
&\ \ \ \ \ \ \ \ \ \ -\sin^2\sigma_1\sin^2\sigma_2\cos^2\frac{\phi_1-\phi_2}{2}].
\label{eq:orbitrig}
\end{align}
In the last line, we have evaluated 
the Green's function for the sphere 
by writing the chordal distance between, e.g., $Q_1$
and $Q_2$ in terms of the spherical
coordinates $(\alpha_{1,2},\theta_{1,2})$, 
$D^2=2[1-\cos\theta_1\cos\theta_2-
\sin\theta_1\sin\theta_2\cos(\alpha_1-\alpha_2)]$ (see \cite{lube92}),
and then combining the two terms together.
To evaluate the integral in Eq. (\ref{eq:selfconsistent}), we have
to note that
the area element of this integral is $d^2\mathbf{u}=\frac{1}{2}
\sin\sigma d\sigma d\phi$.  The area of a region
on the pointed sphere is the \emph{same} as the area
$\sin\theta d\theta d\alpha$ of the
corresponding region
on the original sphere, and the factor of $\frac{1}{2}$ results from
how the angles are related, $\alpha=\frac{\phi}{2}$, see 
Eq. (\ref{eq:half-rations}).
Now the integral on the right-hand side of Eq. (\ref{eq:selfconsistent}) can
be shown to be equal to the left-hand side using the identities
\begin{eqnarray}
\int_0^{2\pi}\ln|A+B\cos t|dt
&=&2\pi\ln\frac{A+\sqrt{A^2-B^2}}{2}\ \mathrm{if}\ B<A\nonumber\\
&=&2\pi\ln\frac{B}{2}\ \mathrm{if}\ B>A\nonumber.
\end{eqnarray}

We have now derived one formulation of the energetics in terms of $\Gamma_s$
and $U_s$, the corresponding geometric potential.
Let us contrast this isometric mapping method with
the conformal mapping method in order to illustrate how different
approaches can naturally lead to different delineations
between vortex-vortex and
vortex-curvature interactions.  (The net result is of course the
same from either point of view.)
As a result of the isometric mapping
each point is doubled, whereas the distance-distorting conformal mapping 
transforms each point on the
pointed sphere to \emph{one} point on the reference sphere.

We first use Eq. (\ref{eq:deimosmap}) to find that 
the conformal map is given by
\begin{equation}
\tan\frac{\Theta}{2}=\tan^2\frac{\sigma}{2}.
\end{equation}
Comparing the conformal mapping results, Eqs. (\ref{eq:conformalself-energy})
and
(\ref{eq:conformalpair-energy}) to the Green's function formulation,
Eqs. (\ref{eq:greenself-energy}) and (\ref{eq:greenpair-energy})
suggests the following identification of
the interaction potential (or Green's function)
and single-particle potential:
\begin{eqnarray}
&&\Gamma_c(\mathbf{u}_1,\mathbf{u}_2)
=\Gamma^{sphere}(\Theta(\sigma_1),\phi_1;\Theta(\sigma_2),\phi_2)\nonumber\\
&&U_c(\mathbf{u})=\omega=\ln\frac{2\sin\sigma}{1+\cos^2\sigma}.
\label{eq:confnoodle}
\end{eqnarray}
These expressions differ from Equations (\ref{eq:standnoodleg}) and
(\ref{eq:standnoodlev}).
Nevertheless, as promised,
the net energy is the same whether the pairs $(\Gamma_s,U_s)$
or $(\Gamma_c,U_c)$ are used in place of $\Gamma$ and $U_G$.
In fact,
\begin{eqnarray}
&&\Gamma_c(\mathbf{u}_1,\mathbf{u}_2)=\Gamma_s(\mathbf{u}_1,\mathbf{u}_2)
-f(\mathbf{u}_1)-f(\mathbf{u}_2)\nonumber\\
&&U_c(\mathbf{u})=U_s(\mathbf{u})+4\pi f(\mathbf{u})
\label{eq:cesnoodle}
\end{eqnarray}
where $f(\mathbf{u})=-\frac{1}{4\pi}\ln(1+\cos^2\sigma)$. This transforms
the energy from the single-particle to the interaction terms consistently
as described at the beginning of the section. Appendix \ref{app:CG}
shows that the Green's function formulation is generally
equivalent to the conformal mapping result derived in Section \ref{sec:map},
even when there is no method of images 
that can be used to determine the Green's function explicitly in general.

\section{\label{sec:geomineq}
Limits on the Strength and Range of Geometrical Forces}
Geometrical forces are limited in strength due to the nonlinear relation
between the curvature and the geometric potential.  Curvature
affects both the \emph{source} of the geometrical force and the \emph{force law},
as illustrated in the examples of Secs. \ref{sec:bumps} and \ref{sec:zucchini}. 
As a consequence,
even on a wildly distorted surface (with planar topology), there is
a precise
limit on the strength of the force on
a single vortex.  This result has the character of a geometrical
optimization problem, like maximizing the capacitance of a solid
when the surface area is given.
Consider a vortex located at the center
of a geodesic disk of radius $R$.  Assume that the
Gaussian curvature is zero within the disk, but may be different
from zero elsewhere.
Then the force $\mathbf{F}$ due to
the curvature satisfies
\begin{equation}
|\mathbf{F}|\leq\frac{4\pi Kn_1^2}{R}.
\label{eq:limit1}
\end{equation}
where $n_1$ is the number of circulation quanta in the vortex.  This relation
between $R$ and $\mathbf{F}$ is proven in Appendix \ref{app:geomineq}.
%
%

If
one warps a surface in a vain attempt to overcome the limit, but
the force gets diluted
because the distortion of the region around the curvature pulls the
force-lines apart, 
as we can understand
from the simple example of vortices on 
cones. 

\begin{figure}
\psfrag{1}{$1$}
\psfrag{2}{$2$}
\psfrag{beta}{$\beta$}
\includegraphics[width=.47\textwidth]{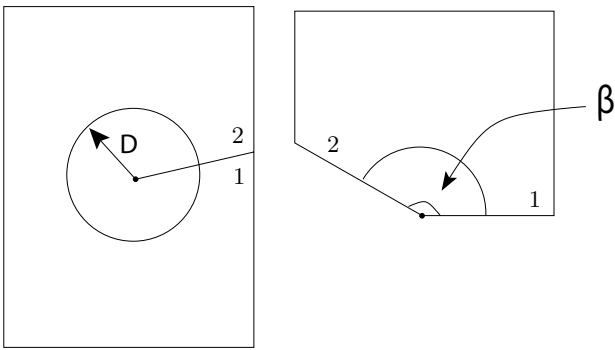}
\caption{\label{fig:firstcontortion}How to form cones of 
 negative curvature. One complete sheet of paper is slit and
an angle is cut out of an additional sheet of paper.
The edges labeled 1 are taped together, and then the
edges labeled 2 are taped. The circular arcs join together
to form an extra-large circle.  The cone angle is $\theta=2\pi+\beta$,
and cones with even larger cone angles can be formed by using additional
sheets.} 
\end{figure}

A cone of \emph{cone-angle} $\theta$
is obtained by taking a segment of paper with an angle  $\theta$ and gluing
the opposite edges of the angle together.  This is most familiar when 
$\theta<2\pi$.
If $\theta=2\pi m+\beta$, such a cone
can be produced by adding $m$ extra sheets of paper, 
as illustrated in Fig. \ref{fig:firstcontortion}.  
We slit the $m$ sheets of
paper and put them together with an angle of size
$\beta$ cut out of an additional sheet.  By gluing the edges of the
slits
together cyclically, a cone of arbitrary angle $\theta$ is made.

A cone has a delta function of curvature at its apex, but
no Gaussian curvature elsewhere
because the surface can be formed from
a flat piece of paper without stretching.
The weight $2 \pi - \theta$ of the delta function
is expressed, according to the Gauss-Bonnet theorem, as an integral of the Gaussian curvature in any region
containing the apex\cite{Kami02}
\begin{equation}
\iint G(\mathbf{u})d^2\mathbf{u}=2\pi-\oint \kappa ds
\label{eq:paralleltrans}
\end{equation}
where $\kappa$ is 
the geodesic curvature along the boundary of the region and $s$ its arc length.
Apply this formula to the circle of radius $D$
centered at the apex of the cone. Imagine the circle as it would appear on the original sheets
of paper, as in Fig. \ref{fig:firstcontortion}.  Its measure
in radians is $\beta+2\pi m=\theta$ since it consists of $m$
 complete circles together with an additional arc.  The length is therefore $S=D\theta$.
The \emph{geodesic} curvature of the circle does not change when the
cone is unfolded, so it is equal to $\frac{1}{D}$.
Upon substituting in  Eq. (\ref{eq:paralleltrans}), we obtain 
\begin{equation}
\iint G(\mathbf{u})d^2\mathbf{u}=2\pi-S\frac{1}{D}=2\pi-\theta.
\label{eq:conepoint}\end{equation}
When $\theta>2\pi$ the curvature is negative.

Now imagine a vortex (with $n_1=\pm 1$, say) at a distance $D$ from
the cone point, on the circle of circumference $S$ just considered. 
The arbitrarily large negative curvature which is possible by making $m$
large seems
to defy the general upper bound on the geometric force.  According
to Newton's theorem, applied to the radius $D$ circle centered
at the cone's apex and passing through the vortex, 
the force on the vortex is $F=\frac{\pi K\iint G d^2\mathbf{u}}{S}$.  Since
the circumference $S=D\theta$ is larger than it would be in the plane, the force
is diluted; substituting the integrated curvature from Eq. (\ref{eq:conepoint}),
we find that it is given by
\begin{equation}
F=\pi \frac{K}{D} \frac{2\pi-\theta}{\theta}.
\label{eq:cone}
\end{equation}
This satisfies Eq. (\ref{eq:limit1}) for all negatively curved cones ($\theta>2\pi$); even
when $\theta\rightarrow\infty$
the magnitude of the force is less than $4\pi\frac{K}{D}$
because the large circumference in the denominator of the Newton's theorem expression cancels
the large integrated curvature in the numerator.

In the opposite limit $\theta\rightarrow 0$, the theorem described by Eq. (\ref{eq:limit1})
is still correct of course.  One has to be careful about applying it, however. The force on
a vortex at radius $D$ (given by Eq. (\ref{eq:cone})) is \emph{not} bounded
by $\frac{4\pi K}{R}$ with $R$ set equal to $D$ when $\theta$ is small enough (in fact, for an
extremely pointed cone, $\theta\ll 1$, the force given by Eq. (\ref{eq:cone}) diverges), 
but this does
not contradict the inequality because the circle of radius $D$ centered at the vortex
is pathological:  although it does not contain any curvature, the circle wraps around the cone and intersects itself.  Taking $R$ to be the radius of the largest circle centered at the
defect which does not intersect itself, one finds that the inequality \emph{is} satisfied,
with room to spare, for all values of the cone angle $\theta$ (see Appendix \ref{app:geomineq}).
One can describe
a more awkwardly shaped surface such that the force on a singly-quantized 
vortex is arbitrarily close to the upper
bound $\frac{4\pi K}{R}$ (see Appendix \ref{app:geomineq}).



One can also provide limits to the strength of the geometric
force from a localized source of curvature.
Rotationally symmetric surfaces such as the Gaussian bump
have force fields that do not extend beyond the bump, since
the net Gaussian curvature is zero, and Newton's theorem says that only
the $integrated$ Gaussian
curvature can have a long range effect for a rotationally
symmetric surface. To
get a longer-range force, one must focus on non-symmetric surfaces, like
the saddle surface of Sec. \ref{BS}.
The integration methods of Appendix \ref{app:calmseas} can
be used to show that this surface's potential has a quadrupole form at
long distance.  Let us consider, more generally, a plane which is flat
except for a non-rotationally symmetric deformation
confined within radius $R$ of the origin. (The result
will not apply directly to the saddle
surface since its curvature extends
out to infinity.) In this case, the total integrated Gaussian curvature is zero,
implying that the long-range force law cannot have any monopole component.
A dipole component is not ruled out by this simple reasoning,
but Appendix \ref{app:geomineq} shows that the
 limiting form of the potential is at least a \emph{quadrupole}
 (or a faster
decaying field),
\begin{equation}
E(\mathbf{r})\sim n_1^2\frac{\mu_2\cos(2\phi-\gamma_2)}{r^2},
\label{eq:quadrupole}
\end{equation}
where $r$ and $\phi$ are the polar coordinates of the vortex relative
to the origin, and
$\mu_2$ and $\gamma_2$ are constants that
depend on the shape of the deformation in the vicinity of the origin.
As in the previous case, there is an upper limit
on the quadrupole moment $\mu_2$, no matter how strong the curvature of
the deformation is:
\begin{equation}
\mu_2\leq \pi K R^2.
\label{eq:limitq}
\end{equation}
For electrostatics in the plane, the 
maximum quadrupole moment of $N$ particles with charge $2\pi$
and $N$ with charge $-2\pi$
in a region of radius $R$ is at most of the order of $KNR^2$,
which has the same form as the bound in Eq. (\ref{eq:limitq}), except
for the factor of $N$.
Because of the
nonlinearity of the 
 geometrical force and 
restrictions on how much positive and negative curvature can be separated
from each other, the quadrupole moment is bounded no matter how drastically 
curved the surface is.


These results describe key physical \emph{differences} (resulting from the
fact that the curvature cannot be adjusted without changing the surface)
between the geometrical forces discussed in this work and their electrostatic counterparts despite the close resemblance from a formal viewpoint. 

\section{Conclusion}
\begin{table*}
\begin{tabular}{c}
\includegraphics[width=\textwidth]{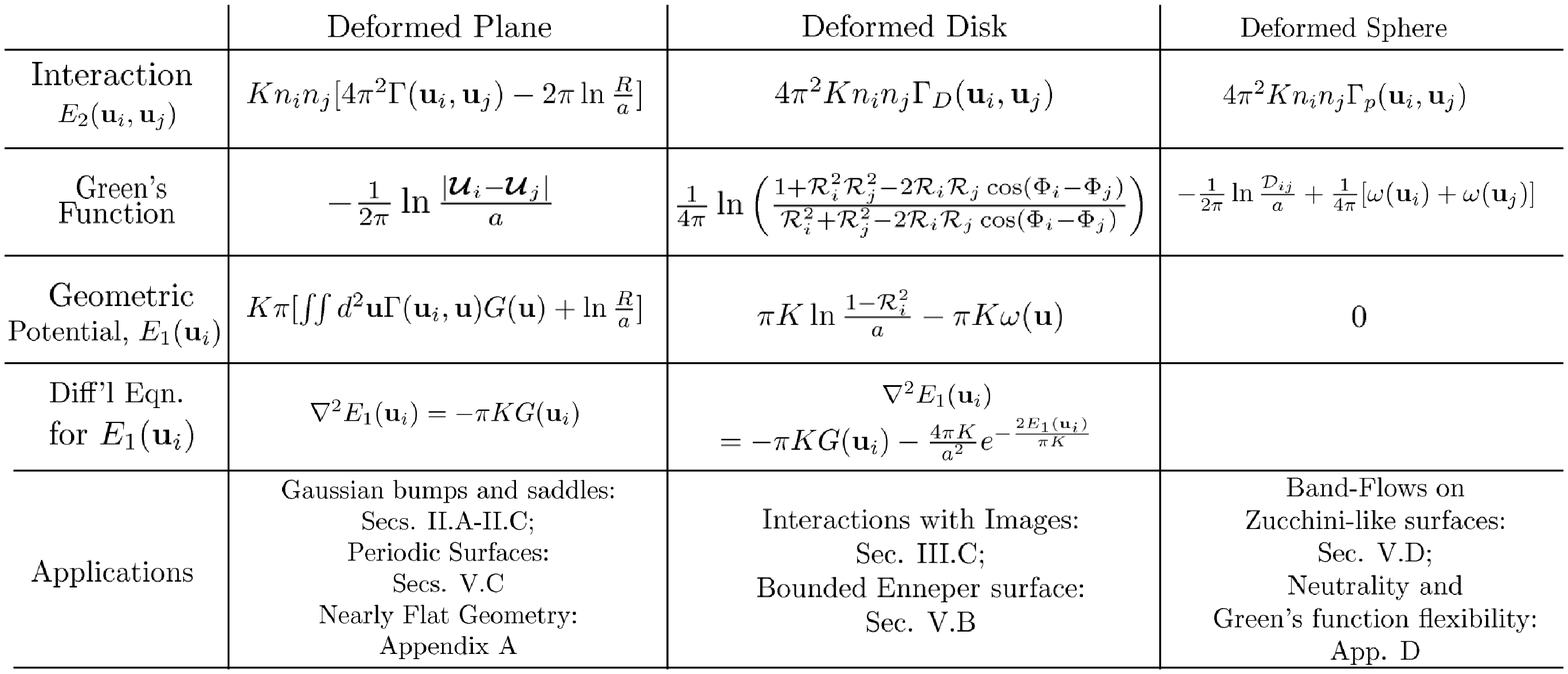}
\end{tabular}
\caption{\label{table:chair}An outline of vortex interactions on curved surfaces.
The net energy of a set of vortices on a surface with the topology of
a plane, disk, or sphere is given by $\sum_i n_i^2E_1(\mathbf{u}_i)+\sum_{i<j}E_2(\mathbf{u}_i,\mathbf{u}_j)$, where simple expressions for
the single-particle (or geometric) potential
and two-particle potentials are given in the table.  A conformal mapping
is necessary for evaluating some of these expressions.  For example,
$\bm{\mathcal{U}}_i$ (in the expression for the Green's function on a deformed
plane)
is the Cartesian coordinates of the conformal image of vortex $i$.
}
\end{table*}

In this article, we have laid out a mathematical formalism based on
the method of conformal mapping that allows one to calculate the energetics
of topological defects on arbitrary deformed substrates with a focus
on applications to superfluid helium films. The starting point of
our approach is the
observation that upon a change of coordinate the metric tensor of a complicated
surface can be brought in the diagonal form $g_{ab}=e^{2 \omega(\mathbf{u})}
\delta_{ab}$. This corresponds to the metric of a \emph{flat} 
plane which is locally
stretched or compressed by the conformal factor  $e^{2 \omega(\mathbf{u})}$.
Many of the geometric interactions experienced by topological defects on curved
surfaces are simply determined once the function $\omega(\mathbf{u})$ is known.
Vortices in thin helium layers wetting a curved surface are a natural arena to
explore this interplay between geometry and physics but our
approach is of broader applicability.

The curved geometry results in a modified law for defect interaction as well as
in a
one body geometric potential. On a deformed plane, the latter is obtained by
solving a covariant Poisson equation with the Gaussian curvature as a source.
Table \ref{table:chair} presents a summary of the general form that the defect
interaction (first row) and the
geometric potential (third row) take up in curved spaces with the topology of a
deformed plane (first column), disk (second column) and
sphere (third column). These results can be 
derived starting from the differential
equations that the geometric potential
satisfies or the appropriate Green's functions that we list in the second and
fourth row respectively for each of the three
surface topologies. The fifth row of Table \ref{table:chair} directs the
reader towards the relevant sections and appendices
of the paper where he will be able to find
some concrete applications
of the formalism and technical derivations.

For example, the geometric potential of an Enneper disk (a minimal surface with
negative curvature described in Sec. \ref{sec:bubble}) 
is given by the conformal factor
$\omega({\bf u})$ evaluated at the point $P=\{u_1,u_2\}$ where the vortex is
located combined with an ``electrostatic-like" interaction
with an image defect located at the inverse of $P$ with respect to the circular
boundary. The geometric potential
satisfies the Liouville (non-linear differential) equation that reduces to the
Poisson equation derived for the plane in the limit of
an infinitely large disk.  In the case of deformed spheres, we showed in 
Appendix \ref{app:CG}
that one can make a convenient choice of
Green's function so that \emph{all} the geometric effects are included in the
defect-defect interactions without introducing
a one-body geometric potential explicitly. An interesting application
naturally arises on vesicles deformed into an elongated shape, like
a zucchini. 
The range of
 the defect interaction becomes much longer and its functional form different
from the logarithmic dependence expected in flat two dimensional spaces.

We hope that the discussion of the geometric effects presented in this work may
pave the way for their observation in thin superfluid or liquid crystal layers
on a curved substrate. A useful starting point could be the design of
experiments to detect the geometric potential by balancing it with forces
exerted on the defects by external fields or rotation of the sample as
discussed in Section \ref{sec:Rotation}.  Such experiments should focus
on single vortices, or on situations where the separation between vortices
is comparable to the length scale of the geometry.
Signatures of the geometric interactions described
here may also survive in defect pinning experiments carried out in some bounded
three dimensional geometries \cite{zieve}.

\section{Acknowledgments}
We thank B. Halperin, R. D. Kamien, S. Trugman and  
R. Zieve for helpful suggestions. 
AMT, VV and DRN acknowledge financial support from the National Science Foundation, 
through Grant DMR-0654191, and through
the Harvard Materials Research Science and Engineering Center through 
Grant DMR-0213805.  VV acknowledges financial support from NSF Grant DMR05-47230. It is a pleasure
to acknowledge the Aspen Center for Physics for providing an interactive research environment where this article was completed.
\appendix
\section{\label{app:calmseas}Nearly Flat Surfaces}
The calculations in Section
 \ref{BS} are based on perturbations about near flatness (see
\cite{Davidreview} and references therein). 
The perturbation theory will be in powers of an aspect ratio, $\alpha$, which
measures the ratio of surface height to width of the landscape features.  
(We imagine
that the height of the surface is given in the form $\alpha m(x,y)$ where 
$m$ is a fixed function.) The leading corrections to the flat space energies
are 
second-order in $\alpha$.  There
are two of these; one is the geometric potential. When there are at least two
vortices present, there is also a second order correction to the Green's
function, which ought to be retained since it is comparable to the
geometric potential. The latter could be calculated by expanding the metric
in  Eq. (\ref{eq:BasicGreen}) in powers of $\alpha$. Nevertheless,
because the perturbations
are singular, we prefer to use conformal mapping for this step just
as we use in Sec. \ref{sec:map} to derive
the geometric potential.  Our
calculations are limited to the case of an infinite deformed plane.

We use the $x$ and $y$ coordinates of a plane parallel to the surface
for our coordinate system (the ``Monge Gauge"). The metric is then
$ds^2=dx^2+dy^2+dz^2=(1+h_x^2)dx^2+(1+h_y^2)dy^2+2h_xh_ydxdy$. Subscripts on
$h$ indicate derivatives, so that $h_{xx}=\partial_x^2h$ etc.
Upon calculating the curvature tensor we find the Gaussian curvature in
the second order approximation \cite{Davidreview}
\begin{equation}
G(x,y)=h_{xx}h_{yy}-h_{xy}^2.
\label{eq:smallcurv}
\end{equation}
The geometric potential is found
by approximating the exact expression Eq. (\ref{eq:curvature-defect}). 
The Green's function 
may be replaced by
the planar one since $G$ is already quadratic in $\alpha$:
\begin{eqnarray}
U_G(x,y)&=&-\frac{E}{\pi K}\nonumber\\
&\approx&-\iint dx'dy'{\Gamma_{flat}(x,y;x',y')(h_{xx}h_{yy}-h_{xy}^2)}\nonumber\\
\label{eq:smallgeom}
\end{eqnarray}

To do a conformal mapping to an equivalent flat space problem, 
we must solve the curved space generalization of the
Cauchy-Riemann equations\cite{Davidreview}
which define isothermal coordinates, $\mathcal{X}$,$\mathcal{Y}$, namely
\begin{equation}
 \nabla^a\mathcal(Y)=-\gamma^a_b\nabla^b\mathcal(X).
\label{eq:tack}
\end{equation}
where $\gamma^{a}_b=g^{ac}\sqrt{g}\epsilon_{cb}$.
(In words, the gradients of $\mathcal{X}$ and $\mathcal{Y}$ are at right angles
to each other and have equal magnitudes at every point.) 
We insert the expression for the metric in terms of $h$ into Eq. (\ref{eq:tack})
and expand to second order
under the assumption that $\mathcal{X}=x+\xi$,$\mathcal{Y}=y+\eta$
where the deformation parameters
 $\xi$ and $\eta$ are second order in $\alpha$:
\begin{eqnarray}
\eta_x+\xi_y\approx h_xh_y\label{eq:antidiag}\\
\eta_y-\xi_x\approx\frac{1}{2}(h_y^2-h_x^2)\label{eq:diag}.
\label{eq:smallmess}
\end{eqnarray}
By taking the derivative of Eq. (\ref{eq:antidiag})
 with respect to $x$ and  Eq. (\ref{eq:diag})  with
respect to $y$ and adding the results, we obtain
\begin{equation}
\nabla_{flat}^2\eta\approx h_y\nabla_{flat}^2 h,
\end{equation}
which may be solved by means of the Green's function
(keeping in mind the boundary condition that the
conformal map must approach the identity at infinity; i.e., $\eta\rightarrow 0$), giving
the result
\begin{equation}
\eta(x,y)\approx -\iint{dx'dy'\Gamma_{flat}(h_y \nabla_{flat}^2h)}.\label{eq:eta}
\end{equation}
Similarly, we may solve for $\xi$ (and then check that (\ref{eq:smallmess}) is
satisfied). 

We can
use our expression for the conformal mapping in conjunction
with the conformal invariance of the Green's function
Eq. (\ref{eq:greentransf}), which implies that
$\Gamma(x_1,y_1;x_2,y_2)=
\Gamma_{flat}(x_1+\xi_1,y_1+\eta_1;x_2+\xi_2,y_2+\eta_2)$. Upon expanding 
the flat space Green's function in $\xi$ and $\eta$, we have
\begin{equation}
\Gamma(x,y;x',y')\approx-\frac{1}{4\pi}\ln(\Delta x^2+\Delta y^2)-\frac{\Delta\xi\Delta x+\Delta\eta\Delta y}{2\pi(\Delta x^2+\Delta y^2)},
\label{eq:smallgreen}
\end{equation}
where $\Delta Q$ indicates taking the difference between the
values of $Q$ at the two points where
the Green's function is evaluated.  Eq. (\ref{eq:eta}), and its analogue
for $\xi$ now show that Eq. (\ref{eq:smallgreen}) gives the interaction eneragy
to the same order as the geometric potential.

Two vortices of opposite signs that
 are very near to one another (a distance $l$ such that $a<<l<<r_0$ for
a Gaussian bump, say)
cannot tell whether they are in curved or flat space. Indeed, the flow fields cancel
one another outside a range moderately greater than $l$. 
If $l$ is much less than
the curvature scale, the effects of curvature are negligible. The energy
of the vortices should therefore be $2\pi K\ln \frac{l}{a}$ as in
flat space. This conclusion can be checked explicitly in
the small aspect ratio approximation by combining
the expressions for the geometric interaction
and the correction to the Green's function
Eq. (\ref{eq:smallgeom}) and Eq. (\ref{eq:smallgreen}). (One can take
the $\Delta x,\Delta y\rightarrow 0$
limit of $\frac{\Delta\xi\Delta x+\Delta\eta\Delta y}{(\Delta x^2+\Delta y^2)}$
with the help of Eqs. (\ref{eq:antidiag}),(\ref{eq:diag}) and (\ref{eq:eta}).)
As expected, the
dependence on the surface profile $h$ cancels.
This consistency check
was behind our original suspicion about the existence of
a geometric interaction.
The energy of two vortices at $\mathbf{u}_1,\mathbf{u}_2$,
without the geometric interaction
 included is 
$-4\pi^2 K\Gamma(\mathbf{u}_1,\mathbf{u}_2)$. 
This energy differs from the flat space energy even
when the vortices are very close to one
another by a position
dependent contribution:
\begin{equation}
E_{int}(\mathbf{u}_1,\mathbf{u}_2)\approx
2\pi K\ln s_{12}-4\pi^2 K g(\frac{1}{2}(\mathbf{u}_2+\mathbf{u}_1)).
\label{eq:inequality}
\end{equation}
This cannot be the correct expression, since as just argued,
the energy should be the same as in flat space.
Single particle energies give a simple resolution.
If the total energy were 
\begin{equation}
E=-4\pi^2 K\Gamma(\mathbf{u}_1,\mathbf{u}_2)
+2\pi^2 K g(\mathbf{u}_1)+2\pi^2 K g(\mathbf{u}_2),
\label{eq:democratic}
\end{equation}
then all the $g$'s will cancel when
$\mathbf{u}_1\rightarrow\mathbf{u}_2$.  The Green's function
calculations in \cite{geomgenerate}
and the conformal mapping calculations in Section \ref{sec:map} 
show that this is actually the correct resolution, and that
$g(\mathbf{u})=-\frac{U_G(\mathbf{u})}{2\pi}$.

\section{\label{app:multipole}The Saddle Surface's Potential}
For the saddle surface with a small aspect ratio (see Eq. (\ref{eq:saddlemesa}),
we may determine the entire
geometric potential analytically as a function of position.  We will
only outline the procedure here.
We would like to evaluate
\begin{equation}
U_{\rho}(\mathbf{r})=-\int\frac{1}{2\pi} \ln|\mathbf{r}-\mathbf{r'}|\rho(\mathbf{r'})dx'dy'
\label{eq:coulrho}
\end{equation}
when $\rho(x,y)=G(x,y)$ 
is the curvature of the surface (at the point vertically
above $(x,y)$; we are using the small-aspect ratio approximation of Appendix
\ref{app:calmseas}).  
Some thought shows that the curvature given by
Eq. (\ref{eq:smallcurv}) for the surface Eq. (\ref{eq:saddlemesa}) 
takes the form of a
polynomial times $G_0=e^{-\frac{x^2+y^2}{r_0^2}}$.  We will therefore
discuss how
to evaluate the potential $U_{\rho}$ for
``charge" distributions of the form
\begin{equation}
\rho(x,y)=P(x,y)e^{-\frac{x^2+y^2}{r_0^2}}.
\label{eq:gaussy}
\end{equation}

We start with $\rho=G_0$; as discussed above, the azimuthal of this
distribution
symmetry allows its potential to be determined by Gauss's Law:
\begin{equation}
-\nabla U_{G_0}=\frac{\mathbf{r}}{2\pi r^2}\int_0^r 2\pi r'G_0(r')dr'
\label{eq:vrho0}
\end{equation}
This integral is elementary and $U_{G_0}$ can be evaluated by one further
integration, although this cannot be done in closed form.  

Conveniently,
the potential due to a distribution of the form (\ref{eq:gaussy}) can be
determined from the special case  of $\rho=G_0$ by differentiation.
(Intuitively, derivative charge distributions such as $\partial_xG_0$ 
are superpositions of infinitesimally shifted copies of $G_0$.
We can therefore apply superposition to find potentials for such
distributions.  This is analogous to finding the 
electric fields of multipoles by differenting the monopole
field.) 
To this end, we rewrite Eq. (\ref{eq:coulrho}) for the special case $\rho=G_0$
as
\begin{equation}
U_{G_0}(\mathbf{r})=-\int\frac{1}{2\pi} \ln |\mathbf{\Delta}|
G_0(\mathbf{r}-\mathbf{\Delta}) d\Delta_x d\Delta_y.
\label{eq:coulrho2}
\end{equation}
where $(\Delta_x,\Delta_y)$ are the components of $\mathbf{\Delta}=\mathbf{r}
-\mathbf{r'}$.  It follows that
\begin{equation}
\partial_{x}^n\partial_{y}^mU_{G_0}(\mathbf{r})=
-\int\frac{1}{2\pi} \ln |\mathbf{\Delta}|
\partial_{x}^n\partial_{y}^mG_0(\mathbf{r}
-\mathbf{\Delta}) d\Delta_x d\Delta_y.
\label{eq:mess}
\end{equation}
The right hand side represents the potential corresponding to the source in
Eq. (\ref{eq:gaussy}) with a special degree $k$
polynomial in place of $P$. This polynomial, obtained by multiple
differentiations of a Gaussian, is very complicated, but we will show that
polynomials of this specific form
can be superimposed to give any desired polynomial 
(including the degree $8$ polynomial appropriate to our Gaussian 
saddle-surface).  We will then, in principle, be able to express $U_G$ as a
superposition of $U_{G_0}$ and its derivatives.

The expansion of the charge distribution of Eq. (\ref{eq:gaussy})
in terms of the derivatives of $G_0$ can be carried out with
the help of Fourier integrals.   Our goal is to find an expression
of the form 
\begin{equation}
\rho(x,y)=P(x,y)e^{-x^2-y^2}=Q(\partial_x,\partial_y) e^{-x^2-y^2}
\label{eq:opeq}
\end{equation}
where $P(x,y)$ is the polynomial appearing in Eq. (\ref{eq:gaussy}). 
We have to determine a polynomial operator 
$Q(\partial_x,\partial_y)=\sum_{n,m}q_{nm}\partial_x^n\partial_y^m$ so that
Eq. (\ref{eq:opeq}) is true.
We have set $r_0=1$ for convenience. 
Applying the  Fourier transform to both sides of Eq. (\ref{eq:opeq}) gives
\begin{equation}
P(i\partial_{p_x},i\partial_{p_y})e^{-(\frac{p_x^2+p_y^2}{4})}=Q(i p_x,i p_y)e^{-(\frac{p_x^2+p_y^2}{4})},
\end{equation}
or (by substituting $u=ip_x$,$v=ip_y$),
\begin{equation}
Q(u,v)=e^{-(\frac{u^2+v^2}{4})}P(-\partial_u,-\partial_v) e^{(\frac{u^2+v^2}{4})}
\label{eq:herm}
\end{equation}
The operator $Q(\partial_x,\partial_y)$ which satisfies
Eq. (\ref{eq:opeq}) can be 
produced by working out the derivatives in this expression and replacing 
$u$ and
$v$ by $\partial_x$ and $\partial_y$.  
Now the potential can be worked out using
\begin{equation}
U_{G}(\mathbf{r})=Q(\partial_x,\partial_y) U_{G_0}(r).
\label{eq:multipole2}
\end{equation}
In fact, multiplying Eq. (\ref{eq:mess}) by $q_{nm}$ and summing over 
$n$ and $m$ shows (with the help of Eq. (\ref{eq:opeq}) that
$Q(\partial_x,\partial_y)U_{G_0}(\mathbf{r})=
-\frac{1}{2\pi}\iint \ln|\bm{\delta}|\rho(\bm{r}-\bm{\Delta}|)
d\Delta_xd\Delta_y$.

Since all \emph{derivatives} of $U_{G_0}$ can be calculated 
analytically starting
from Eq. (\ref{eq:vrho0}), Eq. (\ref{eq:multipole2}) will yield
an analytic
expression for $U_{G}$, provided we can
show that $Q$ has no constant term.  To show this, we integrate both sides of
(\ref{eq:opeq}) to see that
\begin{equation}
\pi Q(0,0)=\int P(x,y)e^{-(x^2+y^2)}dx\ dy=\int G(x,y)dxdy.
\label{eq:sumrule}
\end{equation}
That is, $Q$'s constant term is proportional to the net Gaussian
curvature; since the net curvature is zero for any surface which flattens
out at infinity, $Q$ has no constant term.

The potential of the saddle surface can thus be determined in closed form
by the following procedure: expand the curvature to determine the polynomial
$P$.  Calculate $Q$ from (\ref{eq:herm}).  Since (\ref{eq:sumrule}) guarantees
that $Q$ has no constant term, we may calculate the geometrical potential by
differentiating (\ref{eq:vrho0}) repeatedly.  This method 
is not much more practical for human calculations
than is numerically integrating (\ref{eq:coulrho}) by hand.  
A computer program, like
Mathematica (which produced 272 terms), can use Eqs. (\ref{eq:multipole2})
and (\ref{eq:herm}) to calculate
the values rapidly and make the graphs shown in Figs. \ref{fig:dangerous} and
\ref{fig:peacefulvalley}.  There is one comprehensible
consequence of these calculations:
the long distance potential is dominated
by a quadrupole, attracting vortices from some directions and repelling
them towards others. Hence there are four additional local
minima \emph{outside} of the central trap!

\section{\label{app:vdW}Van der Waals Attraction on a Curved Surface}
\begin{figure}
\psfrag{xi3}{$\xi_3$}
\psfrag{xi1}{$\xi_1$}
\psfrag{xi2}{$\xi_2$}
\psfrag{S'}{$S_{tan}$}
\psfrag{S}{$S$}
\includegraphics[width=.47\textwidth]{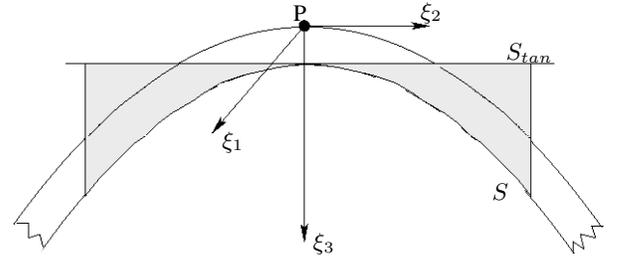}
\caption{\label{fig:curvwet}The disjoining pressure is the interaction between
$P$ and the substrate, which is below the surface $S$ in
the figure. This can be calculated by
using the result for the interaction with an imaginary
flat substrate (below the
tangent plane $S_{tan}$)
and subtracting
the very narrow shaded region.}
\end{figure}

Because the van der Waals force is very short-ranged, falling off like
$\frac{1}{r^6}$, one can
approximate the integral expression for 
the disjoining pressure $\Pi({\bf x})$ in
Eq. (\ref{eq:wet-4}) by corrections depending only on the local curvature of
the substrate. The integral is the total van der Waals interaction energy 
between a point $P$ at $\mathbf{x}$ which is above the helium film and
all
the atoms in the substrate:
\begin{equation}
\Pi(\mathbf{x})=\int w(|\mathbf{x}-\mathbf{x'}|)d^3\mathbf{x'}
\end{equation}
We now choose a simpler coordinate system (see Fig. \ref{fig:curvwet}) 
by rotating
space so that the tangent plane to
the substrate at the point of the substrate closest to $P$ becomes
horizontal. Let us take the point of tangency to be the origin of
our new coordinates, $(\xi_1,\xi_2,\xi_3)$
In this coordinate system, $P$ is  the point $(\xi_1,\xi_2,\xi_3)=(0,0,D)$
(where $D$ is the thickness of the film at $P$).
The
rotated substrate can be described by its height above 
the new ``horizontal plane" (an arbitrary plane parallel to
the tangent plane) 
using the equation $\xi_3=h_{rot}(\xi_1,\xi_2)$. 
The disjoining pressure is
\begin{widetext}
\begin{equation}
\Pi(P)=\iint {d\xi_1d\xi_2\int_{-\infty}^{h_{rot}(\xi_1,\xi_2)}{d\xi_3
w(\sqrt{\xi_1^2+\xi_2^2+(D-\xi_3)^2})}}
=-\frac{\pi \alpha}{6 D^3}-\iint d\xi_1d\xi_2\int_{h_{rot}(\xi_1,\xi_2)}^{h_{rot}(0,0)}
{d\xi_3 w(\sqrt{\xi_1^2+\xi_2^2+(D-\xi_3)^2})}
\end{equation}
\end{widetext}
where we have first integrated over the entire region below the plane $\xi_3=0$,
thereby getting the van der Waals interaction between a point 
and a flat substrate as the first term. 
We then subtract the surplus energy that has been
included by integrating over the shaded region (see Fig. \ref{fig:curvwet}).

Since the force is short-ranged, we use the quadratic approximation to 
$h_{rot}$, $h_{rot}(\xi_1,\xi_2)=h_{rot}(0,0)
-\frac{\xi_1^2}{2}\kappa_1-\frac{\xi_2^2}{2}\kappa_2$ 
where we have assumed the axes to be aligned with the principle curvatures. 
Finally since $\kappa_{1,2}D\ll 1$ 
this wedge-shaped region is extremely thin close to
the origin and the remainder term can therefore be approximated by ignoring
the dependence of $w$ on $\xi_3$:
\begin{equation}
\Delta\Pi=-\iint {d\xi_1d\xi_2  (\frac{\xi_1^2}{2}\kappa_1
+\frac{\xi_2^2}{2}\kappa_2)w(\sqrt{\xi_1^2+\xi_2^2+D^2})}.
\end{equation}
This integral can be evaluated in polar 
coordinates:
\begin{eqnarray}
\Delta \Pi&=&\alpha\iint {\frac{rdrd\phi}{(r^2+D^2)^3}
(\frac{\kappa_1}{2}r^2\cos^2\phi+\frac{\kappa_2}{2}r^2\sin^2\phi)}\nonumber\\
&=&\alpha\frac{\kappa_1+\kappa_2}{2}\pi\int_0^\infty{\frac{r^3 dr}{(r^2+D^2)^3}}\nonumber
\\
&=&\alpha\frac{H\pi}{4D^2},
\end{eqnarray}
since the mean curvature $H$ is given by $\frac{1}{2}(\kappa_1+\kappa_2)$.
Upon combining this expression with the flat substrate result, we obtain
Eq. (\ref{eq:wet-7}).

\section{\label{app:CG}Consumer's Guide to Green's Functions on Compact Surfaces}
The ambiguity in the one-vortex
energy (Eq. (\ref{eq:ces}))
on the sphere also implies that there is no particularly
natural choice of a Green's function on the sphere. 
With so many choices
out there, you'll be greatful for this
 friendly guide to  help you focus on the important features
and possible pitfalls  of these different functions.

The first point you need to know is that all of them work pretty much
just as well, provided they are used consistently; one should not use
the single-vortex energy Eq. (\ref{eq:greenself-energy}) designed
to work with a different Green's function from the one used to calculate the
pair interaction Eq. (\ref{eq:greenpair-energy}).
The general definition of
of a Green's Function, broadened from Eq. (\ref{eq:BasicGreen}),
is that it is a symmetric function of two points
on the deformed sphere satisfying the equation
\begin{equation}
\nabla_{\mathbf{x}}^2\Gamma(\mathbf{x},\mathbf{y})=-\delta(\mathbf{x},\mathbf{y})+F(\mathbf{x}).
\label{eq:generalgreen}
\end{equation}
The only restriction on the function $F$ is that its integral over the
deformed sphere must equal $1$. (Integrating the Laplacian on the left shows
that there is no solution unless the right-hand side integrates to zero.)
The Green's function on a sphere,
$-\frac{1}{2\pi}\ln\frac{\mathcal{D}(\mathcal{X},\mathcal{Y})}{a}$ has 
$\frac{1}{4\pi}-\delta(\mathcal{X},\mathcal{Y})$ as its Laplacian\cite{lube92}.
Eq. (\ref{eq:generalgreen}) is a more versatile vision of what a Green's 
function should be, using a function $F$ in place of the constant.

That Eqs. (\ref{eq:greenself-energy}) and (\ref{eq:greenpair-energy})
give the correct net energy follows from the result proven in Sec.
\ref{sec:map} by conformal mapping to the unit sphere:
\begin{equation}
E(\{n_i,\mathbf{u}_i\})=\sum_{i<j}4\pi^2Kn_in_j
\Gamma_{sphere}(\mathbf{\mathcal{U}}_i,\mathbf{\mathcal{U}}_j)
-\sum_i\pi Kn_i^2\omega(\mathbf{u}_i)
\label{eq:confenergy}
\end{equation}
The interaction potential in this equation
\begin{equation}
\Gamma_c(\mathbf{x},\mathbf{y})=-\frac{1}{2\pi}\ln|\mathbf{\mathcal{X}}-
\mathbf{\mathcal{Y}}|
\label{eq:hildegaard}
\end{equation}
satisfies
\begin{eqnarray}
\nabla_{\mathbf{u}}^2\Gamma_c(\mathbf{u},\mathbf{u'})&=&e^{2\omega(\mathbf{u})})\nabla_{\mathbf{\mathcal{U}}}^2
\Gamma_{sphere}(\mathbf{\mathcal{U}},\mathbf{\mathcal{U'}})\nonumber\\
&=&e^{2\omega}(-\delta_{R}(\mathbf{\mathcal{U}},\mathbf{\mathcal{U'}})+
\frac{1}{4\pi})\nonumber\\
&=&-\delta_T(\mathbf{u},\mathbf{u'})+\frac{e^{2\omega(\mathbf{u})}}{4\pi}.
\label{eq:e2om}
\end{eqnarray}
In the first step, the scale factor is introduced to compensate for
the change from the reference to the target surface. In the second
step, the Laplacian of the sphere's Green's function $-\frac{1}{2\pi}\ln\frac{\mathcal{D}_{ij}}{a}$ is substituted.
In the third step, the $\delta$-function is transformed back to the target
surface.
The last line shows that $\Gamma_c$ is a Green's function as set out by
Eq. (\ref{eq:generalgreen}), which we
call the ``conformal Green's function."  The $F$-function that goes with
this Green's function
gets its spatial
dependence from the conformal factor.

The single particle potential $\omega$ in Eq. (\ref{eq:confenergy})
satisfies
\begin{equation}
\nabla_{\mathbf{u}}^2\omega=G_T(\mathbf{u})-e^{2\omega(\mathbf{u})}
\label{eq:antarctica}
\end{equation}
which follows from Eq. (\ref{eq:Greenland}) with the curvature of the
unit sphere, $G_R=1$, substituted.

Now any Green's function $\Gamma$ can be used to solve
Poisson's equation for any \emph{net-neutral} function $\rho$
on the target surface,
\begin{equation}
\nabla_{\mathbf{u}}^2
\iint d^2\mathbf{u'}\Gamma(\mathbf{u},\mathbf{u'})\rho(\mathbf{u'})=
-\rho(\mathbf{u})
\label{eq:poissonsphere}.
\end{equation}
This follows from Eq. (\ref{eq:generalgreen}).  It can be used to
derive Eqs. (\ref{eq:greenpair-energy}) and (\ref{eq:greenself-energy})
from their special case, the energy derived by conformal mapping. We
first use the Poisson-like integral to 
``solve" two special cases of Poisson's equation,
Eqs. (\ref{eq:e2om}) and Eq. (\ref{eq:antarctica}) in terms of the
arbitrary Green's function $\Gamma$.
Regarding $\mathbf{u'}$ as a constant in the former equation, we find
that
\begin{alignat}{1}
\Gamma_c(\mathbf{u},\mathbf{u'})=\Gamma(\mathbf{u},\mathbf{u'})-
\iint d^2&\mathbf{u''}
\Gamma(\mathbf{u},\mathbf{u''})\frac{e^{2\omega(\mathbf{u''})}}{4\pi}\nonumber\\
&+f(\mathbf{u'})
\label{eq:cvsgamma}
\end{alignat}
where $f(\mathbf{u'})$ is the constant left undetermined by the Poisson
equation.  Since both $\Gamma$ and $\Gamma_c$ are symmetric in $\mathbf{u},
\mathbf{u'}$,
$f(\mathbf{u'})=-\iint d^2\mathbf{u''}\Gamma(\mathbf{u'},\mathbf{u''})
\frac{e^{2\omega(\mathbf{u''})}}{4\pi}+C_1$ where $C_1$ is
a constant. Again applying Eq. (\ref{eq:poissonsphere}), this time
to Eq. (\ref{eq:antarctica}), implies that
\begin{equation}
\omega(\mathbf{u})=-\iint d^2{\mathbf{u''}}\Gamma(\mathbf{u},\mathbf{u''})
G_T(\mathbf{u''})-4\pi f(\mathbf{u})+C_2.
\label{eq:omvsgamma}
\end{equation}
(This is not really a \emph{solution}
 of the nonlinear Eq. (\ref{eq:antarctica})
since $\omega$ still appears on both sides of the equation.)
Rewriting the previous equations implies that
\begin{eqnarray}
&&\Gamma(\mathbf{u},\mathbf{u'})=\Gamma_c(\mathbf{u},\mathbf{u'})-f(\mathbf{u})-f(\mathbf{u'})-C_1\nonumber\\
&&U_G(\mathbf{u})=\omega(\mathbf{u})+4\pi f(\mathbf{u})-C_2\nonumber,
\end{eqnarray}
namely that $U_G$ and $\Gamma$ are related to $\omega$ and $\Gamma_c$
according to the energy-shuffling transformation
Eq. (\ref{eq:ces}) so that the more general expressions of Eqs. (\ref{eq:greenself-energy}) and (\ref{eq:greenpair-energy}) can be used in place
of Eq. (\ref{eq:confenergy})
to determine the energy.  The sum of the energies from
Eqs. (\ref{eq:greenself-energy}), (\ref{eq:greenpair-energy}) is
equal to the correct energy, Eq. (\ref{eq:confenergy}) up to a constant.
The arbitrary 
Green's function can also be used to find 
the flow pattern according to the formula
\begin{equation}
\chi(\mathbf{u})=\sum_{i=1}^N\frac{hn_i}{m}\Gamma(\mathbf{u},\mathbf{u}_i).
\label{eq:green-stream}
\end{equation}

There \emph{are} some advantages and disadvantages of different choices
for $F$ in Eq. (\ref{eq:generalgreen}).  Let us focus on the most popular
choices.  The ``standard Green's function"
is defined with $F=\frac{1}{A}$ ($A$ is the area of the surface)
 and is simply related to the eigenfunctions of the
Laplacian,
$\nabla^2\Psi_{\lambda}=-\lambda \Psi_{\lambda}$:
\begin{equation}
\Gamma_{s}(\mathbf{x},\mathbf{y})=
\sum_{\lambda\neq 0}\frac{1}{\lambda}\Psi_{\lambda}(\mathbf{x})^*
\Psi_\lambda(\mathbf{y}).
\label{eq:Gstandard}
\end{equation}
The ``pair Green's function" is defined via conformal mapping,
\begin{equation}
\Gamma_p(\mathbf{x},\mathbf{y})=\Gamma_{sphere}(\mathbf{\mathcal{X}},
\mathbf{\mathcal{Y}})+\frac{1}{4\pi}(\omega(\mathbf{x})+\omega(\mathbf{y}))
\label{eq:Gpair}
\end{equation}
and incorporates all the single-particle energy into the interaction
energy, so that $U_{pair}=0$.   This Green's function satisfies the most
elegant differential equation,
\begin{equation}
\nabla^2\Gamma_p(\mathbf{x},\mathbf{y})=
-\delta(\mathbf{x},\mathbf{y})+\frac{G(\mathbf{x})}{4\pi}
\label{eq:curvatureneut}
\end{equation}
Last, the ``conformal Green's function" (which was
our starting point) has $F=\frac{e^{2\omega}}{4\pi}$ (see Eq. (\ref{eq:confenergy})).

If you are looking for style in your Green's functions, I would choose
the pair Green's function. It is easy to calculate by conformal mapping
(Eq. (\ref{eq:Gpair})) but it can be defined without referring to $\omega$,
Eq. (\ref{eq:curvatureneut}), much preferable 
to the haphazard looking Eq. (\ref{eq:e2om}) defining the conformal Green's 
function.
The standard Green's function is stodgier
and does not handle well.  The methods for finding the standard Green's
function, Eq. (\ref{eq:Gstandard}) are more limited and, if one wants to use it,the best option
might be to derive it by using conformal mapping anyway:
\begin{equation}
\Gamma_s(\mathbf{x},\mathbf{y})=\Gamma_c(\mathbf{x},\mathbf{y})
-\frac{1}{A_{tot}}\iint
\Gamma_{c}(\mathbf{x},\mathbf{u})+\Gamma_c(\mathbf{y},\mathbf{u})d^2\mathbf{u}
+C_3.
\label{eq:conftostand}
\end{equation}
(This equation is derived analogously to Eq. (\ref{eq:cvsgamma}).)
On the other hand, there are always advantages to familiarity. In particular,
in the limit where
part of the deformed sphere is stretched out to infinity so that it actually
becomes a deformed plane, $\Gamma_s$ converges to the ordinary
Green's function of a noncompact surface, since
$\frac{1}{A}$ tends to zero. For a short summary of all the Green's
functions features and failings, see Table \ref{table:table}.

\begin{table}
\begin{tabular}{l|c|c|c}
&Calculability&Neutralizer&Limit\\ \hline
Conformal&+&-&-\\ \hline
Pair &+&+&-\\ \hline
Standard &-&+&+\\ \hline
\end{tabular}
\caption{\label{table:table}The advantages
and disadvantages of the Green's functions, as far as their ease of
calculation, simplicity of the neutralizing function $F$, and limiting behavior in
case the deformed sphere is stretched into a deformed plane.}
\centering\end{table}

\section{Approximations for Long Surfaces of Revolution\label{app:deimos}}
Let us start by determining the conformal map from the surface of revolution defined
by the equation $r=r(z)$, $z_s\leq z\leq z_n$, to the unit sphere.
We use the coordinates
$\phi,\sigma$ introduced in Sec. \ref{sec:zucchini} to parameterize the surface;
$\sigma$ is given by:
\begin{equation}
\sigma=\int_z^{z_n} \sqrt{1+(\frac{dr}{dz})^2}dz.
\label{eq:thread}
\end{equation}
The Cartesian coordinates are
 $x=r(\sigma)\cos\phi,y=r(\sigma)\sin\phi,z=z(\sigma)$, and hence
the metric is
\begin{equation}
ds^2=dx^2+dy^2+dz^2=d\sigma^2+r(\sigma)^2d\phi^2.
\end{equation}
If the map $(\sigma,\phi)\rightarrow(\Theta,\Phi)$ to the unit
sphere is to be conformal, then according to Eq. (\ref{eq:confdef}),
\begin{equation}
d\sigma^2+r(\sigma)^2d\phi^2=e^{-2\omega}(d\Theta^2+\sin^2\Theta d\phi^2).
\label{eq:confsphere}
\end{equation}
By symmetry, $\Phi=\phi$ and $\Theta=\Theta(\sigma)$ and is
independent of $\phi$ (see \cite{geomgenerate}
for the analogous use of symmetry on a rotationally symmetric bump on
a plane). By matching the coefficients of $d\phi$ and $d\sigma$ one finds
that $\frac{d\sigma}{r(\sigma)}=\frac{d\Theta}{\sin\Theta}$, or (after
integration):
\begin{equation}
\sin\Theta=\mathrm{sech}(\int_{\sigma}^{\sigma_{0}}
{\frac{d\sigma'}{r(\sigma')}})
\label{eq:smoosh}
\end{equation}
where $\sigma_{0}$ can be an arbitrary arc length. 
According to Eq. (\ref{eq:confsphere}), $\omega=\ln\frac{d\Theta}{d\sigma}$,
or
\begin{equation}
\omega=\ln\frac{1}{r(\sigma)}\mathrm{sech}\int_{\sigma}^{\sigma_{eq}}
\frac{d\sigma'}{r(\sigma')}.
\label{eq:omsphere}
\end{equation}

To determine the energies and flow patterns on a rotationally
symmetric surface, we use the ``Pair Green's function" Eq. (\ref{eq:Gpair}),
the Green's function which incorporates all of the energy into
interaction-energy terms. This Green's function can be found
using Eqs. (\ref{eq:omsphere}) and (\ref{eq:smoosh}); adding
$\omega$ at the sites of the vortices to
the Green's function on the sphere,\\
$-\frac{1}{2\pi}\ln\sqrt{2[1-\cos\Theta_1\cos\Theta_2-
\sin\Theta_1\sin\Theta_2\cos(\Phi_1-\Phi_2)]}$, and rearranging, gives
\begin{equation}
\Gamma_{pair}=-\frac{1}{2\pi}\ln\sqrt{\frac{2r(\sigma_1)
r(\sigma_2)}{a^2}[\cosh\int_{\sigma_1}^{\sigma_2}\frac{d\sigma}{r}-\cos(\phi_1-
\phi_2)]}
\label{eq:zucpair}
\end{equation}
The energy of a set of vortices is simple using the pair
Green's function (see the previous appendix), $E=\sum_{i<j} 4\pi^2n_in_j
K\Gamma_{pair}(\mathbf{u}_i,\mathbf{u}_j)$.  As an example
the energy of a vortex-antivortex pair at opposite sides 
of a circle of latitude
($\phi_1=\phi_2+\pi$ and $\sigma_1=\sigma_2=\sigma$) is
\begin{equation}
E=2\pi\ln\frac{2r(\sigma)}{a},
\label{eq:sqrt2}
\end{equation}
showing that the energy grows logarithmically with
the distance between the vortices in this case, as in Eq. (\ref{eq:opposite}).

To prove the azimuthal symmetry of the flows, note that according to
Eq. (\ref{eq:green-stream}), the flow velocity at $\mathbf{u}$ is
\begin{equation}
\mathbf{v}=\nabla_{\mathbf{u}}\times \sum_i\frac{n_i h}{m}\Gamma_{pair}
(\mathbf{u},\mathbf{u}_i) .
\label{eq:curlstream}
\end{equation}
Now if the vortices are all far from $\mathbf{u}$,
then the integral in (\ref{eq:zucpair})
is very large. Since $\ln (A+\epsilon)\approx A+\frac{\epsilon}{A}$ for
large $A$, the cosine term, the only one which depends on
the azimuthal angles, gives exponentially small contributions.
Therefore the flows can be calculated as in Sec. \ref{sec:zucchini},
by using the circulation quantization and the approximate
 azimuthal symmetry to determine the flow speeds. Alternatively,
we may calculate the velocity
 directly from Eq. (\ref{eq:zucpair}) with the help of
 the further approximation
that $\cosh x\approx\frac{1}{2}e^{|x|}$,
yielding
\begin{equation} \Gamma(\mathbf{u},\mathbf{u}_i)\approx -\frac{1}{4\pi}|\int_{\sigma}^{\sigma_i}\frac{d\sigma'}{r}|
-\frac{1}{4\pi}\ln\frac{r(\sigma)r(\sigma_i)}{a^2}.
\label{eq:pseudonewton}
\end{equation}
The first term gives the flow pattern of Eq. (\ref{eq:bandspeed}), after
a brief calculation using the neutrality constraint, while
the second term, when summed as in Eq. (\ref{eq:zucpair}), cancels
out also by neutrality.


For a surface (such as an ellipsoid)
where the xy-plane is a plane of symmetry, our results will
simplify if we make the choice $\sigma_0=\sigma_{eq}$ in
Eq. (\ref{eq:smoosh}) where $\sigma_{eq}$ is the arclength corresponding
to the equator, at $z=0$. In this case, the conformal
map takes pairs of
antipodal points on the deformed
surface to antipodal points on the sphere. (Since antipodal
points $(\sigma_1,\phi_1)$, $(\sigma_2,\phi_2)$
are points at opposite ends of a diameter of the surface, 
$\sigma_2=2\sigma_{eq}-\sigma_1,\phi_2=\pi+\phi_1$.)
If we consider the
interaction energy of a pair of antipodal points, we find
according to Eqs. (\ref{eq:conformalpair-energy}) and 
(\ref{eq:conformalself-energy}),
\begin{equation}
\frac{E_{antipodal}}{K}=2\pi\ln\frac{2}{a}-2\pi\omega(z_1)
\label{eq:worldview}
\end{equation}
Whether the two vortices are at opposite tips or at opposite
ends of the equator, their image vortices are always at the same distance
on the unit sphere, so the first term, the interaction energy
of the images, is a constant.
This gives another illustration of 
the folly of making a strict separation between intervortex 
and curvature-vortex
interactions.
One would like to think that the
growth of the energy as the two vortices are separated on an elongated
surface is 
due to the attraction between them. But Eq. (\ref{eq:worldview}) 
shows that it can also be interpreted as resulting from the single
particle potential $\omega$.

Let us now turn to the problem of describing the equilibrium positions of
a pair of vortices on a rotating ellipsoid. 
Both the transitions at $\Omega_a$ and $\Omega_b$
can be understood only with a more accurate version of the force than
the band-force approximation, Eq. (\ref{eq:bandforce1}).
The error in the approximation is important when the vortices are near
the poles of the ellipsoid or, as just illustrated, near each other. 
We will assume
that 
the aspect ratio of the ellipsoid $\alpha=\frac{H}{R}$ is very large.
The equation for the ellipsoid
can be expressed in terms of $\alpha$ in the form:
\begin{equation}
r=R\sqrt{1-\frac{z^2}{\alpha^2 R^2}}.
\label{eq:tubey}
\end{equation}
The energy of a vortex-antivortex pair according to Eq. (\ref{eq:Erot}) is
\begin{equation}
E=E_{rest}+E_{\Omega}.
\label{eq:contributions}
\end{equation}
Here
the energy of the flow pattern, or ``resting energy," is the energy
of the vortices on a stationary ellipsoid,
$E_{rest}=-4\pi^2 K\Gamma_{pair}(\sigma_1,\phi_1;\sigma_2,\phi_2)$.
The ``rotation energy"
 $E_{\Omega}$ is
given by Eq. (\ref{eq:Lzvortex}). Both energies are functions of a single
variable, the
distance $s$ between the two vortices along the surface, if we 
assume that the vortices are at $(\sigma_1,\phi_1)
=(\sigma_{eq}-\frac{s}{2},0)$ and 
$(\sigma_2,\phi_2)=(\sigma_{eq}+\frac{s}{2},0)$. 
These relationships assume
that the vortices are situated symmetrically about the $xy$-plane;
the 
vortices will have equal azimuthal
angles in order to minimize the energy of the flow pattern.

\begin{figure}
\includegraphics[width=.47\textwidth]{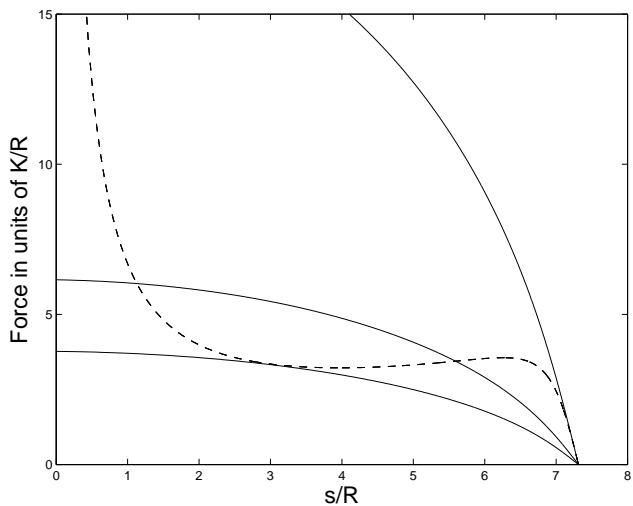}
\psfrag{oma}{$\Omega=\Omega_a$}
\psfrag{omb}{$\Omega=\Omega_b$} 
\caption{\label{fig:ski}The resting force (dashed line) and minus the
rotational force (solid
lines) between two vortices on an ellipsoid of aspect ratio $3.5$.
The rotational forces are illustrated for $\Omega_a$ (top
curve),$\Omega_b$ (the bottom curve) and
an intermediate value of $\Omega$. For smaller values of $\Omega$, the
attraction between the vortices always overcomes the rotational force,
causing them to annihilate.
For larger values of $\Omega$, the rotational force overcomes the attraction,
causing the vortices to move to opposite poles.}
\end{figure}

The equilibrium position of a pair of vortices is determined by balancing
the rotational force and resting force
acting on one of them.  The resting 
force $F_{rest}$ on vortex $1$
is derived from the kinetic energy of the flow pattern and is
positive since it pulls
the vortices toward each other in order to decrease the width of
the band of moving fluid between them.  The rotational force $F_{rot}$
is negative since it pulls the vortices
toward the poles of the ellipsoid (and away from each other)
in order to increase the total
angular momentum of the flow.
At equilibrium the rotational and resting forces on the vortices balance,
as we can see by differentiating Eq. (\ref{eq:contributions}) to obtain
$0=\frac{dE_{rest}}{ds}+\frac{dE_{\Omega}}{ds}$ or equivalently
\begin{equation}
F_{rest}(s)=-F_{\Omega}(s)
\label{eq:forcebalance}
\end{equation}
where $F_{rest}$ and $F_{\Omega}$ are the resting and rotational
forces on the vortices. (A short calculation shows that the force
on one of the vortices $-\frac{dE}{d\sigma_1}$ is equal to $\frac{dE}{ds}$,
since the energy change produced by moving one vortex an infinitesimal
distance is the same as the energy change produced by moving both
vortices half the distance.)  The equilibrium positions can be found by
graphing $F_{rest}$ and $-F_{\Omega}$ as in Fig. \ref{fig:ski} and
finding the intersection points. 
The exact expression
for the resting force can be found by differentiating
Eq. (\ref{eq:zucpair}) to obtain
\begin{equation}
F_{rest}=\frac{\pi K}{r(\sigma_1)}\left(\mathrm{coth}[\int_{\sigma_1}^
{\sigma_{eq}}\frac{d\sigma'}{r(\sigma')}]
-\frac{1}{\sqrt{1+(\frac{dz}{dr})^2}}\right).
\label{eq:crazyforce}
\end{equation}
The rotational force is given exactly by Eq. (\ref{eq:Frotsequel}). 

Fig. \ref{fig:ski} shows the resting force and minus the rotational force 
 on one of the vortices for $\Omega=\Omega_b$, $\Omega=\Omega_a$ and
for an intermediate value of the frequency. 

Stability of the equilibria illustrated in Fig. \ref{fig:ski}
can be determined by considering the
direction in which
the resting force curve crosses the rotational force curve. The
middle point of the three equilibrium
points at the intermediate frequency is a stable equilibrium because
the resting force curve
crosses the rotational force curve from bottom to top. This implies that
 if the vortices
fluctuate away from each other (increasing $s$), 
then the resting force becomes stronger
than the rotational force and pulls them back together.

Let us consider how the stable equilibrium disappears at $\Omega_b$.
As $\Omega$ is lowered the stable and unstable equilibrium come
together and then ``annihilate" when the rotation-force curve detaches
from the resting force curve,
as illustrated by the lowest curve in Fig. \ref{fig:ski}, which
corresponds to $\Omega=\Omega_b$. Since the rotation force
and resting force
curves are tangent
at $\Omega_b$, the frequency $\Omega_b$ and separation of the
vortices $s_b$ at this transition point can be 
determined by solving Eq. (\ref{eq:forcebalance})
simultaneously with
\begin{equation}
F_{rest}'(s_b)=-F_{\Omega}'(s_b).
\label{eq:jerkbalance}
\end{equation}
When $\alpha\gg 1$, we will be able to avoid solving
simultaneous equations since the value of $\Omega_b$
is already determined by Eq. (\ref{eq:newadventure}).
Using this result, we will be able to
solve Eq. (\ref{eq:jerkbalance}) for $s_b$.

The simple band approximation to the force, Eq. (\ref{eq:bandforce1}),
suggests that the vortices move continuously toward one another as
$\Omega$ is decreased, annihilating at the equator.  
Substituting the expression for the critical
frequency that is implied by the band
model, Eq. (\ref{eq:newadventure}), into Eq. (\ref{eq:prize})
in fact implies that $s_b=0$, which is incorrect.
The band approximation fails because it 
implies that the force between
the vortices decreases \emph{monotonically} as the vortices approach one 
another.
Though in conflict with our intuition from the plane, this result
is correct over the large middle range of the resting force curve
in Fig. \ref{fig:ski}. As the rotational confinement weakens,
the vortices get closer together, and the resting force weakens too, preserving
the equilibrium.
However the resting force starts \emph{increasing}
 strongly as the
vortices approach one another, because the vortices start to feel
one anothers' asymmetric flow fields.  This force will certainly
overcome the rotational force when the rotational confinement decreases further.
(Actually,
Eq. (\ref{eq:jerkbalance}) implies that $s_b$ does not correspond
exactly to the maximum of $F_{rest}$ because the 
rotational confinement is not a constant force field.)


We can derive the corrections to the force
 from Eq. (\ref{eq:crazyforce});
if $\alpha$ is large, we may neglect the second term and
assume that
\begin{equation}
\int_{\sigma_1}^{\sigma_{eq}}\frac{d\sigma'}{r(\sigma')}\approx
\frac{\sigma_{eq}-\sigma_1}{R}
\label{eq:deltafn}
\end{equation}
 since
the radial profile of the ellipsoid, Eq. (\ref{eq:tubey}), is
slowly varying. We then obtain the approximation that is valid when the
vortices are close (compared to $\frac{r}{r'}$, the characteristic
distance for variation of the radius).
%
%
%
\begin{equation}
F_{rest}\approx \frac{\pi K}{r(\sigma_1)}\mathrm{coth}(\frac{s}{2R})
\label{eq:uncoth}
\end{equation}
Notice that the force diverges as $\frac{2\pi K}{s}$ when the vortices
are close together (as in the plane)
and approaches Eq. (\ref{eq:bandforce1}) exponentially fast
as the vortices move apart; this generalizes the band model
approximation to the case where the two vortices may be close together.
As we will see, for a large value of $\alpha$, $s_b>>R$
at the moment when the vortices annihilate. We therefore
simplify Eq. (\ref{eq:uncoth}) by making another approximation,
$\mathrm{coth\ } x\approx 1+2e^{-2x}$. 
Then an approximate version of Eq. (\ref{eq:jerkbalance}) that
is derived from Eqs. (\ref{eq:uncoth}), (\ref{eq:Frotsequel})
reads
\begin{equation}
\frac{\pi K}{2}\frac{1}{R^2}\frac{dr}{d\sigma_1}|_{s_b}-
2\pi K\frac{e^{-\frac{s_b}{R}}}{R^2}
=-\frac{\pi\Omega_b\hbar\rho_s}{m}\frac{dr}{d\sigma_1}|_{s_b}
\label{eq:erggh1}
\end{equation}
The first term describes the decrease of the resting force due to
the variation in $r(z)$. The second term results from the exponentially
decaying portions of the flow fields.
(We are replacing $r(\sigma_1)$ by $R$ whenever that is accurate enough
since the width
of the ellipsoid is slowly varying.  Of course, the
slow variation of $r(\sigma_1)$ is important in some terms; 
the resting force initially decreases as $s$ decreases because the 
band approximation to the force
decreases with increasing circumference.)
Using Eq. (\ref{eq:newadventure}) for $\Omega_b$ in Eq. (\ref{eq:erggh1})
gives
\begin{equation}
\frac{dr}{d\sigma_1}|_{s_b}=2e^{-\frac{s_b}{R}}
\label{eq:erggh2}
\end{equation}
In order to evaluate the left-hand side, we note that 
$\sigma_1=\sigma_{eq}-\frac{s}{2}\approx\sigma_{eq}-z$ 
aside from terms of order $\frac{1}{\alpha^2}$
since the sides of the ellipsoid are nearly vertical near the equator.
Therefore Eq. (\ref{eq:tubey}) implies that 
$\frac{dr}{d\sigma_1}=\frac{s_b}{2\alpha^2R}$.

Rearranging Eq. \ref{eq:erggh2} now gives
\begin{equation}
s_b=R\ln\frac{4\alpha^2 R}{s_b}.
\label{eq:erggh5}
\end{equation}
which can be solved by substituting it into itself.
The first iteration gives
\begin{equation}
s_b=R\ln 4\alpha^2-R\ln\ln\frac{4\alpha^2 R}{s_b}.
\label{eq:erggh6}
\end{equation}
Since the second term has two logarithms in it, it is smaller than the first in
the limit where $\alpha\rightarrow\infty$, so finally 
\begin{equation}
z_b\approx R\ln\alpha,
\label{eq:erggh7}
\end{equation}
(since $z_b$, the distance from a vortex to the equatorial plane,
is approximately half the distance between the vortices). We
have justified Eq. (\ref{eq:logs}). 
Two iterations of Eq. (\ref{eq:erggh5}) give $z_b=R\ln\alpha-\frac{1}{2}\ln
\frac{\ln\alpha}{2}$; the error for \emph{this} approximation actually
approaches $0$ for large $\alpha$.
The exact result can be found
by computer, but the approximate result is reasonable even at
 $\alpha=5$, where $\frac{z_b}{R}=1.8\approx\ln 5=1.6$.

The height $z_b$ depends only logarithmically on $\alpha$  because
the extra short-distance vortex-vortex interaction decays exponentially
and would not be strong enough to pull the vortices together if $z_b$ were
very large.
(Check this by substituting our final result, Eq. (\ref{eq:erggh7}),
into Eq. (\ref{eq:erggh1}).   All the terms, the ones from
the band model as well as the exponential correction, 
have the same basic dependence on $\alpha$.)
To see that the approximations we have made are valid, one has to
calculate $F_{rest}'(s)$ from the exact expression Eq. (\ref{eq:crazyforce}).
The resulting expression can be simplified by dropping various terms, which
mostly have a relative size of 
$\frac{1}{\alpha^2}$ and $(\frac{\ln\alpha}{\alpha})^2$;
the reason is that $\frac{z_b}{H}=\frac{\ln\alpha}{\alpha}$ so the
vortices are proportionally very close to the equator, and again $r(z)$ can
be replaced by $R$. (This also justifies
the approximation
in Eq. (\ref{eq:deltafn}) where the integrand is replaced by
a constant.) One particularly large term, resulting
from the second term of Eq. (\ref{eq:crazyforce}), has been neglected
in Eq. (\ref{eq:erggh1}), 
but the neglected term, $\frac{K\pi r''(\sigma_1)}{2r(\sigma_1)}$
is still of relative order $\frac{1}{\ln\alpha}$.

\begin{figure}
\includegraphics[width=.47\textwidth]{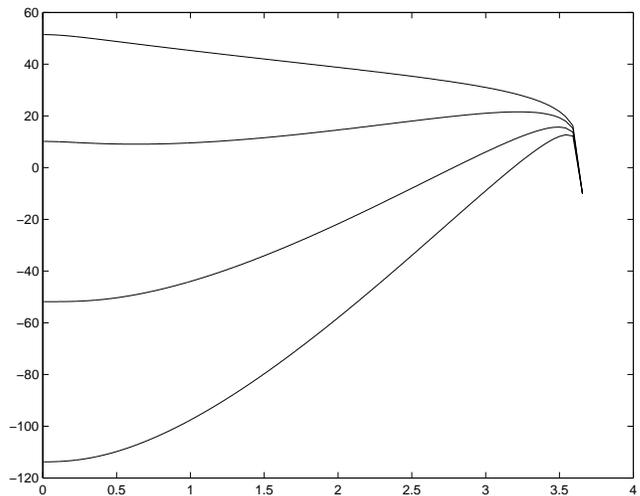}
\caption{\label{fig:ruffle}
Illustration of the combined rotational and
resting energies near $\Omega_a$,
for $H=3.5 R$ and
$\frac{\hbar\Omega}{m}=0,1.2,3.1,4.9 R^{-2}$. The first of these
is the uppermost curve. The third curve corresponds to $\Omega=\Omega_a$.
Although the second curve looks practically flat on this scale, it has
a curvature of about $\frac{10K}{R^2}$ at its off-center minimum.}
\end{figure}

Now we turn to the critical frequency $\Omega_a$ where the
vortices move to the poles. The band model also
requires a correction in order for this transition
to be described correctly.  In fact  Eq. (\ref{eq:prize}) would imply
that the vortices never exactly reach the tips of the ellipsoid even
as $\Omega\rightarrow \infty$, and thus $\Omega_a=\infty$. In fact,
the band force on the left hand side of Eq. (\ref{eq:heartgrowsfonder}),
which
approaches infinity at the poles, cannot be balanced by the
rotational force at a finite frequency. Of course, the exact force
approaches zero rather than infinity at the pole (see Fig. \ref{fig:ski}).
The value of $\Omega_a$ may be derived from the condition
that the actual resting force curve and the rotation force curve have
to be tangent at the origin, as for the uppermost ($\Omega=\Omega_a$)
curve in
Fig. \ref{fig:ski}. We therefore have to find when $s=2\sigma_{eq}$
satisfies
Eq. (\ref{eq:jerkbalance}).
Linearizing Eq. (\ref{eq:crazyforce}) near $\sigma_1=0$ to 
find the derivative of the force implies that
$\Omega_a$ is given by
\begin{equation}
\frac{m\Omega_a}{\hbar}=\frac{\alpha^2\kappa^2}{4}
+\frac{1}{R^2}e^{-2\int_0^R(\frac{d\sigma}{dr}-1)\frac{dr}{r}}
\label{eq:oma}
\end{equation}
where $\kappa$ is the curvature at the tip of the ellipsoid.
The critical frequency is larger than the result $\hbar\frac{\kappa^2}{4m}$,
derived in Section \ref{subsec:single},
for a bump with the same curvature
 because the rotational confinement must overcome the mutual
attraction of the vortices as well as the repulsion of the vortices from
the curvature. For an ellipsoid with a large value of $\alpha$, the
correction term is unimportant, so
\begin{equation}
\Omega_a\approx\frac{\alpha^2\hbar}{4mR^2}
\label{eq:BOring}
\end{equation}
The transition can be visualized using the energy curves
illustrated in Fig. 
\ref{fig:ruffle}, where the local minimum of the energy function moves
away from the axis as $\Omega$ is decreased through $\Omega_a$.

\begin{figure}
\includegraphics[width=.45\textwidth]{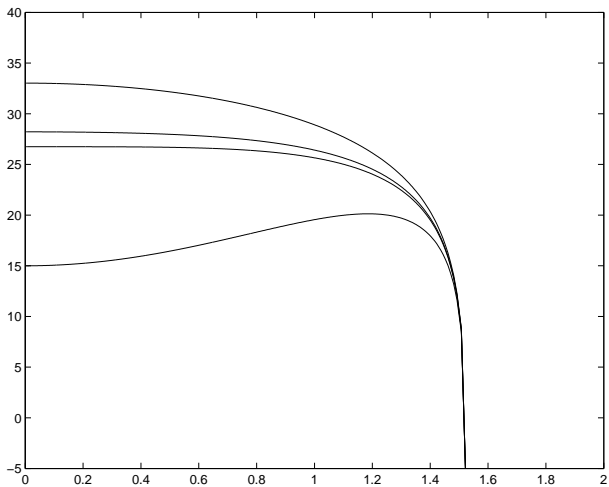}
\caption{\label{fig:puffy} The combined rotational and kinetic energy
for vortices on a sphere, where there are no
stable off-axis positions. The energy is
graphed as a function of $\sigma_1=2\sigma_{eq}-\sigma_2$
in units of $R$. From the top, the
curves correspond to $\Omega=0,.4,.52,1.5 R^{-2}$. The \emph{third} of these 
corresponds to $\Omega=\Omega_a=\Omega_b$.}
\end{figure}
There is an aspect ratio $\alpha_c$ below which there are no off-center local
minima, for any rotation speed. That is, when the angular
velocity is decreased enough,
a vortex-antivortex pair initially at the poles \emph{immediately} moves
to the equator and annihilates.
This situation is illustrated for a sphere in Fig. \ref{fig:puffy}.
The value of $\alpha_c$ can be determined numerically, and is $1.33$.
One simply graphs the total energy at $\Omega=\Omega_a$ (as given
by the exact expression, Eq. (\ref{eq:oma}))
and checks whether there is an energy barrier
or not. At $\Omega>\Omega_a$, a pair of vortices at the poles will be
stable. If there is no barrier, as in Fig. \ref{fig:puffy},
 slightly decreasing $\Omega$ will cause
these vortices to leave the poles and annihilate each other. If there
is a barrier, as in Fig. \ref{fig:ruffle}, 
slightly decreasing $\Omega$ will create an off-center
local minimum. This can be seen from the energy curves: there is a local
maximum at the origin, because $\Omega<\Omega_a$, and there is also
a local maximum at the top of the barrier. Therefore 
the vortex can find a local minimum somewhere in between.

\section{\label{app:geomineq}Derivations of Bounds Valid Even for Strong Distortions}

The results of Sec. \ref{sec:geomineq} can be derived from theorems on
``univalent" analytic functions. We will state these theorems here and
derive the limits on the geometric force from them.
(See \cite{rudin} for the proofs.)
An abstract example
of the type of question
these theorems address is the following. Let $f(t)$ be an analytic
function defined by the following series:
\begin{equation}
f(t)=t+a_2 t^2+a_3 t^3+\cdots
\label{eq:Taylor}
\end{equation}
Suppose this series converges out to radius 1, at least.
If one of the coefficients, maybe $a_{6}$, is much larger than the
rest, then the function is dominated by the $t^{6}$ behavior, and most points
in the range of the function will occur six times as values of the function.
Therefore, if one is looking for a univalent function (a function which is one-to-one inside the unit circle) then there will be upper limits on the sizes of
the $a_n$.  A challenging mathematical problem
is ``What are the maximum sizes for the $a_n$'s?"
The answer
(proved by De Branges) is that $|a_n|\leq n$, and that the function
$\frac{t}{(1-t)^2}$ attains the maximum value for every Taylor series
coefficient
simultaneously.
To find the upper bound on the vortex force in a flat disk, we will
use only the bound
\begin{equation}
a_2\leq 2
\label{eq:univalent}
\end{equation}
which has a simpler proof \cite{rudin}. Note that the conditions
of this theorem do
not require that the function remains one-to-one \emph{outside} the unit
circle.  For example, 
the function $t+.1 t^2$ satisfies the conditions of the theorem 
although it
takes on the value zero at $t=0,-10$. The analyticity of $f(t)$ is allowed to
break down as well beyond a radius of 1.

 Similar problems can be stated for functions $g(t)$ defined \emph{outside}
 of
the unit circle, with expansions of the form
\begin{equation}
g(t)=t+\frac{b_1}{t}+\frac{b_2}{t^2}+\cdots.
\label{eq:recip}
\end{equation}
To make the predictions about the quadrupole force due to a bump in a plane
we will use the Area Theorem \cite{rudin} which
states that, if $g$ is one-to-one and analytic outside the unit circle,
and
\begin{equation}
g(t)=t+\frac{a_1}{t}+\dots
\label{eq:areanorm}
\end{equation}
then
\begin{equation}
a_1\leq 1.
\label{eq:areatheorem}
\end{equation}

To prove Eq. (\ref{eq:limit1}), one
just realizes that the assumption means that a part of the surface
has the same geometry as a radius $R$ disk \emph{in the plane}
with a vortex at the center.
We can introduce a coordinate system on this portion of the surface by
introducing Cartesian coordinates $u,v$ (with $w=u+iv$)
on the disk in the plane, and then
mapping these coordinates isometrically to the surface.  This mapping
is different from the conformal mapping $C$ used to calculate vortex
energies.  To relate them, let $\mathcal{Z}=\mathcal{X}+i\mathcal{Y}$
where $\mathcal{X},\mathcal{Y}$ are the coordinates of the conformal image of
the surface. Then Eq. (\ref{eq:confdef}) takes the form
$du^2+dv^2=e^{-2\omega}(d\mathcal{X}^2+d\mathcal{Y}^2)$ and it follows
that $\mathcal{Z}(w)$ is a conformal map from a portion of the plane to
itself, hence an analytic function of $w$
on the circle of radius $R$ (say $\mathcal{Z}=c_1w+c_2w^2+\dots$).
 Furthermore, rewriting the
expression for the scaling of lengths as $|dw|^2=e^{-2\omega}
|d\mathcal{Z}|^2$, we see that
\begin{equation}
\omega=\ln|\frac{d\mathcal{Z}}{dw}|.
\label{eq:factorcplex}
\end{equation}

 We now define
\begin{equation*}
f(t)=\frac{\mathcal{Z}(Rt)-\mathcal{Z}(0)}{Rc_1}.
\end{equation*}
Then $f$ is a one-to-one analytic function on the unit circle (which
is scaled by $t\rightarrow Rt$ into the radius $R$ circle).
Since $f(0)=0,f'(0)=1$, $f$ has
the form of Eq. (\ref{eq:Taylor}), so Eq. (\ref{eq:univalent}) implies
\begin{equation}
2\geq 
R\frac{c_2}{c_1}
\label{eq:coefficients}
\end{equation}
Now the force on the vortex is
$\pi K\nabla\omega(0)$ which can be expressed in terms of the coefficients
of $\mathcal{Z}$'s Taylor series by means of Eq. (\ref{eq:factorcplex}):
$\mathbf{F}=2\pi K(\Re \frac{c_2}{c_1},-\Im \frac{c_2}{c_1})$.
The upper bound, Eq. (\ref{eq:limit1}), follows from Eq. 
(\ref{eq:coefficients}).

\begin{figure}
\psfrag{theta}{$\theta$}
\psfrag{D}{$D$}
\psfrag{Dsin}{$R_{max}=D\sin\frac{\theta}{2}$}
\includegraphics[width=.45\textwidth]{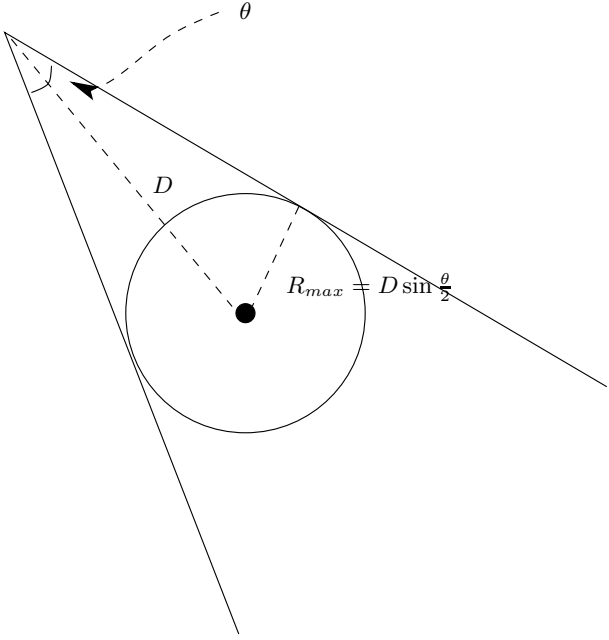}
\caption{\label{fig:sonicboom}Construction of the 
largest circle centered at a point on a cone with
$\theta<\pi$.  The cone is cut open and flattened so that
the center is on the bisector of the angle.}
\end{figure}

Let us now see whether the bound just proven can be improved at all; i.e., whether the
ratio of the force
on a singly quantized vortex to $\frac{K}{R}$ can ever be as big as
$4\pi$.
For example,
for vortices on cones, the ratio of the force to $\frac{K}{R}$
is maximal in the limit where the cone angle $\theta\rightarrow 0$. 
To find this ratio, we must take $R$ to be the radius of a disk
centered at the vortex which is \emph{flat} and \emph{non-self-intersecting}; taking
$R=R_{max}$, the radius of a disk which is as large as possible, maximizes the ratio
we are interested in.
The radius $R_{max}$
can be found by imagining the disk expanding out from the vortex.
If $\theta>\pi$, $R_{max}$=$D$ because
the first calamity that befalls the disk as it expands is that it starts overlapping
the cone's apex. But if $\theta<\pi$, then the disk
overlaps itself before this as one can see on the unfolded
version of the cone illustrated in Fig. \ref{fig:sonicboom}. Some
simple trigonometry shows that
$R_{max}=D\sin\frac{\theta}{2}$.  
Eq. (\ref{eq:cone}) shows that for small $\theta$, $\frac{|\mathbf{F}|R}{K}\rightarrow \pi^2$, which
is a little less than $4\pi$.

\begin{figure}
\includegraphics[width=.47\textwidth]{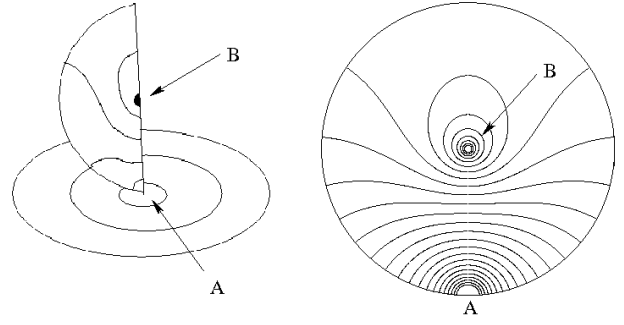}
\caption{\label{fig:fakecoin} (a) The surface which contains an isometric
disk of radius $R$ and has the maximum geometrical force. A semicircle (with
films on both sides) is connected by a neck $A$ to a plane.
If a single
vortex is placed at $B$, the force on the vortex approaches $\frac{4\pi K}{R}$
as the edges of the surface becomes sharper. (b) An unfolded image of the
flow pattern.}
\end{figure}

There \emph{is} a surface that saturates the original bound, though; this surface
is illustrated in
Fig. \ref{fig:fakecoin}. The surface is obtained
by folding a disk in half and sealing it shut except for a very
small opening at one end of the diameter. This opening is then connected
to an infinite plane. The top part of the substrate is a
semi-circular slab with the superfluid layer laminating both sides so that the
helium spreads out to the plane.  The topology of the helium film
is still that of a plane.
A vortex placed at the center of the disk, $B$, saturates the bound;
it is attracted by the negative curvature of the neck joining the plane
and the disk and is repelled by the positive curvature at the top of the fold.

To show that this surface (the ``calzone surface") saturates the bound,
we will find the force on the vortex using conformal
mapping.  Instead of mapping the entire surface to a reference plane, we can
just map the folded disk portion of the surface. The flow patterns
on the two portions of the surface are uncorrelated
when the neck becomes infinitely small, aside from requirements imposed
by the circulation's invariance.
The circulation around
any curve on the plane enclosing the neck will be $\frac{h}{m}$ because
the vortex in the disk region is inside it, and the flow pattern
on the planar base will not be sensitive to the location of this vortex
because the neck is so narrow.  It will consist of a set of concentric
circles, representing a flow whose energy is independent
of the position of the vortex.
The force does not depend on this portion of the flow,
so  the two portions may be dismantled at the neck. As illustrated
in Fig. \ref{fig:fakecoin}, the neck now turns into the core of a second
vortex, at
point $A$ of the folded disk.  The folded disk now has the topology of
a sphere, satisfying the neutrality condition because the two vortices
are equal and opposite.
The
map $\mathcal{Z}=\frac{R^2w}{(R-w)^2}$ on the radius $R$ disk can be used
to relate the folded disk to a reference plane, since the points on
the circle which fold together,
$w=Re^{\pm i\phi}$, both map to the same point of the real axis in the plane.
Since
the vortex at $B$ maps to infinity, the force on the vortex at $A$ can
be calculated from the geometric potential alone (without any interaction
terms), giving
$\pi K \frac{\mathcal{Z}''^*(0)}{\mathcal{Z}'^*(0)}=\frac{4\pi K}{R}$. Also
the flow
pattern illustrated in the figure
can be found by mapping the concentric circles
centered around the origin in the $\mathcal{Z}$ plane to the disk
using the function $w(\mathcal{Z})$.

The result Eq. (\ref{eq:limitq})
about the long range force due to a bump contained
inside of a radius $R$ but with an arbitrary height and arbitrary
curvatures follows (by an argument similar to the one used for
the first inequality) from Eq. (\ref{eq:areathm}).
The conformal mapping takes the flat part of the surface (a plane
with a radius $R$ hole parameterized by the complex variable $w$) in
a one-to-one fashion to  a reference plane with a hole of some
distorted shape.  As above,
this function is analytic and $\omega=\ln|\frac{d\mathcal{Z}}{dw}|$.
By rescaling one can ensure that $\mathcal{Z}\sim w$ at infinity.
Applying the area theorem to
$g(t)=\frac{1}{R}\mathcal{Z}(Rt)$
shows that
\begin{equation}
\mathcal{Z}(w)=w+R^2\frac{a_1}{w}+\dots
\end{equation}
where $a_1\leq 1$. Expand $E=\frac{\pi K}{2}\Re\ln\frac{d\mathcal{Z}}{dw}$
for large $w$ to find the large distance form of the energy:
\begin{equation}
E\sim -\pi K R^2\Re \frac{a_1}{w^2};
\label{eq:quadrupolecplex}
\end{equation}
it follows that $\mu_2=\pi K R^2|a_1|$ and $\gamma_2=\arg(a_1)+\pi$ in Eq.
(\ref{eq:quadrupole}) and the bound on the quadrupole
moment $\mu_2\leq \pi K R^2$  follows from the bound on $a_1$.

Now we can also ask what type of bump maximizes the quadrupole moment. It turns
out that the value $\pi K R^2$ \emph{cannot} be attained by any surface which
is flat outside a circle of radius $R$. There \emph{is} a surface which consists
of a bump surrounded by a surface \emph{isometric} but not \emph{congruent}
to $K$, the plane with a circle of radius $R$ removed. This surface
is gotten from $K$ by sealing opposite sides of the circle together to make
a mountain ridge.

The reason this surface has the biggest quadrupole moment is because
its conformal mapping to the plane is the function that maximizes $a_1$.
According to the area theorem
the only one-to-one analytic function on $K$ for which $a_1=1$ is
\begin{equation}
\mathcal{Z}(w)=w+\frac{R^2}{w}
\label{eq:areathm}
\end{equation}
This function maps both points
$Re^{\pm i\phi}$ to the same point in the reference
plane, so any flow pattern on the target plane will still be continuous
when the two edges of $K$ are sealed.

This quadrupole-maximizing surface does not contain a \emph{flat}
copy of $K$.  Hence an open question is to find the largest value of $\mu_2$ for a bump
in a plane which is actually flat outside a radius of $R$, as well as
the shape of the bump which has this maximum quadrupole moment.

\end{document}